\documentclass[12pt,a4wide,epsf]{article}

\usepackage{epsfig}
\usepackage{graphicx}
\usepackage[german, english]{babel}
\usepackage{amsmath}
\usepackage{amssymb}
\usepackage{ifthen}
\usepackage{epsfig}

\newcounter{fig}   \newcommand{\lbfig}[1]{\refstepcounter{fig}
\label{#1} } 
\newcommand{\vphi}{\varphi}

\begin{document}

\title{
Gravitating Monopole-Antimonopole Systems
at Large Scalar Coupling
\vspace{0.0truecm}
\author{
\vspace{0.5truecm}
{\bf Jutta Kunz, Ulrike Neemann, Yasha Shnir}\\
Institut f\"ur Physik, Universit\"at Oldenburg, Postfach 2503\\
D-26111 Oldenburg, Germany\\
}
}

\date{\today}

\maketitle
\vspace{0.0truecm}

\begin{abstract}
We discuss static axially symmetric 
solutions of $SU(2)$ Einstein-Yang-Mills-Higgs theory
for large scalar coupling $\lambda$. 
These regular asymptotically flat solutions represent 
monopole-antimonopole chain and vortex ring solutions,
as well as new configurations, present only for larger values of $\lambda$.
When gravity is coupled to the Yang-Mills-Higgs system, 
branches of gravitating solutions emerge from the flat-space solutions,
and extend up to critical values of the gravitational coupling constant.
For small scalar coupling only two branches of gravitating solutions exist, 
where the second branch connects to a generalized Bartnik-McKinnon solution.
For large scalar coupling, however, a plethora of gravitating branches can be
present and indicate the emergence of new flat-space branches. 
\end{abstract}

\vspace{1.0truecm}

\vfill\eject

\section{Introduction}

The 't Hooft-Polyakov monopole \cite{mono} represents
but the simplest of a rich variety of regular classical
non-perturbative finite energy solutions
of $SU(2)$ Yang-Mills-Higgs (YMH) theory.
Besides this spherically symmetric solution,
axially symmetric multimonopoles \cite{WeinbergGuth,RebbiRossi,mmono,KKT},
monopole-antimonopole pairs (MAPs) \cite{Taubes,Rueber,mapKK},
monopole-antimonopole chains (MACs) \cite{KKS}, 
and vortex ring solutions \cite{KKS,KNS} are known,
as well as multimonopole solutions with only discrete symmetries
\cite{plato}.

In the Bogomol'nyi-Prasad-Sommerfield (BPS) limit 
of vanishing Higgs potential,
the repulsive and attractive forces between two monopoles exactly compensate
for any separation, thus BPS monopoles experience no net interaction 
\cite{Manton77}. 
As the scalar field becomes massive,
the fine balance of forces between the monopoles is broken,
since the attractive Yukawa interaction becomes short-ranged.
Consequently non-BPS monopoles experience repulsion \cite{KKT}.
This repulsion can be overcome by the inclusion of gravity,
which allows for bound multimonopoles \cite{HKK}.

In monopole and multimonopole solutions, the nodes of the Higgs field
are associated with the location of the magnetic charges.
This also holds for monopole-antimonopole pair and chain solutions,
where $m$ monopoles and antimonopoles
are located symmetrically on the symmetry axis in alternating order
\cite{Taubes,Rueber,mapKK,KKS}.
When the magnetic charge $n$ of the individual
monopoles and antimonopoles becomes large,
the balance between the repulsive and attractive
interactions of the monopoles and antimonopoles is shifted.
For vanishing Higgs potential,
the structure of the solutions then changes completely,
and the zeros of the Higgs field form one or more rings,
centered around the symmetry axis.
Thus vortex ring solutions arise \cite{KKS}.

When the scalar coupling $\lambda$ is small,
the transition from MAPs and MACs
to vortex ring solutions appears for magnetic charge $n=3$ \cite{KKS}.
For larger scalar coupling $\lambda$, however, bifurcations arise,
where new pairs of solutions appear \cite{KNS}.
Thus MAP or MAC solutions and vortex ring solutions may coexist 
beyond some critical value of $\lambda$. 
Moreover new configurations with mixed node structure arise,
possessing isolated nodes on the symmetry axis as well
as vortex rings \cite{KNS}.
The strength of the Higgs self-interaction is clearly critical for the
existence and the properties of these new equilibrium configurations.

We here address such monopole-antimonopole systems in flat-space,
reviewing and supplementing previous results \cite{KNS}.
In particular, we obtain new sets of solutions associated with further
bifurcations at larger values of the scalar coupling,
both in the topologically trivial and the non-trivial sector.
While the mass of these configurations increases with the scalar coupling, 
it is largely determined by their internal structure.
We confirm our previous conjecture, that for large scalar coupling,
MAC solutions appear to be the energetically most 
favourable configurations.

The balance of forces changes again, when gravity is coupled to YMH theory.
This makes the investigation of self-gravitating MAPs, MACs
and vortex ring solutions of Einstein-Yang-Mills-Higgs (EYMH) theory 
very interesting. Depending on the strength of gravity,
the structure of the solutions may change again \cite{KKN}.
Also new solutions may arise, which do not possess a flat-space limit
\cite{gmono,MAP,KKSg}.
But gravity cannot become too strong, for such regular
equilibrium configurations to exist,
since beyond a certain gravitational coupling strength,
a horizon is expected to form \cite{gmono}.

For gravitating monopoles and multimonopoles a degenerate horizon
forms indeed
at a critical value of the gravitational coupling \cite{gmono,Lue,HKK}.
For monopole-antimonopole systems, however, 
no horizon is seen to form.
Instead the flat-space branch of regular graviting monopoles bifurcates 
with a second branch
at a maximal value of the gravitational coupling.
This second 
branch then extends backwards and, in the limit of vanishing   
coupling constant $\alpha$, it becomes linked to a generalized
Bartnik-McKinnon (BM) solution of Einstein-Yang-Mills (EYM)
theory \cite{BM,KK,IKKS}. 

Anticipating new phenomena \cite{Lue,IKKS},
we here address the effect of gravity on 
mono\-pole-antimonopole systems
at large values of the scalar coupling.
Indeed, as $\lambda$ becomes large,
we observe a rich pattern of bifurcating branches of solutions
arising at finite gravitational coupling strength,
which encompasses all possible combinations of isolated nodes and vortex rings.
Moreover, certain bifurcations of the gravitating branches of solutions 
turn out to be precursors of the bifurcations of the flat-space solutions,
and, in the limiting case of infinite $\lambda$, 
particular bifurcations become linked
to the corresponding generalized BM solutions of EYM theory.

In section II we present the action of EYMH theory, 
the axially symmetric ansatz and the boundary conditions. 
We then discuss the properties of gravitating monopole-antimonopole
paris, chains and vortex rings at large $\lambda$ in section III,
focussing on monopole-antimonopole systems with $m=2,3,4$ and $n=3$.
We give our conclusions in section IV.


\boldmath
\section{$SU(2)$ EYMH Action and Axially Symmetric Ansatz}
\unboldmath

\subsection{Action}

We consider the $SU(2)$ Einstein-Yang-Mills-Higgs 
theory with action 
\begin{equation} 
S =  \int \left\{ \frac{R}{16\pi G} 
-\frac{1}{2} {\rm Tr} 
\,\left( F_{\mu\nu}F^{\mu\nu} \right)
-\frac{1}{4} {\rm Tr}
\left(  D_\mu \Phi\, D^\mu \Phi  \right)
-\frac{\lambda}{4} 
 {\rm Tr}\left[ \left(\Phi^2 - \eta^2 \right)^2 \right]
\right\}
\sqrt{- g} d^4 x
\ , \label{action} \end{equation}
with curvature scalar $R$,
$su(2)$ field strength tensor
\begin{equation}
F_{\mu\nu} = \partial_\mu A_\nu - \partial_\nu A_\mu + i e [A_\mu, A_\nu] \ ,
\end{equation}
gauge potential $A_\mu = A_\mu^a \tau^a/2$,
and covariant derivative of the Higgs field $\Phi = \Phi^a \tau^a$
in the adjoint representation
\begin{equation}
D_\mu \Phi = \partial_\mu \Phi +i e [A_\mu, \Phi] \ .
\end{equation}
Here $G$ and $e$ denote the gravitational and gauge coupling constants,
respectively,
$\eta$ denotes the vacuum expectation value of the Higgs field,
and $\lambda$ represents the strength of the scalar coupling.

Under $SU(2)$ gauge transformations $U$,
the gauge potentials transform as
\begin{equation}
A_{\mu}' = U A_{\mu} U^\dagger + \frac{i}{e} (\partial_\mu U) U^\dagger
\ , \label{gtgen} \end{equation}
and the Higgs field transforms as
\begin{equation}
\Phi' = U \Phi U^\dagger
\ . \label{gtgen2} \end{equation}

The nonzero vacuum expectation value of the Higgs field
breaks the non-Abelian $SU(2)$ gauge symmetry to the Abelian U(1) symmetry.
The particle spectrum of the theory then consists of a massless photon,
two massive vector bosons of mass $M_v = e\eta$,
and a massive scalar field $M_s = {\sqrt {2 \lambda}}\, \eta$.
In the BPS limit the scalar field also becomes massless,
since $\lambda = 0$, i.e., the Higgs potential vanishes.

Variation of the action (\ref{action}) with respect to the metric
$g^{\mu\nu}$ leads to the Einstein equations
\begin{equation}
G_{\mu\nu}= R_{\mu\nu}-\frac{1}{2}g_{\mu\nu}R = 8\pi G T_{\mu\nu}
\  \label{ee} \end{equation}
with stress-energy tensor
\begin{eqnarray}
T_{\mu\nu} &=& g_{\mu\nu}L_M -2 \frac{\partial L_M}{\partial g^{\mu\nu}}
 \nonumber \\
  &=&
      2\, {\rm Tr}\,
    ( F_{\mu\alpha} F_{\nu\beta} g^{\alpha\beta}
   -\frac{1}{4} g_{\mu\nu} F_{\alpha\beta} F^{\alpha\beta}) \nonumber \\
  &+&
      {\rm Tr}\, (\frac{1}{2}D_\mu \Phi D_\nu \Phi
    -\frac{1}{4} g_{\mu\nu} D_\alpha \Phi D^\alpha \Phi)
   -\frac{\lambda}{8}g_{\mu\nu} {\rm Tr}(\Phi^2 - \eta^2)^2
\ .
\end{eqnarray}

Variation with respect to the gauge field $A_\mu$
and the Higgs field $\Phi$
leads to the matter field equations,
\begin{eqnarray}
& &\frac{1}{\sqrt{-g}} D_\mu(\sqrt{-g} F^{\mu\nu})
   -\frac{1}{4} i e [\Phi, D^\nu \Phi ] = 0 \ ,
\label{feqA} \\
& & \frac{1}{\sqrt{-g}} D_\mu(\sqrt{-g} D^\mu \Phi)
+\lambda (\Phi^2 -\eta^2) \Phi  = 0 \ ,
\label{feqPhi}
\end{eqnarray}
respectively.

\subsection{\bf Static axially symmetric Ansatz}

To obtain gravitating static axially symmetric solutions,
we employ isotropic coordinates \cite{KK,HKK,IKKS}.
In terms of the spherical coordinates $r$, $\theta$ and $\vphi$
the isotropic metric reads
\begin{equation}
ds^2=
  - f dt^2 +  \frac{m}{f} d r^2 + \frac{m r^2}{f} d \theta^2
           +  \frac{l r^2 \sin^2 \theta}{f} d\vphi^2
\ , \label{metric2} \end{equation}
where the metric functions
$f$, $m$ and $l$ are functions of
the coordinates $r$ and $\theta$, only.
The $z$-axis ($\theta=0, \pi$) represents the symmetry axis.
Regularity on this axis requires 
\begin{equation}
m|_{\theta=0, \pi}=l|_{\theta=0, \pi}
\ . \label{lm} \end{equation}

We take a purely magnetic gauge field, $A_0=0$,
and parametrize the gauge potential and the Higgs field by the Ansatz
\cite{KKS}
\begin{eqnarray}
A_\mu dx^\mu
& = &
\left( \frac{K_1}{r} dr + (1-K_2)d\theta\right)\frac{\tau_\vphi^{(n)}}{2e}
\nonumber \\
&-& n \sin\theta \left( K_3\frac{\tau_r^{(n,m)}}{2e}
                     + K_4\frac{\tau_\theta^{(n,m)}}{2e}\right) d\vphi
\ , \label{ansatzA} \\
\Phi
& = &
\eta \left( \Phi_1\tau_r^{(n,m)}+ \Phi_2\tau_\theta^{(n,m)} \right) \  .
\label{ansatzPhi}
\end{eqnarray}
where the $su(2)$ matrices
$\tau_r^{(n,m)}$, $\tau_\theta^{(n,m)}$, and $\tau_\vphi^{(n)}$
are defined as products of the spatial unit vectors
\begin{eqnarray}
{\hat e}_r^{(n,m)} & = & \left(
\sin(m\theta) \cos(n\vphi), \sin(m\theta)\sin(n\vphi), \cos(m\theta)
\right)\ , \nonumber \\
{\hat e}_\theta^{(n,m)} & = & \left(
\cos(m\theta) \cos(n\vphi), \cos(m\theta)\sin(n\vphi), -\sin(m\theta)
\right)\ , \nonumber \\
{\hat e}_\vphi^{(n)} & = & \left( -\sin(n\vphi), \cos(n\vphi), 0 \right)\ ,
\label{unit_e}
\end{eqnarray}
with the Pauli matrices $\tau^a = (\tau_x, \tau_y, \tau_z)$, i.e.
\begin{eqnarray}
\tau_r^{(n,m)}  & = &
\sin(m\theta) \tau_\rho^{(n)} + \cos(m\theta) \tau_z \ ,
\nonumber\\
\tau_\theta^{(n,m)} & = &
\cos(m\theta) \tau_\rho^{(n)} - \sin(m\theta) \tau_z \ ,
\nonumber\\
\tau_\vphi^{(n)} & = &
 -\sin(n\vphi) \tau_x + \cos(n\vphi)\tau_y \ ,
\nonumber
\end{eqnarray}
with $\tau_\rho^{(n)} =\cos(n\vphi) \tau_x + \sin(n\vphi)\tau_y $.
For $m=2$, $n=1$ the Ansatz corresponds to the one for
the monopole-antimonopole pair solutions \cite{Rueber,mapKK,MAP},
while for $m=1$, $n>1$ it corresponds to
the Ansatz for axially symmetric multimonopoles \cite{RebbiRossi,KKT,HKK}.
The four gauge field functions $K_i$ and two Higgs field functions 
$\Phi_i$ depend on the coordinates $r$ and $\theta$, only.
All profile functions are even or odd w.r.t.~reflection symmetry, $z \rightarrow -z$.

The gauge transformation
\begin{equation}
U = \exp \{i \Gamma (r,\theta) \tau_\vphi^{(n)}/2\}
\  \end{equation}
leaves the Ansatz form-invariant \cite{BriKu}.
To construct regular solutions we have to fix the gauge \cite{KKT}.
Here we impose the gauge condition \cite{KKS}
\begin{equation}
 r \partial_r K_1 - \partial_\theta K_2 = 0
\ . \label{gauge} \end{equation}

With this Ansatz the equations of motion reduce to a set of 
9 coupled partial differential equations,
to be solved numerically subject to the set of boundary conditions,
discussed below.

\subsection{\bf Boundary conditions}

To obtain globally regular asymptotically flat solutions
with the proper symmetries,
we must impose appropriate boundary conditions \cite{HKK,KKS}.

{\sl Boundary conditions at the origin}

Regularity of the solutions at the origin ($r=0$) 
requires for the metric functions the boundary conditions 
\begin{equation}
\partial_r f(r,\theta)|_{r=0}= 
\partial_r m(r,\theta)|_{r=0}= 
\partial_r l(r,\theta)|_{r=0}= 0
\ , \label{bc2a} \end{equation}
whereas the gauge field functions $K_i$ satisfy
\begin{equation}
K_1(0,\theta)= K_3(0,\theta)= K_4(0,\theta)= 0\ , \ \ \ \
K_2(0,\theta)= 1 \ ,
\end{equation}
and the Higgs field functions $\Phi_i$ satisfy
\begin{equation}
\sin(m\theta) \Phi_1(0,\theta) + \cos(m\theta) \Phi_2(0,\theta) = 0 \ ,
\end{equation}
\begin{equation}
\left.\partial_r\left[\cos(m\theta) \Phi_1(r,\theta)
              - \sin(m\theta) \Phi_2(r,\theta)\right] \right|_{r=0} = 0 \ ,
\end{equation}
i.e.~$\Phi_\rho(0,\theta) =0$, $\partial_r \Phi_z(0,\theta) =0$.

{\sl Boundary conditions at infinity}

Asymptotic flatness imposes on the metric functions of the solutions
at infinity ($r=\infty$) the boundary conditions
\begin{equation}
f \longrightarrow 1 \ , \ \ \ 
m \longrightarrow 1 \ , \ \ \ 
l \longrightarrow 1 \ 
\ . \label{bc1a} \end{equation}
Considering the gauge field at infinity, we require that
solutions in the vacuum sector $Q=0$, where $m=2k$, tend to
a gauge transformed trivial solution,
$$
\Phi \ \longrightarrow \eta U \tau_z U^\dagger \   , \ \ \
A_\mu \ \longrightarrow  \ \frac{i}{e} (\partial_\mu U) U^\dagger \ ,
$$
and that solutions in the sector with topological charge $n$, where $m=2k+1$,
tend to
$$
\Phi  \longrightarrow  U \Phi_\infty^{(1,n)} U^\dagger \   , \ \ \
A_\mu \ \longrightarrow \ U A_{\mu \infty}^{(1,n)} U^\dagger
+\frac{i}{e} (\partial_\mu U) U^\dagger \  ,
$$
where
$$ \Phi_\infty^{(1,n)} =\eta \tau_r^{(1,n)}\ , \ \ \
A_{\mu \infty}^{(1,n)}dx^\mu =
\frac{\tau_\vphi^{(n)}}{2e} d\theta
- n\sin\theta \frac{\tau_\theta^{(1,n)}}{2e} d\vphi
$$
is the asymptotic solution of a charge $n$ multimonopole,
and $U = \exp\{-i k \theta\tau_\vphi^{(n)}\}$, both
for even and odd $m$.

In terms of the functions $K_1 - K_4$, $\Phi_1$, $\Phi_2$ these boundary
conditions read
\begin{equation}
K_1 \longrightarrow 0 \ , \ \ \ \
K_2 \longrightarrow 1 - m \ , \ \ \ \
\label{K12infty}
\end{equation}
\begin{equation}
K_3 \longrightarrow \frac{\cos\theta - \cos(m\theta)}{\sin\theta}
\ \ \ m \ {\rm odd} \ , \ \ \
K_3 \longrightarrow \frac{1 - \cos(m\theta)}{\sin\theta}
\ \ \ m \ {\rm even} \ , \ \ \
\label{K3infty}
\end{equation}
\begin{equation}
K_4 \longrightarrow \frac{\sin(m\theta)}{\sin\theta} \ ,
\label{K4infty}
\end{equation}
\begin{equation}        \label{Phiinfty}
\Phi_1\longrightarrow  1 \ , \ \ \ \ \Phi_2 \longrightarrow 0 \ .
\end{equation}

{\sl Boundary conditions along the symmetry axis}

The boundary conditions along the $z$-axis
($\theta=0$ and $\theta=\pi $) are determined by the
symmetries.
The metric functions satisfy along the axis
\begin{eqnarray}
& &\partial_\theta f = \partial_\theta m =
   \partial_\theta l =0 \ ,
\label{bc4a}
\end{eqnarray}
whereas the matter field functions satisfy
\begin{equation}
K_1 = K_3 = \Phi_2 =0 \ , \ \ \  \
\partial_\theta K_2 = \partial_\theta K_4 = \partial_\theta \Phi_1 =0 \ .
\end{equation}

\subsection{Mass and charge}

Let us introduce the dimensionless coordinate $x$
and the dimensionless coupling constant $\alpha$ \cite{gmono},
\begin{equation}
x = \frac{e\alpha}{\sqrt{4 \pi G}} \, r \ , \ \ \
\alpha = \sqrt{4 \pi G} \eta \   \ \ \
\ . \end{equation}
The limit $\alpha \to 0$ can be approached in two different 
ways: $G \to 0$ 
while the Higgs vacuum expectation value $\eta$ remains finite
(flat-space limit), 
and $\eta \to 0$ 
while Newton's constant $G$ remains finite. 
Corresponding to these two limits
two branches of solutions may exist.

The dimensionless mass $M$ of the solutions is
obtained from the asymptotic expansion of the metric function $f$,
\begin{equation}
M = \frac{1}{2\alpha^2} \lim_{x\to\infty} x^2 \partial_x f
\ . \label{mass} \end{equation}
The asymptotic expansion of the gauge fields yields the
dimensionless magnetic charge $P$ of the solutions,
\begin{equation}
P = \frac{n}{2}\left[1 - (-1)^m\right]
\ , \label{P} \end{equation}
i.e., solutions in the topologically trivial sector have no charge, $P=0$,
whereas solutions in the non-trivial sectors have charge $P=n$.

\section{Numerical Results}

We have constructed numerically regular gravitating 
monopole-antimonopole systems,
including monopole-antimonopole chain and vortex ring solutions.
We here focus on solutions with $m=2$, 3 and 4 and $n=3$.
In particular, we illustrate the plethora of solutions appearing, 
when the scalar coupling $\lambda$ is sufficiently large.
At the same time we study the dependence
of the properties of these solutions on the 
gravitational coupling strength $\alpha$.

We first briefly address the numerical procedure.
Then we present our results for 
monopole-antimonole chains, vortex rings and
configurations with more complicated structure,
beginning with the $m=2$, $n=3$ systems.

\subsection{Numerical procedure}

To construct solutions subject to the above boundary conditions,
we map the semi-infinite range of the radial coordinate $r$
onto the closed unit interval of the new compactified radial variable
$\bar x \in [0:1]$,
$$
\bar x = \frac{r}{1+r}
\ , $$
i.e., the partial derivative with respect to the radial coordinate
changes according to
$$
\partial_r \to (1- \bar x)^2\partial_{\bar x}
\ . $$
The numerical calculations are then performed with help of the 
package FIDISOL,
based on the Newton-Raphson iterative procedure \cite{FIDI}. 
Technically, the system of 
non-linear elliptic partial differential equations of second order is 
first discretized on a non-equidistant grid in 
$\bar x$ and $\theta$. Typical grids used have sizes $60 \times 40$ 
covering the integration 
domain $0 \le \bar x \le 1$ and $0 \le \theta \le \pi/2$. 
The numerical errors are typically of order of $10^{-4}$.

\boldmath
\subsection{MAPs and vortex rings: $m=2$, $n=3$} 
\unboldmath

\subsubsection{Flat space solutions}

Monopole-antimonopole pairs (MAPs) reside in the topologically 
trivial sector. 
These solutions represent unstable equilibrium configurations and 
can be thought of as consisting of two solitons of individual charges $\pm n$.
The position of each constituent can be identified 
according to the location of the nodes of the Higgs field. 
The separation between the poles depends on the subtle interplay of forces
and thus on the strength of the scalar coupling $\lambda$ \cite{mapKK}.

The MAPs with $n=1$ exist for arbitrary values of $\lambda$. 
They have two isolated nodes on the symmetry axis,
indicating the position of the poles,
and their energy density possesses two local maxima,
associated with the nodes of the Higgs field \cite{MAP}. 

As the charge of the poles increases, 
the interaction between the nonabelian matter fields 
becomes stronger than in the case of unit charge constituents,
leading to new types of solutions.
Indeed, for small and vanishing values of $\lambda$, 
MAPs with $n>2$ no longer exist. 
Instead solutions of vortex ring type emerge, 
where the Higgs field vanishes on a ring centred around the symmetry axis 
in the $xy$ plane \cite{KKS}. 
These flat-space vortex ring solutions 
form the fundamental $\lambda$-branches,
which persist for arbitrary values of $\lambda$. 
As the scalar coupling strength increases, however,
bifurcations are observed, 
where pairs of additional flat-space solutions emerge
at critical values of $\lambda$ \cite{KNS}.

For $n=3$, in particular,
a pair of MAP solutions arises at 
the critical value $\lambda_c^1=1.382$.
The $\lambda$-dependence of these flat-space solutions
is demonstrated in Fig.~\ref{f-1}, where we show their mass and the
location of the nodes of their Higgs field configurations.
At the bifurcation, the mass of the new MAP solutions is higher than
the mass of the vortex ring solution. For large values of $\lambda$,
however, the MAP solution on the lower mass branch
becomes energetically favoured.
Note, that the node structure of the solutions on the upper mass branch
changes from MAP to vortex ring, as $\lambda$ increases,
but the radius of the ring remains small.
\begin{figure}[h!]
\lbfig{f-1}
\begin{center}
(a)\hspace{-0.6cm}
\includegraphics[height=.25\textheight, angle =0]{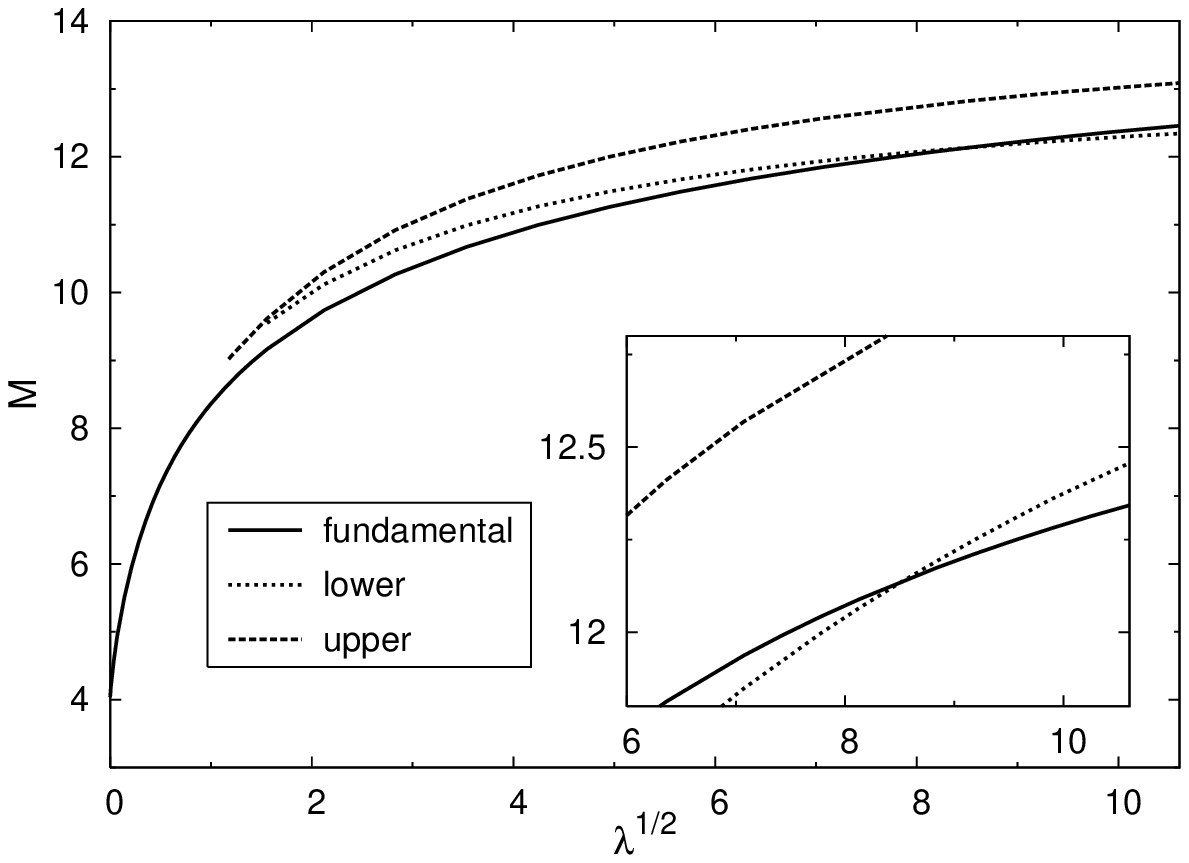}
\hspace{0.5cm} (b)\hspace{-0.6cm}
\includegraphics[height=.25\textheight, angle =0]{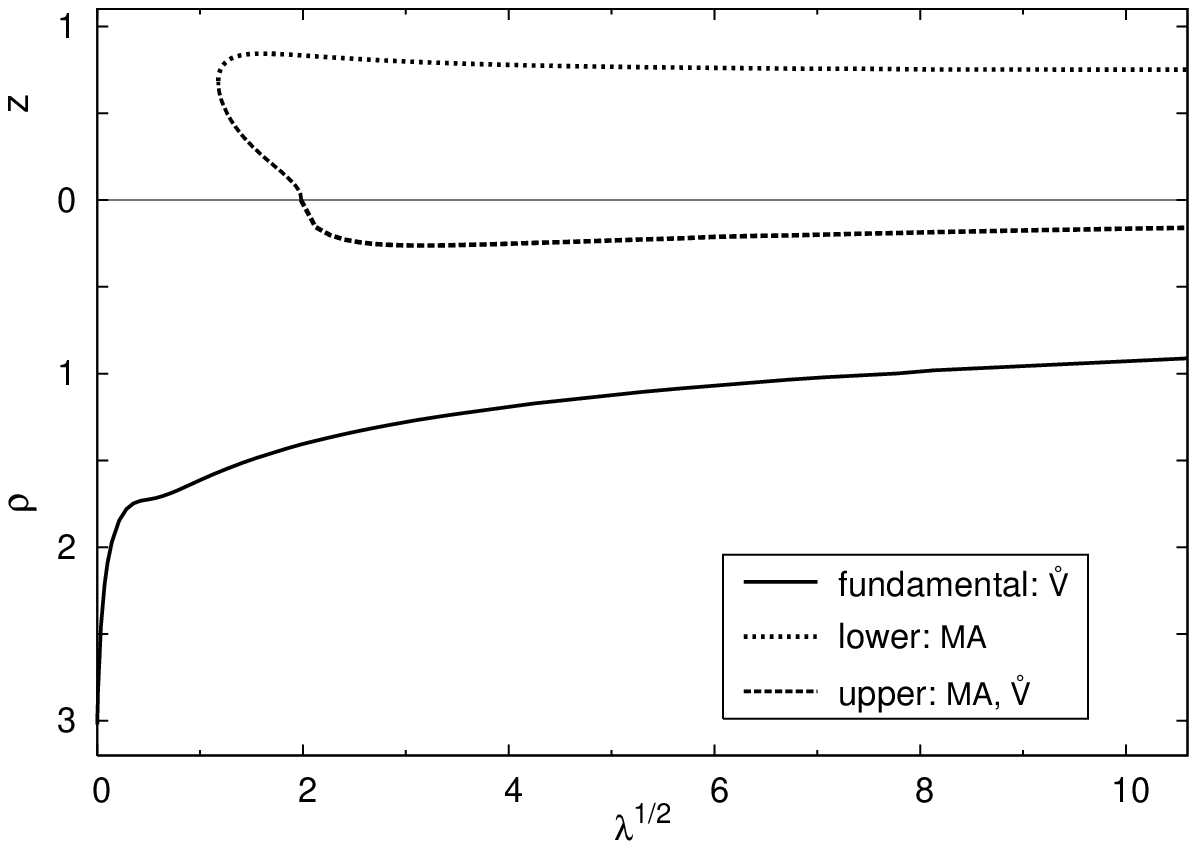}
\end{center}
\vspace{-0.5cm}
\caption{\small
(a)
The mass $M$ of the flat-space solutions on the fundamental $\lambda$-branch
as well as on the two additional $\lambda$-branches
for $m=2$, $n=3$ solutions.
(b)
The location of the isolated nodes $z$ and the radius $\rho$ of the 
vortex ring of the Higgs field
for the same set of solutions. (See Table~1 for the notation of the 
node and vortex ring configurations of
the Higgs field.)
}
\end{figure}
\begin{table}[h!]
\label{t-1}
{\small
\begin{center}
\begin{tabular}{|p{1.6cm}|p{3cm}|p{8cm}|}\hline
\begin{center}\vspace{2mm}{\sf MA}\end{center}
                &
                \vspace{-8mm}\begin{center}\includegraphics[height=.104\textheight, angle =0]{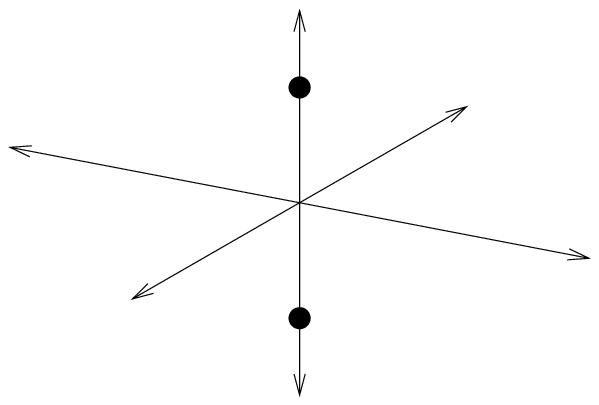}\end{center}\vspace{-11mm}
                &
\begin{center}\vspace{-3mm}
                A monopole-antimonopole pair 
                with a monopole above and an antimonopole below the
                $xy$-plane on the symmetry axis.
\end{center}
         \\ \hline
\begin{center}{\sf \r{V}}\end{center}
                &
                \vspace{-8mm}\begin{center}\includegraphics[height=.104\textheight, angle =0]{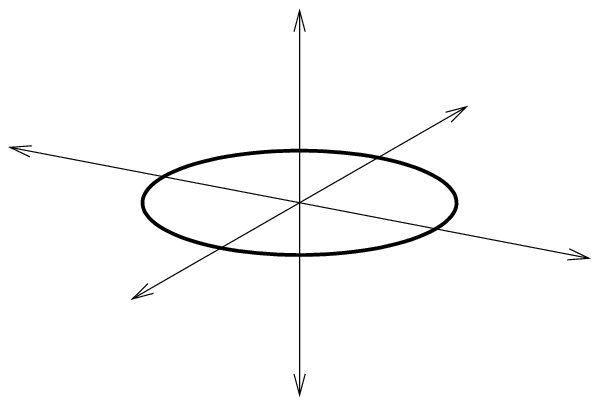}\end{center}\vspace{-11mm} &
 \begin{center}\vspace{0mm}
                A vortex ring in the $xy$-plane.
 \end{center}
\\ \hline
\end{tabular}
\end{center}
}
\vspace{-3mm}
\caption{\small 
Configurations of nodes and vortex rings 
of the Higgs field for $m=2$, $n=3$ solutions.}
\end{table}

\subsubsection{Gravitating solutions} 

When gravity is coupled, 
branches of gravitating solutions arise from these flat-space configurations. 
For sufficiently large gravitational coupling $\alpha$, however,
a gravitational instability should develop.
Therefore regular gravitating solutions will exist only
up to a maximal value of $\alpha$.

The critical behaviour of these $\alpha$-branches,
emerging from flat-space solutions,
is elucidated by numerical investigation. 
As such an $\alpha$-branch approaches its maximal value of $\alpha$, 
it merges with a second $\alpha$-branch of regular gravitating solutions,
where $\alpha_{\rm max}$ is the endpoint of both branches.
The nature of the second branch, however, depends on the nature of the
first branch and on the value of $\lambda$.

{\sl Fundamental solutions}
 
When considering sets of solutions, emerging from 
the fundamental flat-space $\lambda$-branch,
we always find two $\alpha$-branches, where
the solutions on the second $\alpha$-branches shrink to zero size 
in the limit $\alpha \rightarrow 0$,
while their mass diverges.
By scaling the coordinates and the Higgs field
of the solutions via
\begin{equation}
\hat{x} = \frac{x}{\alpha} \ , \ \ \
\hat{\Phi} = \alpha \Phi
\ , \end{equation}
one realizes that the second $\alpha$-branches
connect to limiting EYM solutions,
when $\alpha \rightarrow 0$, 
solutions with finite scaled size and
finite scaled mass $\hat{M}$
\cite{MAP},
\begin{equation}
{\hat{M}}= \alpha {M}
\ . \end{equation}

Indeed, after the scaling, the field equations do not depend on $\alpha$.
Instead $\alpha$ appears in the
asymptotic boundary conditions of the Higgs field,
$|\hat \Phi | \rightarrow \alpha$.
The Higgs field then becomes trivial
on the second $\alpha$-branches as $\alpha \rightarrow 0$,
and the solutions thus approach EYM solutions.
Depending on $n$, 
the limiting solutions correspond to the lowest BM or a generalized BM solution,
i.e., the solutions on these second $\alpha$-branches
possess no flat-space limit.

The two $\alpha$-branches are illustrated in Fig.~\ref{f-2}
for solutions with $m=2$, $n=3$
at scalar coupling strength $\lambda=0$, $\lambda=2.0$ and $\lambda=4.5$.
The figure exhibits the mass and the scaled mass of these solutions
and the location of their nodes. Note, that they always correspond to
vortex ring solutions.

\begin{figure}[h!]
\lbfig{f-2}
\begin{center}
(a)\hspace{-0.6cm}
\includegraphics[height=.25\textheight, angle =0]{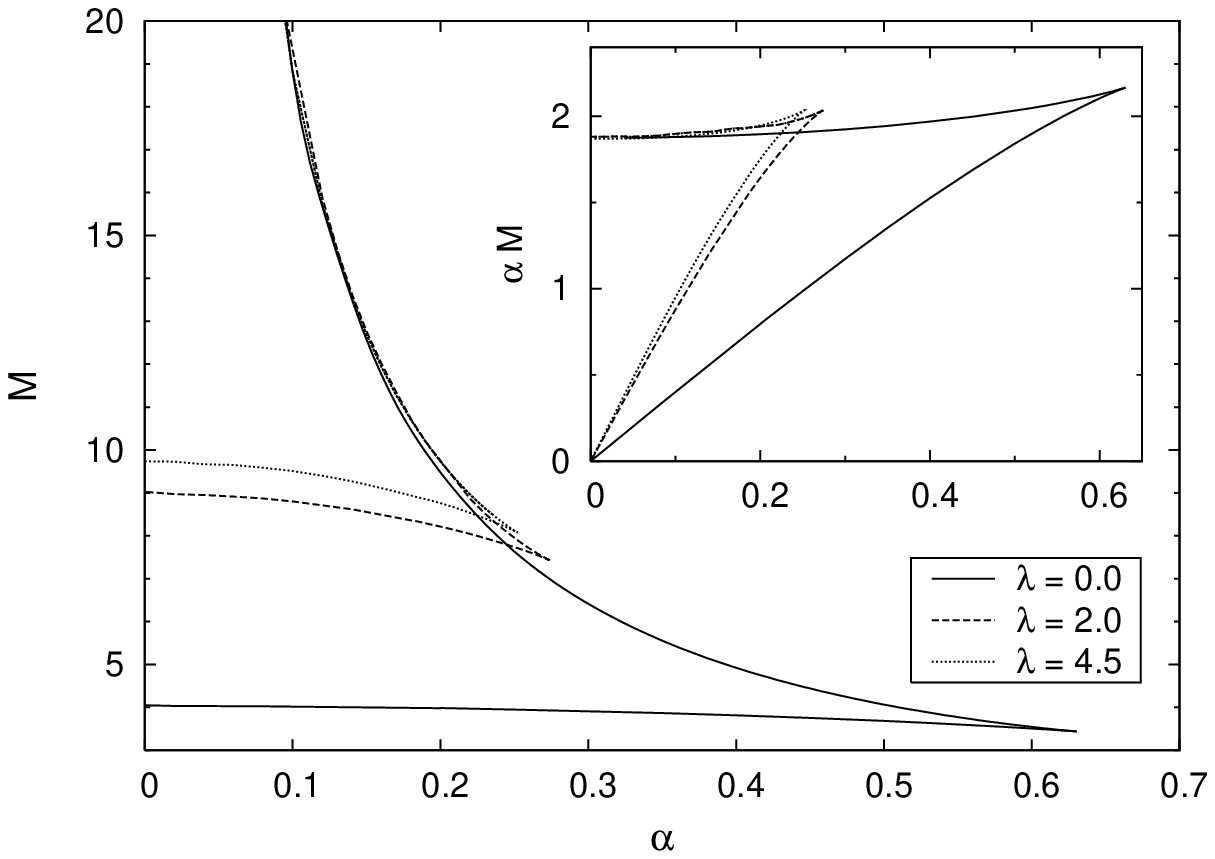}
\hspace{0.5cm} (b)\hspace{-0.6cm}
\includegraphics[height=.25\textheight, angle =0]{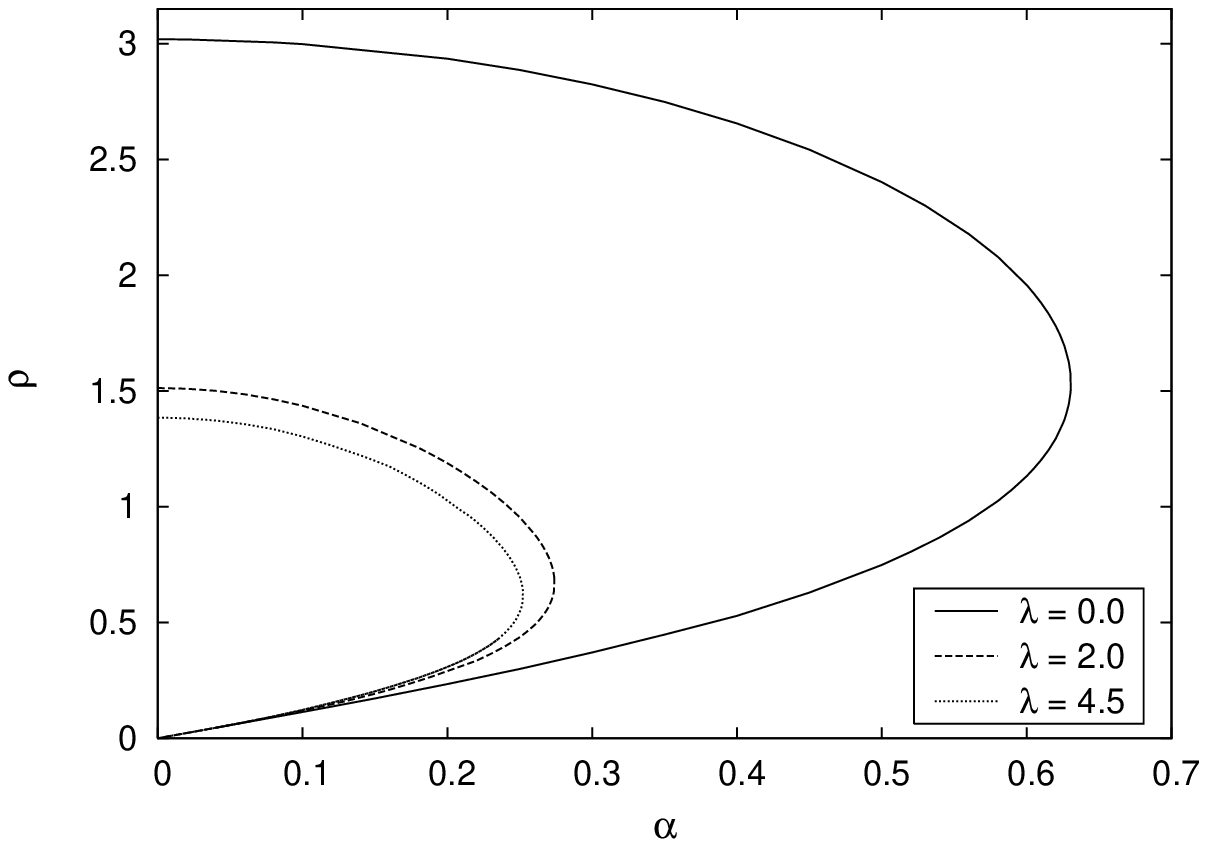}
\end{center}
\vspace{-0.5cm}
\caption{\small
(a)
The mass $M$ and the scaled mass $\alpha M$ of $m=2$, $n=3$ solutions,
emerging from the fundamental flat-space $\lambda$-branch
versus $\alpha$.
(b)
The radius $\rho$ of the vortex ring of the Higgs field for the same
set of solutions.
}
\end{figure}

{\sl New solutions: intermediate $\lambda$}

The second $\alpha$-branch
always connects to an EYM solution, 
when the first $\alpha$-branch emerges from
a fundamental $\lambda$-branch solution. 
But for the $\alpha$-branches of the new flat-space solutions,
which exist only beyond a critical value of $\lambda$,
one may anticipate a different pattern,
because these solutions always come in pairs.

In the presence of gravity, each of the solutions of such a pair
forms an $\alpha$-branch, extending up to a maximal value of $\alpha$.
But the critical behaviour at $\alpha_{\rm max}$ now depends
on the strength of the scalar coupling $\lambda$.
In the simplest case, observed for not too large
values of $\lambda$, the two $\alpha$-branches
simply merge and end at $\alpha_{\rm max}$.
Indeed, this critical behaviour is different from the above,
since both branches merging at $\alpha_{\rm max}$ have flat-space limits.

This is illustrated in Fig.~\ref{f-3},
where we exhibit the mass and the location of the nodes
for $m=2$, $n=3$ configurations
at $\lambda= 2.0$, $\lambda=2.645$, $\lambda=3.645$, $\lambda= 4.5$,
$\lambda=5.25$ and $\lambda=12.5$ 
versus the coupling constant $\alpha$. 
Evidently, for the smaller $\lambda$ values,
a first $\alpha$-branch of solutions
originating from the lower mass flat-space solution,
extends up to a maximal value $\alpha_{\rm max}$, where 
it merges with a second $\alpha$-branch, originating
from the upper mass flat-space solution. 
The solutions on these branches correspond to MAPs.

\begin{figure}[h!]
\lbfig{f-3}
\begin{center}
\hspace{0.0cm} (a)\hspace{-0.6cm}
\includegraphics[height=.25\textheight, angle =0]{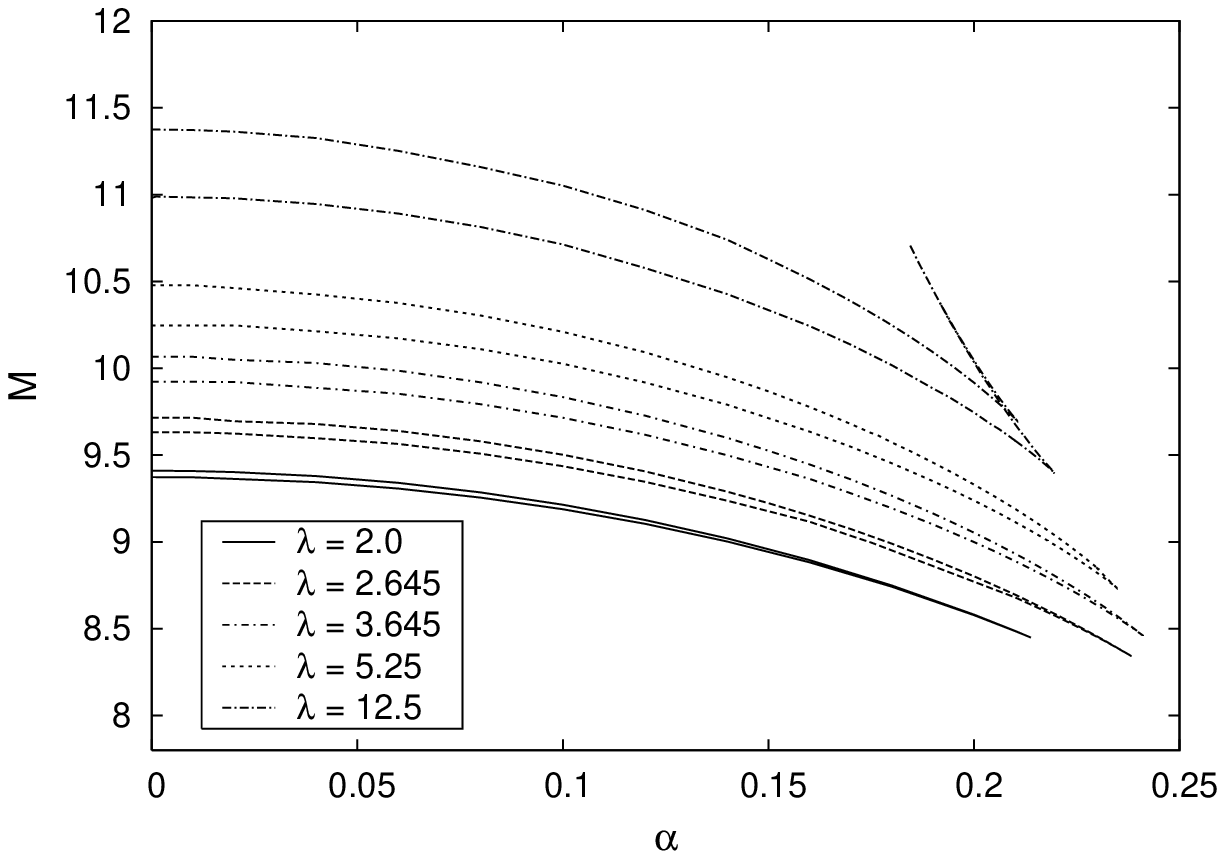}
\hspace{0.5cm} (b)\hspace{-0.6cm}
\includegraphics[height=.25\textheight, angle =0]{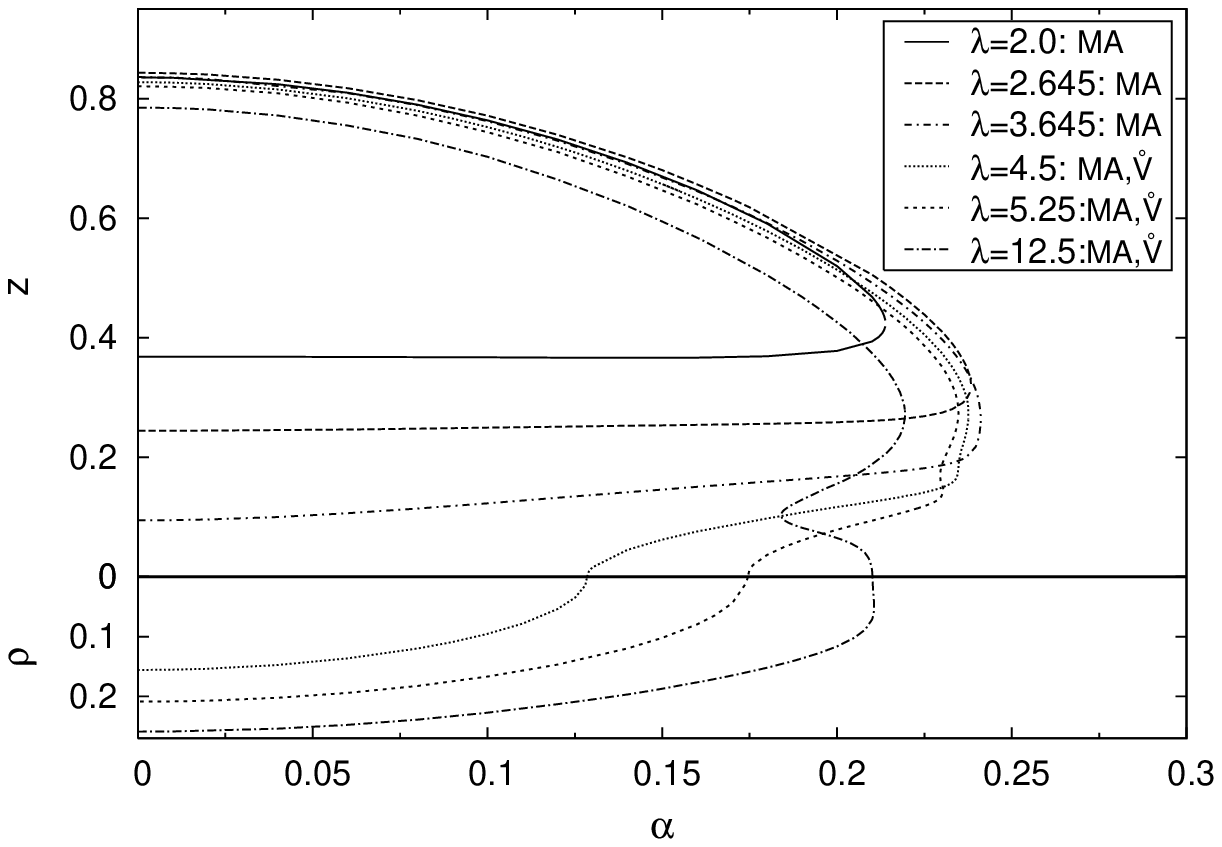}
\end{center}
\vspace{-0.5cm}
\caption{\small
(a)
The mass $M$ of $m=2$, $n=3$ solutions,
related to the new flat-space $\lambda$-branches
versus $\alpha$
for several values of the scalar coupling $\lambda$.
(b)
The location of the isolated nodes $z$ and radius $\rho$ of the vortex ring 
of the Higgs field for the same set of solutions.
(See Table~1 for the notation of the node and vortex ring configurations of
the Higgs field.)
}
\end{figure}

However, as $\lambda$ increases, a new phenomenon arises
on the second $\alpha$-branch,
as it evolves back from $\alpha_{\rm max}$ towards the flat-space solution.
As seen in Fig.~\ref{f-3},
at a critical value of $\lambda$ ($\lambda \approx 4.5$)
a bifurcation appears on the second $\alpha$-branch, giving rise to
two additional $\alpha$-branches of solutions.

These 3rd and 4th $\alpha$-branches exist beyond a minimal value
of $\alpha$, $\alpha_{\rm min}$, and below a respective
maximal value of $\alpha$, where they merge with the first
and second $\alpha$-branch, respectively.
The range $\alpha_{\rm min} \le \alpha_{\rm max}^{(i)}$,
where these additional branches exist, 
increases with increasing $\lambda$.
In Fig.~\ref{f-3} the additional branches
manifest in the mass of the $\lambda=12.5$ solutions as a higher mass spike.
Note, that in this range of $\lambda$ vortex ring solutions
arise on a part of the second $\alpha$-branch.

The emergence and evolution of these new $\alpha$-branches
is illustrated further in Fig.~\ref{f-4},
where we exhibit the values of the metric functions $f$ and $l$ at the origin, 
$f(0)$ and $l(0)$. In particular, $l(0)$ clearly exhibits the various
branches and critical points and therefore represents an instructive
means to clarify the pattern of solutions.

\begin{figure}[h!]
\lbfig{f-4}
\begin{center}
\hspace{0.0cm} (a)\hspace{-0.6cm}
\includegraphics[height=.25\textheight, angle =0]{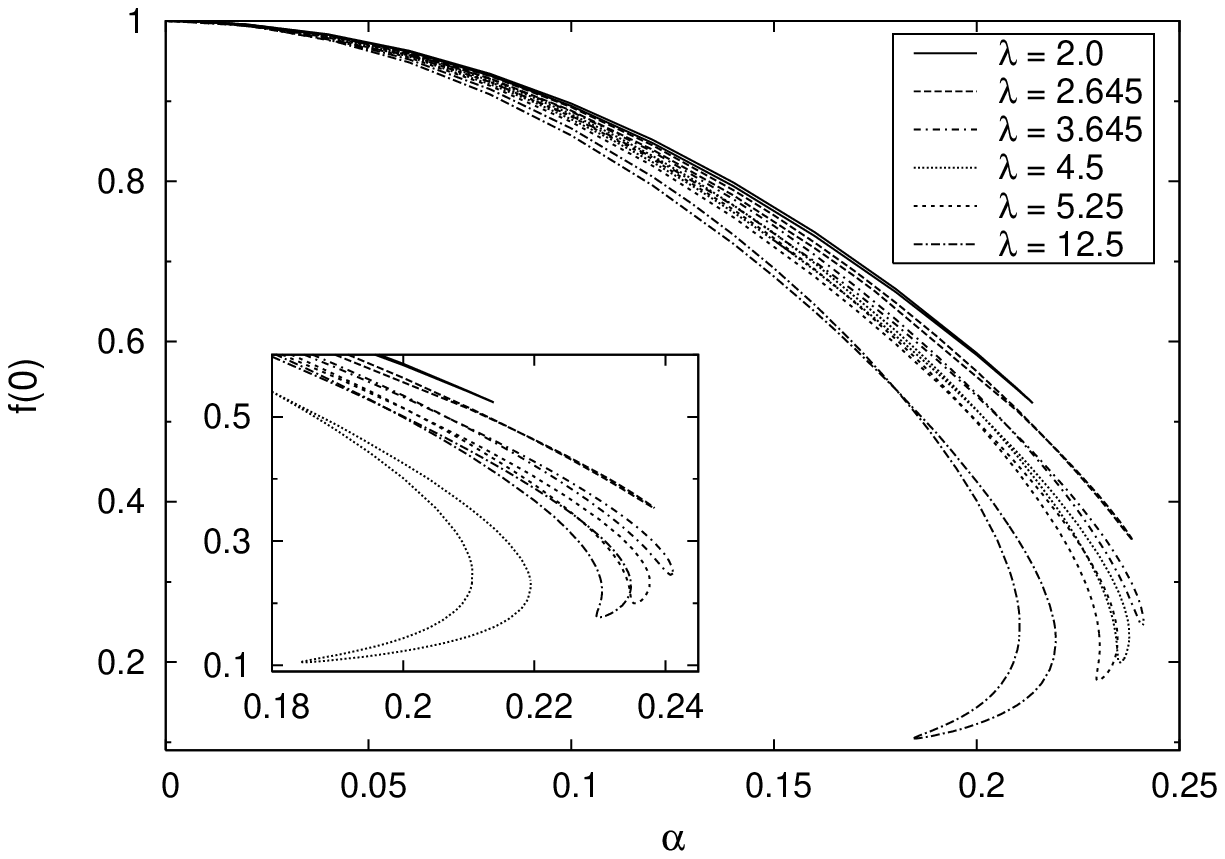}
\hspace{0.5cm} (b)\hspace{-0.6cm}
\includegraphics[height=.25\textheight, angle =0]{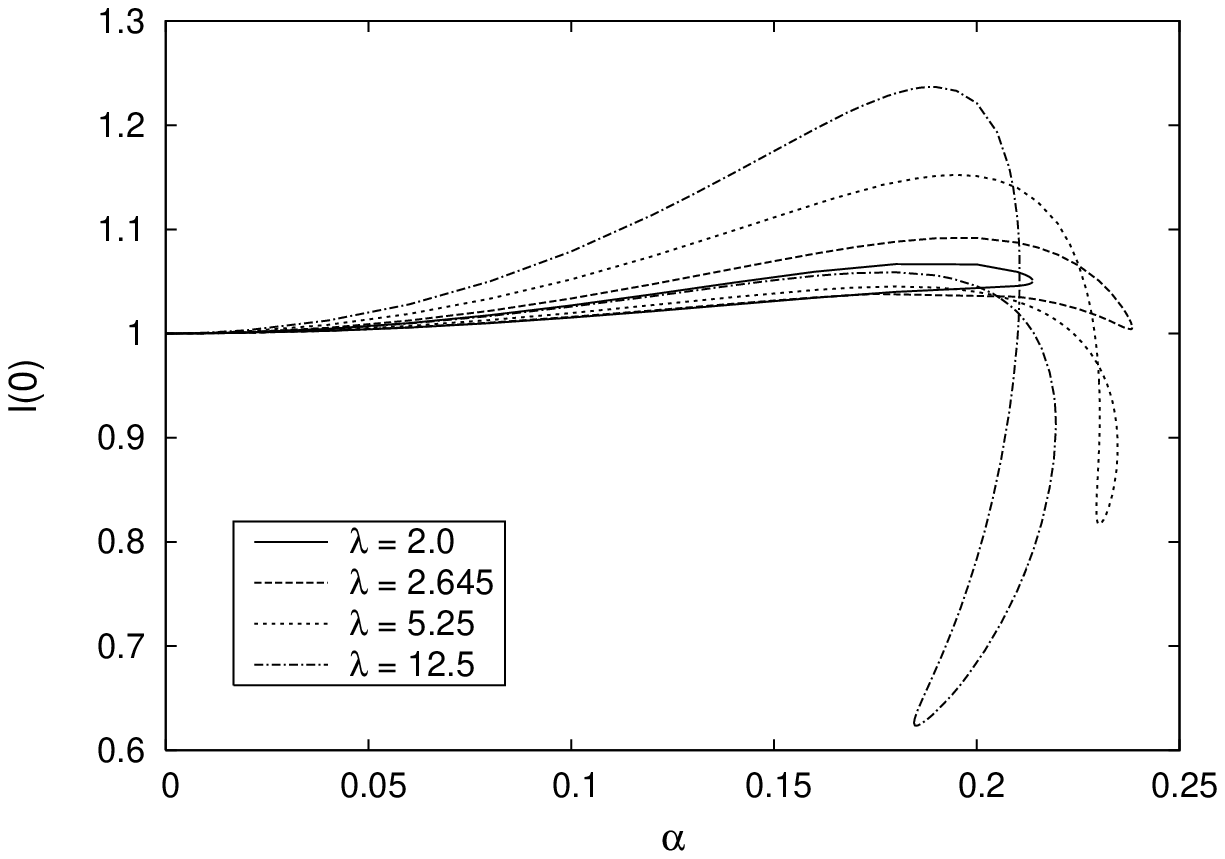}
\end{center}
\vspace{-0.5cm}
\caption{\small
The values of metric functions at the origin, $f(0)$ (a) and $l(0)$ (b)
of $m=2$, $n=3$ solutions,
related to the new flat-space $\lambda$-branches
versus $\alpha$
for several intermediate values of the scalar coupling $\lambda$.
}
\end{figure}

To complement the previous discussion of the new solutions,
let us consider the $\lambda$-dependence of the solutions
for fixed values of $\alpha$.
In Fig.~\ref{f-5} we exhibit
the mass, the location of the nodes,
and the values of the metric functions at the origin
versus $\lambda$
for $m=2$, $n=3$ configurations at
 $\alpha=0.2$,  $\alpha=0.21$, $\alpha=0.22$, 
$\alpha=0.23$ and $\alpha=0.24$.
One clearly recognizes the new solutions,
present in a certain region of parameter space.

\begin{figure}[h!]
\lbfig{f-5}
\begin{center}
\hspace{0.0cm} (a)\hspace{-0.6cm}
\includegraphics[height=.25\textheight, angle =0]{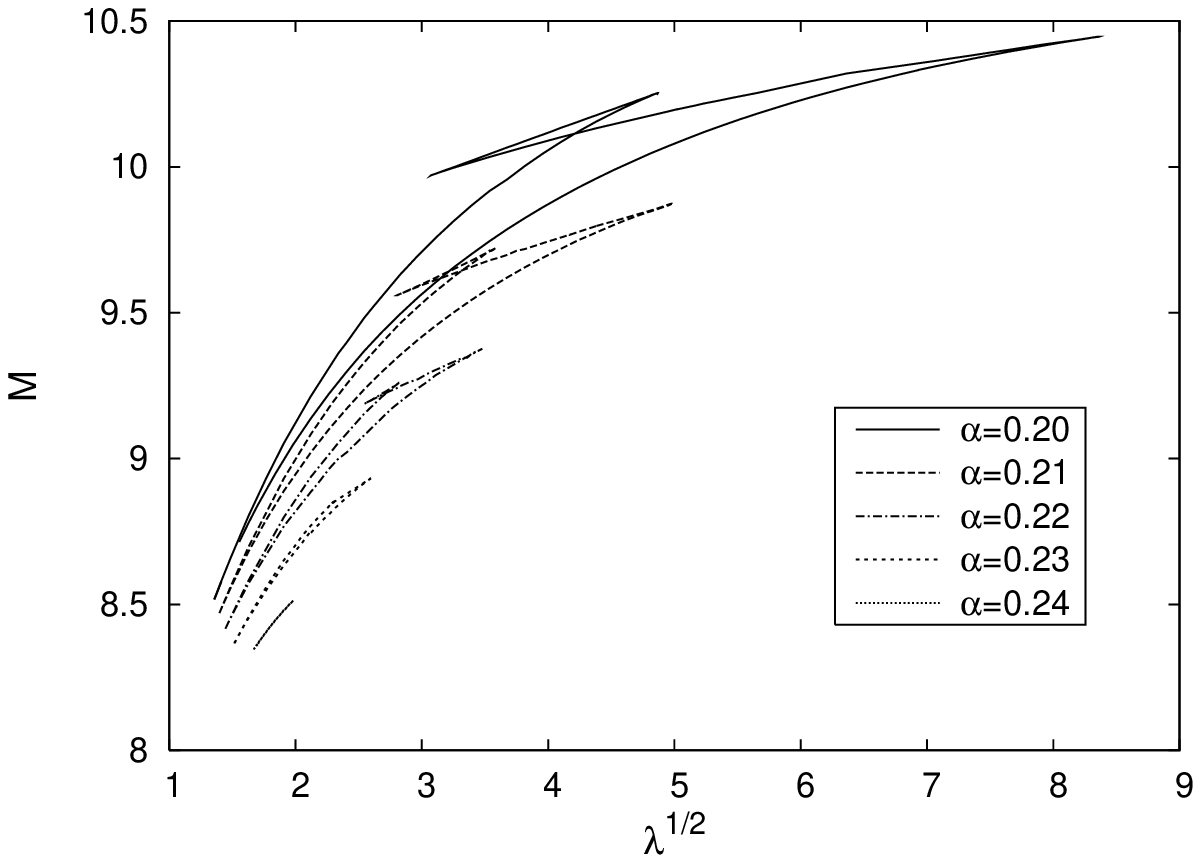}
\hspace{0.5cm} (b)\hspace{-0.6cm}
\includegraphics[height=.25\textheight, angle =0]{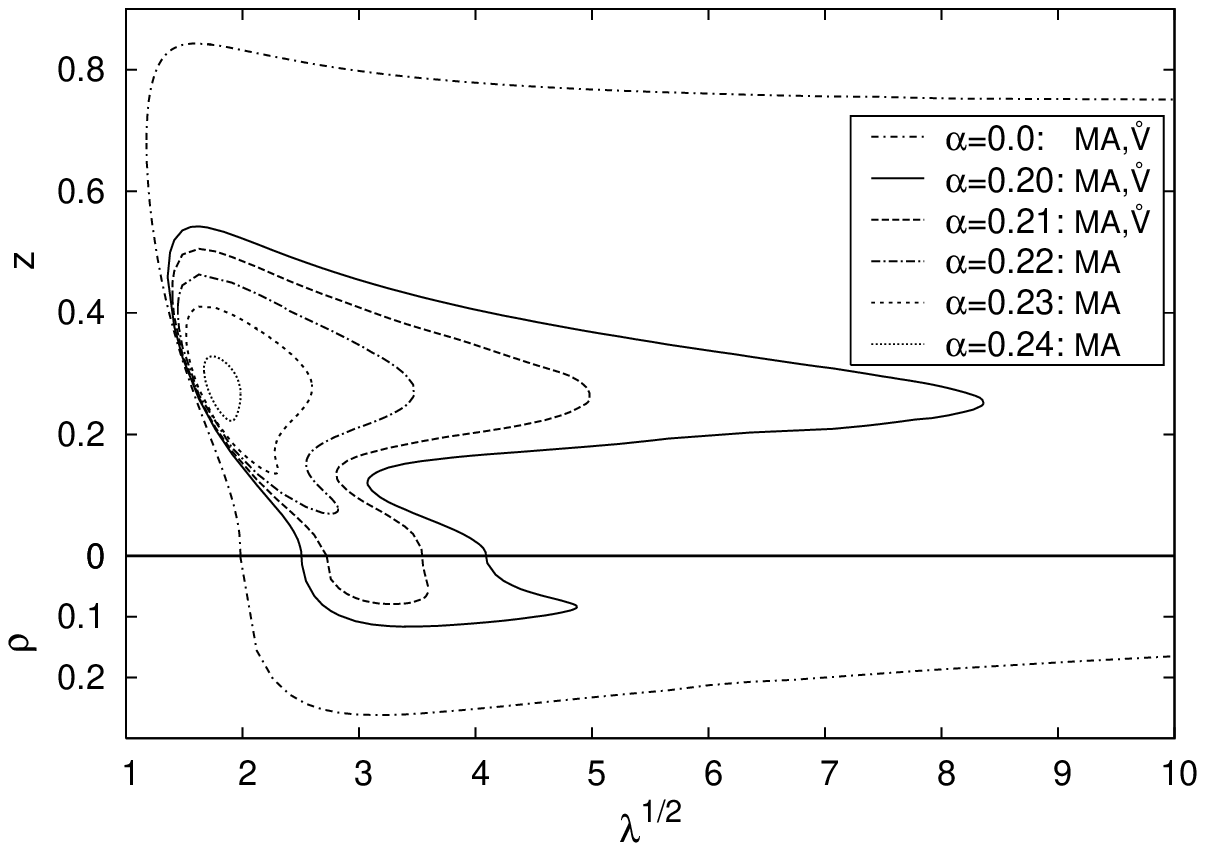}
\\
\hspace{0.0cm} (c)\hspace{-0.6cm}
\includegraphics[height=.25\textheight, angle =0]{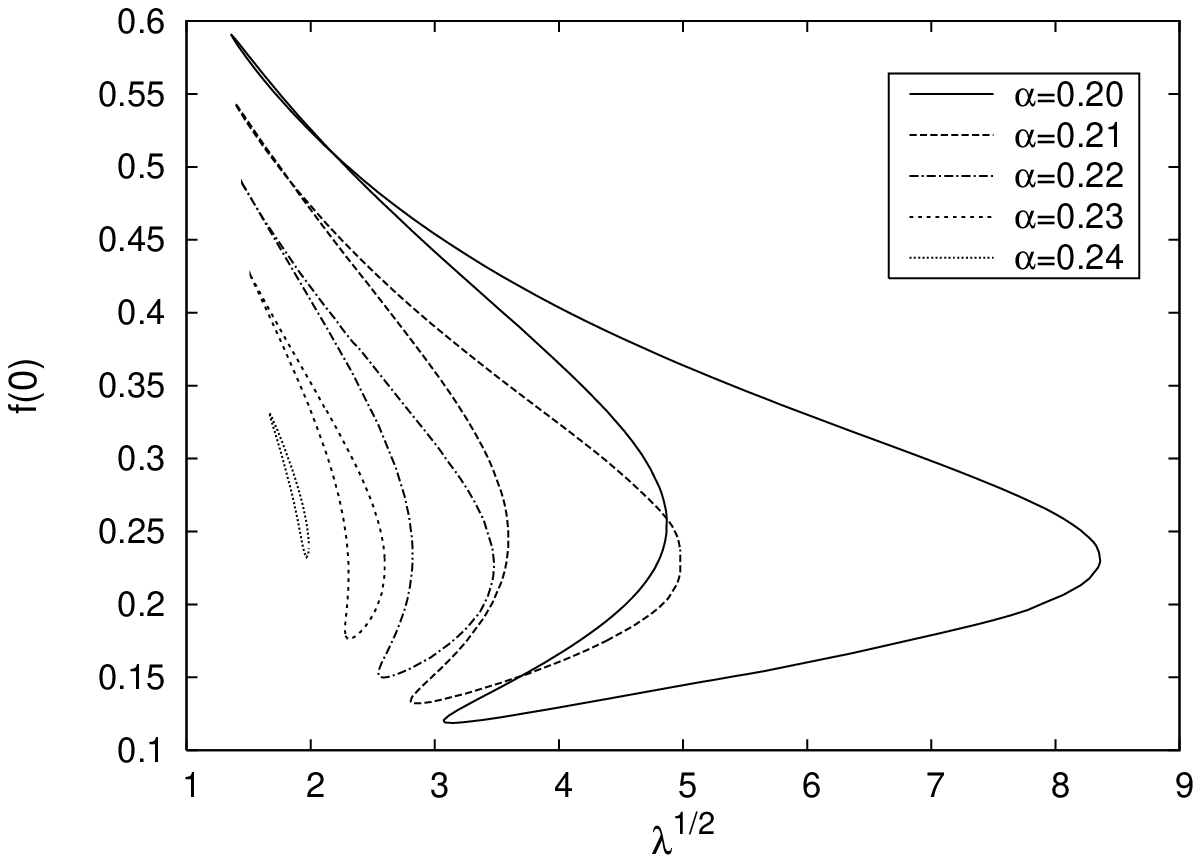}
\hspace{0.5cm} (d)\hspace{-0.6cm}
\includegraphics[height=.25\textheight, angle =0]{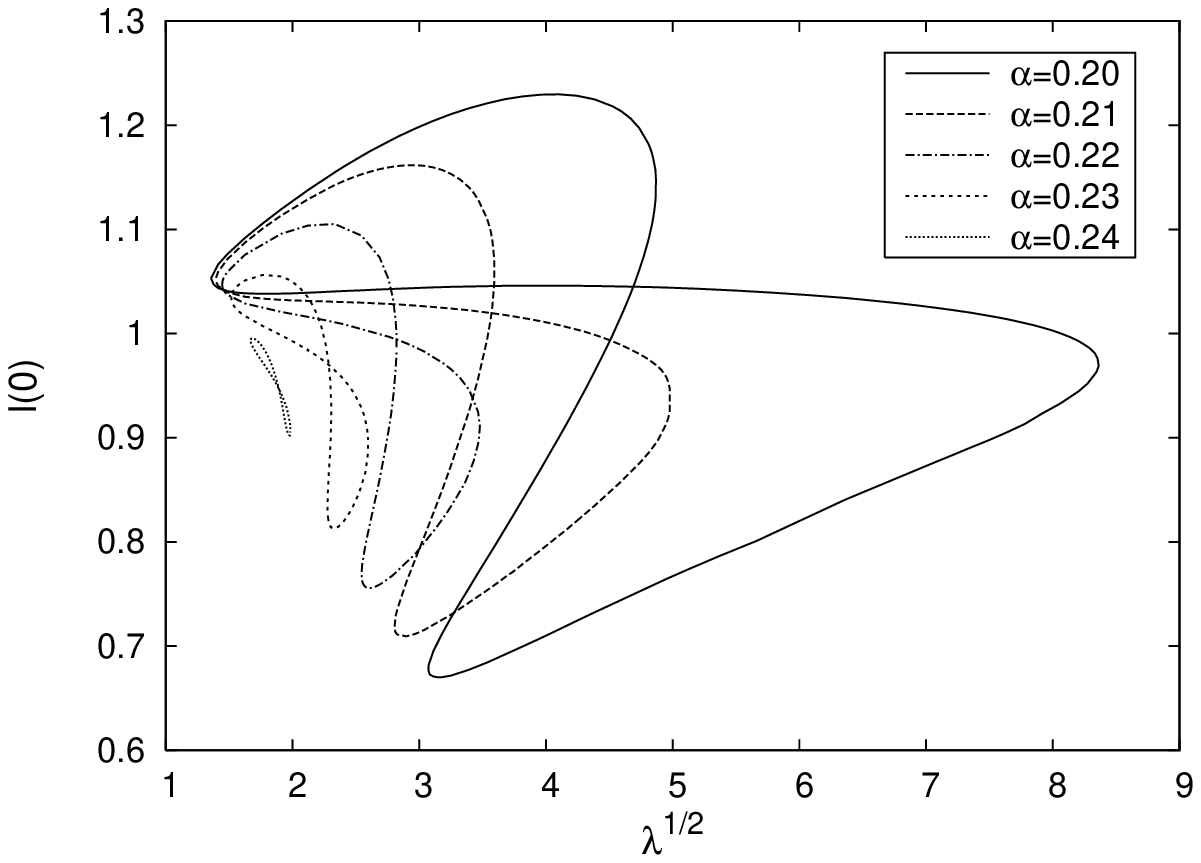}
\end{center}
\vspace{-0.5cm}
\caption{\small
(a)
The mass $M$ of $m=2$, $n=3$ solutions,
related to the new flat-space $\lambda$-branches
versus $\lambda$
for several values of the coupling $\alpha$.
(b)
The location of the isolated nodes $z$ and the radius $\rho$ of the
 vortex ring of the Higgs field ,
and the values of metric functions at the origin, $f(0)$ (c) and $l(0)$ (d)
for the same set of solutions.
(See Table~1 for the notation of the node and vortex ring configurations of
the Higgs field.)
}
\end{figure}

Since both the scalar interaction and gravity are attractive,
one can expect a similar effect 
concerning the changing structure of the nodes,
when $\lambda$ is increased at fixed $\alpha$ as
when $\alpha$ is increased at fixed $\lambda$.
For the new flat-space upper mass $\lambda$-branch,
a transition from MAPs to vortex rings appears at
$\lambda=3.941$ \cite{KNS}.

As the gravitational coupling constant $\alpha$ increases,
the pattern of the nodes of the solutions changes significantly as well,
and there are no solutions with vortex rings 
(except for those connected to the fundamental $\lambda$-branch),
when $\alpha > 0.215$ for any value of $\lambda$.
This pattern is illustrated in Figure \ref{f-5},
where the location of the nodes, both isolated zeros and vortex rings,
is shown versus the scalar coupling for several values of $\alpha$.
Evidently, the coupling to gravity restricts the range of possible values
of $\lambda$, where these solutions can exist,
and this restriction is the stronger, the larger $\alpha$,
shrinking the domain of existence to zero 
at a critical value of $\alpha$.

{\sl New solutions: very large $\lambda$}

As $\lambda$ increases further, 
the new $\alpha$-branches extend further backwards.
Along these branches the mass of the solutions increases further
as $\alpha$ decreases.
At the same time the locations of the nodes of the Higgs field
move towards the origin,
as illustrated in Fig.~\ref{f-6}
for very large values of the scalar coupling $\lambda$.

\begin{figure}[h!]
\lbfig{f-6}
\begin{center}
\hspace{0.0cm} (a)\hspace{-0.6cm}
\includegraphics[height=.25\textheight, angle =0]{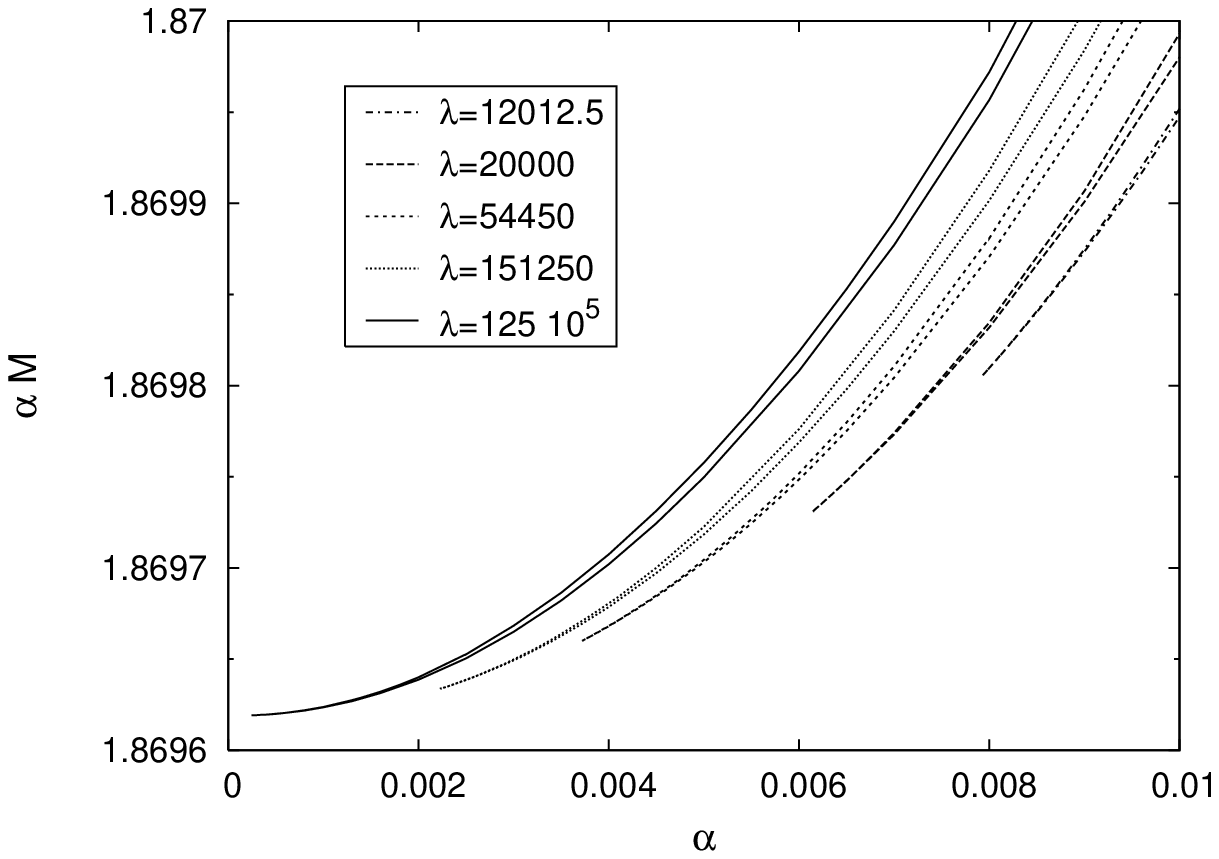}
\hspace{0.5cm} (b)\hspace{-0.6cm}
\includegraphics[height=.25\textheight, angle =0]{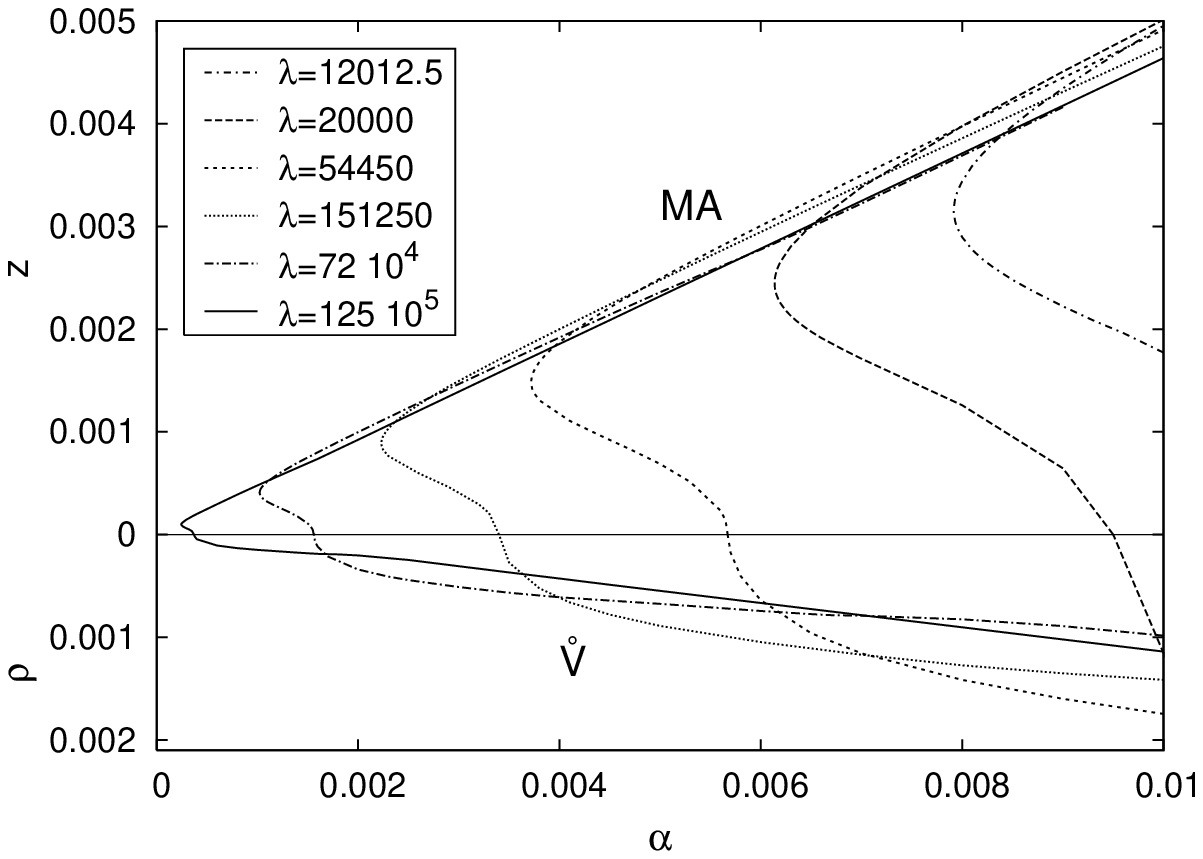}
\\
\hspace{0.0cm} (c)\hspace{-0.6cm}
\includegraphics[height=.25\textheight, angle =0]{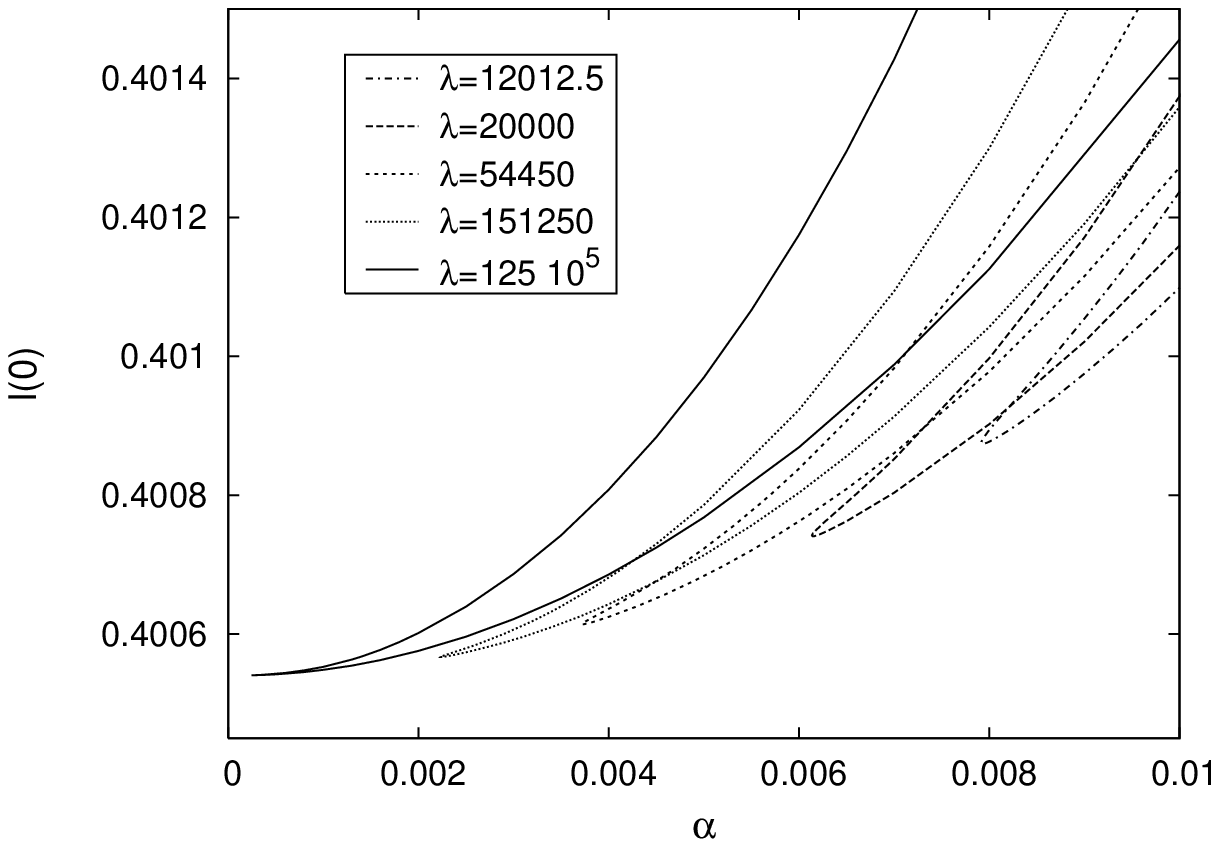}
\hspace{0.5cm} (d)\hspace{-0.6cm}
\includegraphics[height=.25\textheight , angle =0 ]{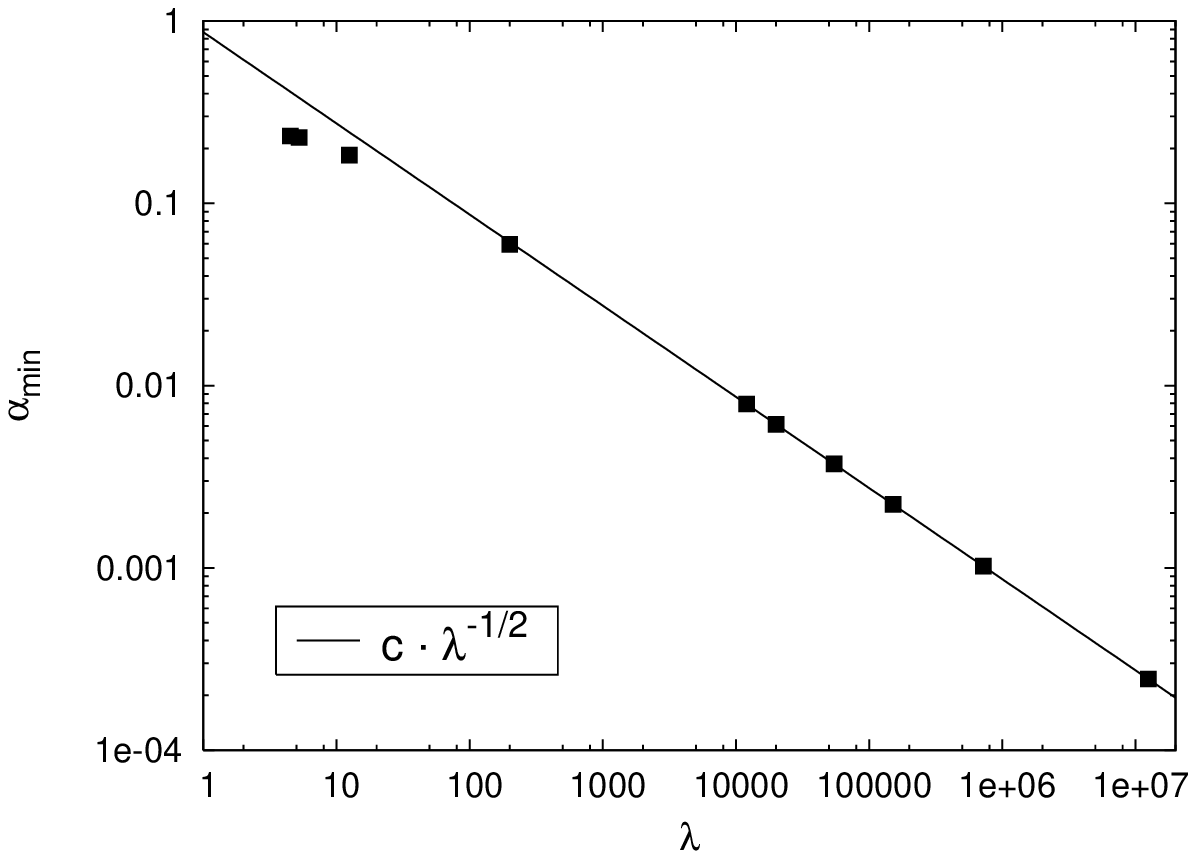}
\end{center}
\vspace{-0.5cm}
\caption{\small
(a)
The scaled mass $\alpha M$ of $m=2$, $n=3$ solutions,
related to the new flat-space $\lambda$-branches
versus $\alpha$
for very large values of the scalar coupling $\lambda$.
(b)
The location of the isolated nodes $z$ and the radius $\rho$ of the 
vortex ring of the Higgs field,
the value of metric functions at the origin, $l(0)$ (c)
for the same set of solutions.
(d) The minimal value $\alpha_{\rm min}$ of the new $\alpha$-branches
of $m=2$, $n=3$ solutions versus $\lambda$.
(See Table~1 for the notation of the node and vortex ring configurations of
the Higgs field.)
}
\end{figure}

With help of Fig.~\ref{f-6} we can now
address the limit $\lambda \rightarrow \infty$
for these new $\alpha$-branches.
As $\lambda$ increases without bound, the minimal value $\alpha_{\rm min}$,
beyond which the new branches exist,
decreases towards zero.
(Note, that $\alpha_{\rm min} \sim 1/\sqrt{\lambda}$ for very large
$\lambda$.)
At the same time the scaled mass of the 
critical solutions at $\alpha_{\rm min}$
decreases towards the mass 
of the generalized BM solution with $n=3$ \cite{KK},
while their mass itself diverges.
Likewise,
the values of their metric functions at the origin 
approach those of the $n=3$ generalized BM solution
\cite{IKKS}.
Their nodes finally move continuously towards the origin,
and the solutions shrink to zero size as $\lambda \to \infty$.
Scaling the coordinates and Higgs field again reveals 
the generalized BM solution as limiting solution
of finite scaled mass and finite scaled size \cite{MAP,KKSg}.

This limiting behavior for the critical solutions
on the new $\alpha$-branches is similar to that observed for 
the solutions on the second $\alpha$-branches of
the fundamental solutions.
In both cases the respective generalized BM solution is approached.

One can understand the reason for this behaviour here,
by noting that, as the scalar field becomes infinitely heavy,
it decouples 
in these solutions
and, in the double limit of infinite $\lambda$ and zero $\eta$, it 
does not affect the dynamics of the gauge sector anymore.
In that sense the EYMH system
is getting actually truncated here to an EYM system.

\boldmath
\subsection{Topologically nontrivial sector: $m=3$, $n=3$}
\unboldmath

\subsubsection{Flat space solutions}

Let us now turn to solutions in the topologically nontrivial sector 
with $m=3$, $n=3$.
In these solutions a triply charged monopole resides at the origin,
providing the topological charge of the solutions.
Again we first address the flat-space solutions,
reviewing and supplementing previous results \cite{KNS}.

For $\lambda \rightarrow 0$,
the $m=3$, $n=3$ solution possesses a triply charged monopole at the origin
and two oppositely oriented vortex rings located symmetrically
above and below the $xy$-plane \cite{KKS}.
For larger values of $\lambda$ one expects a bifurcation to occur,
where new branches of solutions appear, which possess
a different node structure, with three isolated nodes on the symmetry axis.
For high values of $\lambda$ such a monopole-antimonopole chain (MAC)
should then represent the energetically most favourable solution.

Indeed, at $\lambda_{c_1} = 0.8099$ two new branches of
solutions appear which possess the node structure of MACs \cite{KKS}.
But as seen in Fig.~\ref{f-7},
the vortex ring solutions on the first (fundamental) $\lambda$-branch
turn already into MACs before $\lambda_{c_1}$ is reached:
As $\lambda$ increases, 
two nodes emerge from the origin and separate from
each other along the $z$-axis.
The solutions then possess
three nodes on the symmetry axis and two vortex rings,
located symmetrically above and below the $xy$-plane.
As $\lambda$ increases further,
the new nodes move further apart,
while the vortex rings shrink to zero size,
merging with the new nodes on the symmetry axis.

\begin{figure}[h!]
\lbfig{f-7}
\begin{center}
\hspace{0.0cm} (a)\hspace{-0.6cm}
\includegraphics[height=.25\textheight, angle =0]{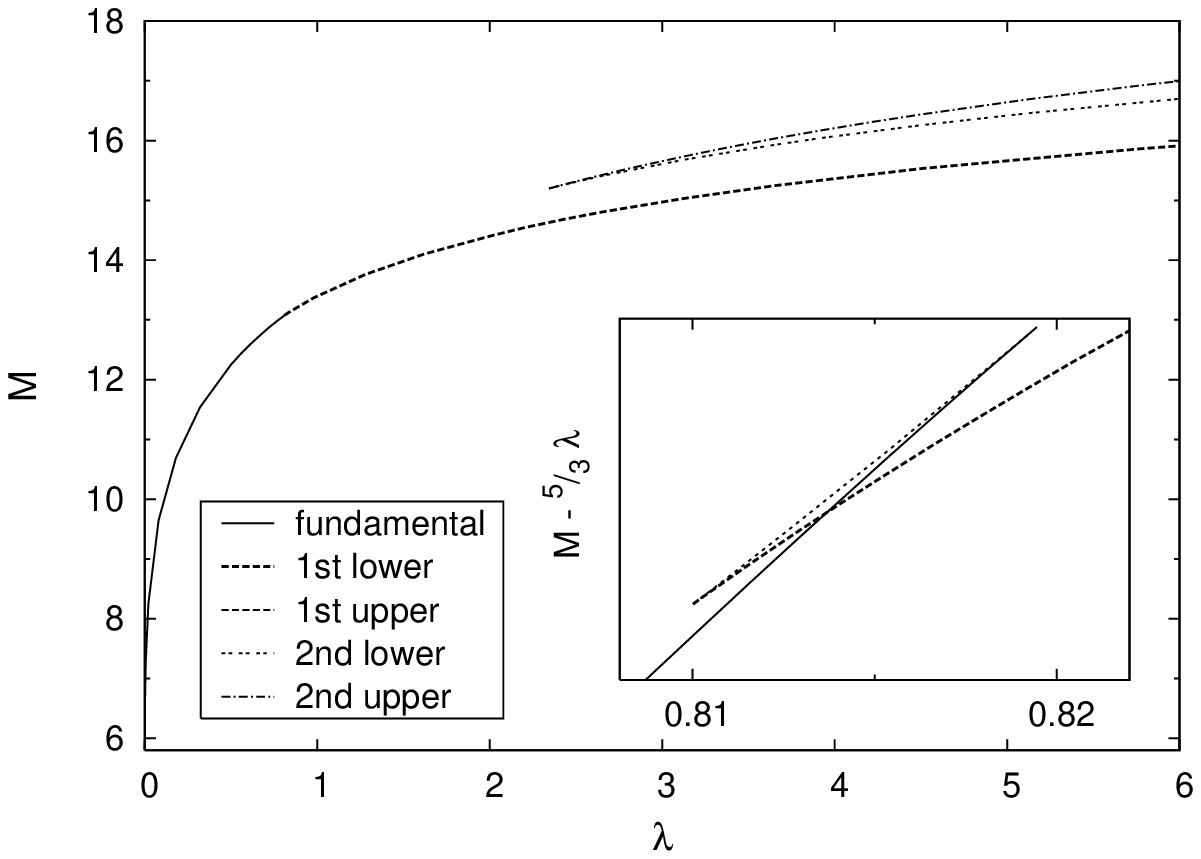}
\hspace{0.5cm} (b)\hspace{-0.6cm}
\includegraphics[height=.25\textheight, angle =0]{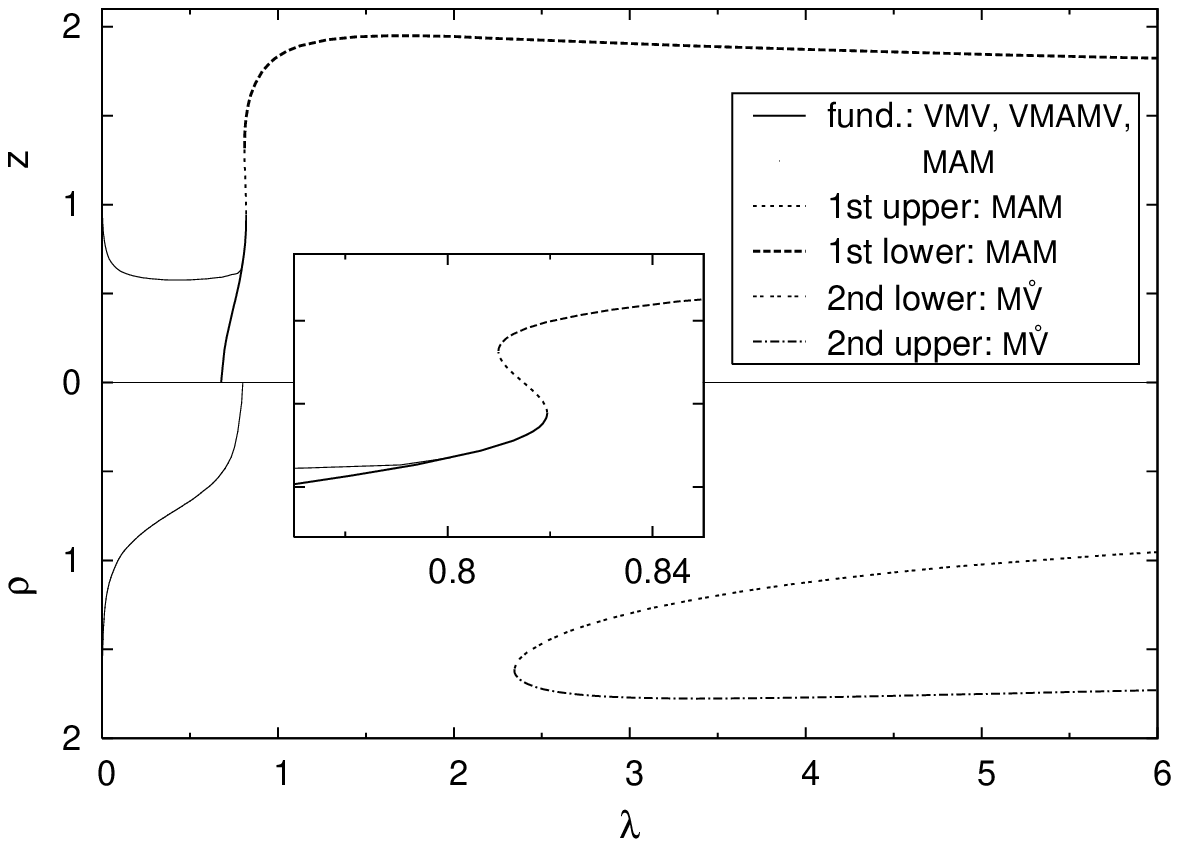}
\end{center}
\vspace{-0.5cm}
\caption{\small
(a)
The mass $M$ of the flat-space solutions on the fundamental $\lambda$-branch
as well as on the additional $\lambda$-branches for $m=3$, $n=3$ solutions.
(b)
The location of the isolated nodes $z$ and vortex rings $(\rho, z)$ 
of the Higgs field for the same set of solutions.
(See Table~2 for the notation of the node and vortex ring configurations of
the Higgs field.)}
\end{figure}

\begin{table}[h!]
\label{t-2}
\vspace{+3mm}
{\small
\begin{center}
\begin{tabular}{|p{1.6cm}|p{3cm}|p{9cm}|}\hline
\begin{center}\vspace{1mm}{\sf MAM}\end{center}
		&
		\vspace{-8mm}\begin{center}\includegraphics[height=.104\textheight, angle =0]{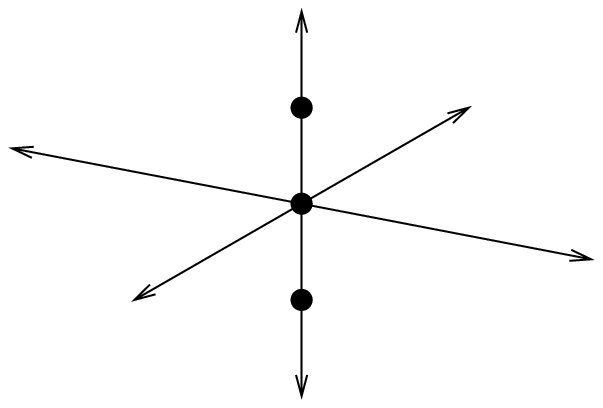}\end{center}\vspace{-11mm}
		&	
\begin{center}\vspace{-4mm}
A monopole-antimonopole
		chain with two monopoles on the symmetry axis, one above and 
		one below the $xy$-plane, and an antimonopole at the origin. 
\vspace{-1mm} \end{center}
	\\ \hline 
\begin{center}\vspace{1mm}{\sf VMV}\end{center}
		& 
		\vspace{-8mm}\begin{center}\includegraphics[height=.104\textheight, angle =0]{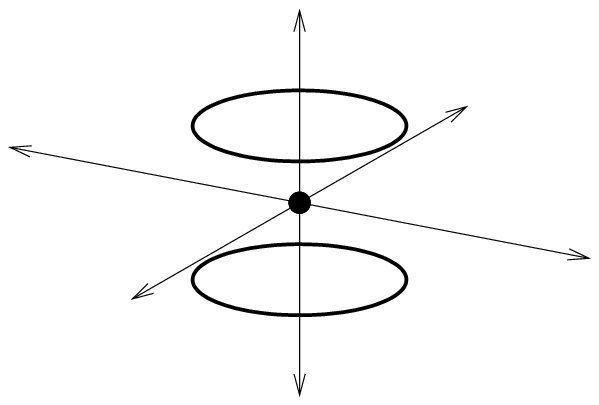}\end{center}\vspace{-11mm}
		& 
\begin{center}\vspace{-2mm}
		A monopole at the origin and two vortex rings,
		one above and one below the $xy$-plane.	
\vspace{-1mm}\end{center}
	\\ \hline
\begin{center}\vspace{0mm}{\sf M\r{V}}\end{center}
		& 
		\vspace{-8mm}\begin{center}\includegraphics[height=.104\textheight, angle =0]{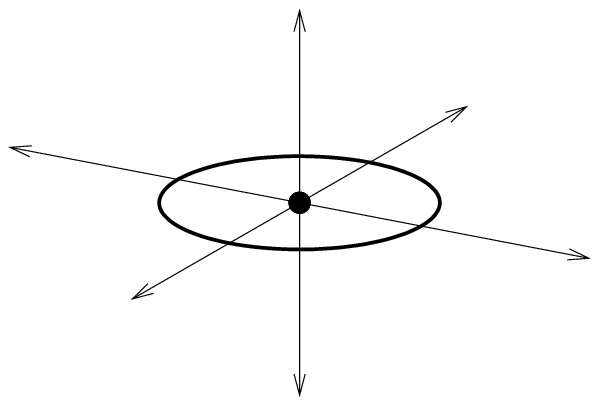}\end{center}\vspace{-11mm}
		& 
\begin{center}\vspace{-2mm}
		A monopole at the origin and a vortex 
		ring in the $xy$-plane. 	
\vspace{-4mm}\end{center}
				\\ \hline
\begin{center}\vspace{1mm}{\sf VMAMV}\end{center}
		& 
		\vspace{-8mm}\begin{center}\includegraphics[height=.104\textheight, angle =0]{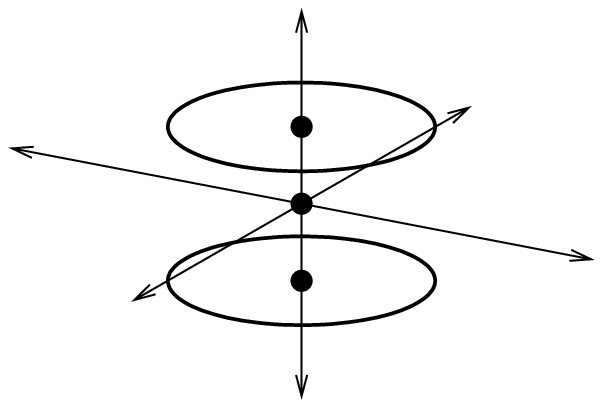}\end{center}\vspace{-11mm}
		& 
\begin{center}\vspace{-4mm}
	Two vortex rings, one above and one below the $xy$-plane, 
	two monopoles on the symmetry axis and an antimonopole at the origin. 
\vspace{-1mm} \end{center}
 \\ \hline
\begin{center}\vspace{0mm}{\sf VM\r{V}V}\end{center}
		& 
		\vspace{-8mm}\begin{center}\includegraphics[height=.104\textheight, angle =0]{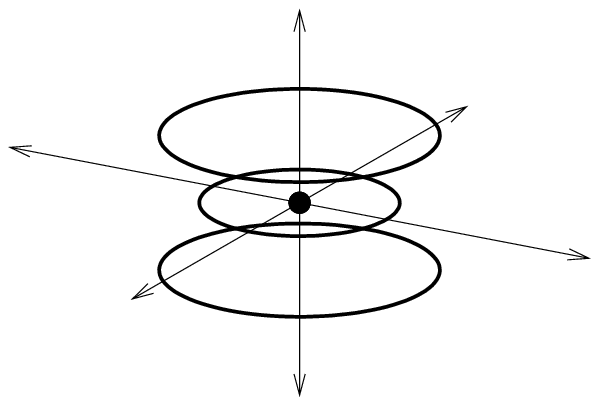}\end{center}\vspace{-11mm}
		& 
\begin{center}\vspace{-4mm}
		Three vortex rings, one above and one below the {$xy$-plane}
	and one in the $xy$-plane, and a monopole at the origin.
\vspace{-1mm} \end{center}
\\ \hline
\end{tabular}
\end{center}
}
\vspace{-3mm}
\caption{\small Configurations of the nodes and vortex rings
 of the Higgs field for $m=3$, $n=3$ solutions.}
\end{table}

Thus beyond a certain value of $\lambda$,
which is smaller than the first critical value $\lambda_{c_1}$,
no vortex ring solutions are present any longer \cite{KKS}.
The fundamental branch and the new upper branch then merge
at $\lambda_{c_2} = 0.8194$,
while the new lower branch continues to large values of $\lambda$,
representing the fundamental branch there.
This is seen in Fig.~\ref{f-7}, where the mass and the node structure
of the $m=3$, $n=3$ $\lambda$-branches are exhibited,
with emphasis on the three branches of MACs
present in the small range of $\lambda$,
$\lambda_{c_1} \le\lambda\le \lambda_{c_2}$.

Interestingly, a further bifurcation arises at $\lambda_{c_3}=2.3436$,
which was missed before.
At $\lambda_{c_3}$ two new branches of vortex ring solutions appear,
whose single ring is located in the $xy$-plane.
Thus beyond $\lambda_{c_3}$, vortex ring solutions are present again,
but possess higher masses than the
fundamental MAC solutions. These new $\lambda$-branches
are also exhibited in Fig.~\ref{f-7}.

\subsubsection{Gravitating solutions}

When gravity is coupled, branches of gravitating solutions arise 
from these $m=3$, $n=3$ flat-space configurations.
Depending on the value of $\lambda$, 
two or more $\alpha$-branches are present,
whose nature is considered in detail in the following.


The $\alpha$-dependence of the $\lambda < \lambda_{c_1}$ solutions
is simple. A first $\alpha$-branch emerges from the flat-space
solution and merges at some $\alpha_{\rm max}$
with a second $\alpha$-branch, which connects to the
generalized BM solution in the limit $\alpha \rightarrow 0$.
Indeed,
for small values of $\alpha$ on the second branch, the solutions
may be thought of as composed of a scaled $n=3$ generalized BM solution
in an inner region and a flat-space charge-3 monopole configuration
in an outer region \cite{KKSg}.
In the limit $\alpha \to 0$, the mass of the solutions diverges,
while their scaled mass approaches the mass of the 
generalized $n=3$ BM solution.       

{\sl Solutions in the range $\lambda_{c_1} \le\lambda\le \lambda_{c_2}$}

While for $\lambda < \lambda_{c_1}$ only a single flat-space solution
is present, there are three flat-space solutions
in the range $\lambda_{c_1} \le\lambda\le \lambda_{c_2}$.
This complicates the picture considerably.
We exhibit in Fig.~\ref{f-8} the node structure of the
$\alpha$-branches, associated with all three flat space solutions,
for several values of $\lambda$ in this critical range.

\begin{figure}[h!]
\lbfig{f-8}
\begin{center}
\hspace{0.0cm} (a)\hspace{-0.6cm}
\includegraphics[height=.25\textheight, angle =0]{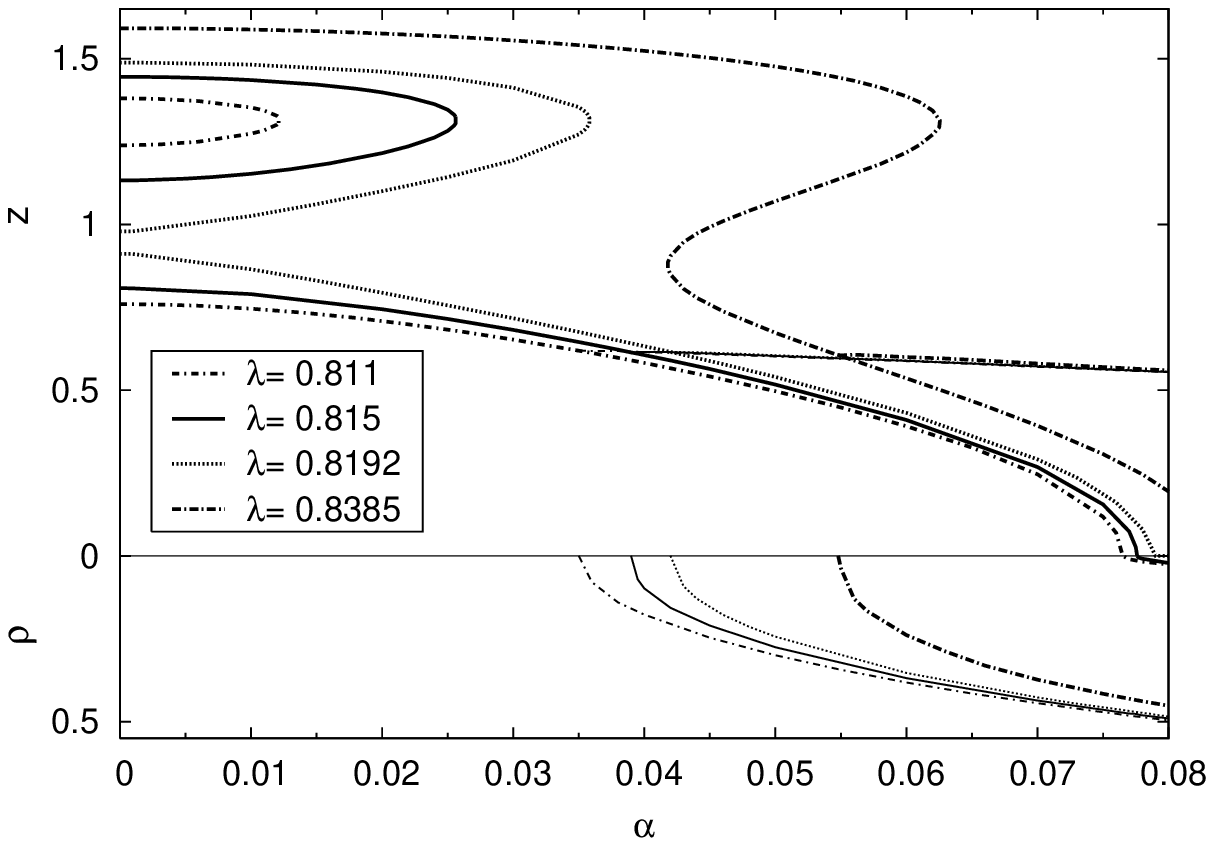}
\hspace{0.5cm} (b)\hspace{-0.6cm}
\includegraphics[height=.25\textheight, angle =0]{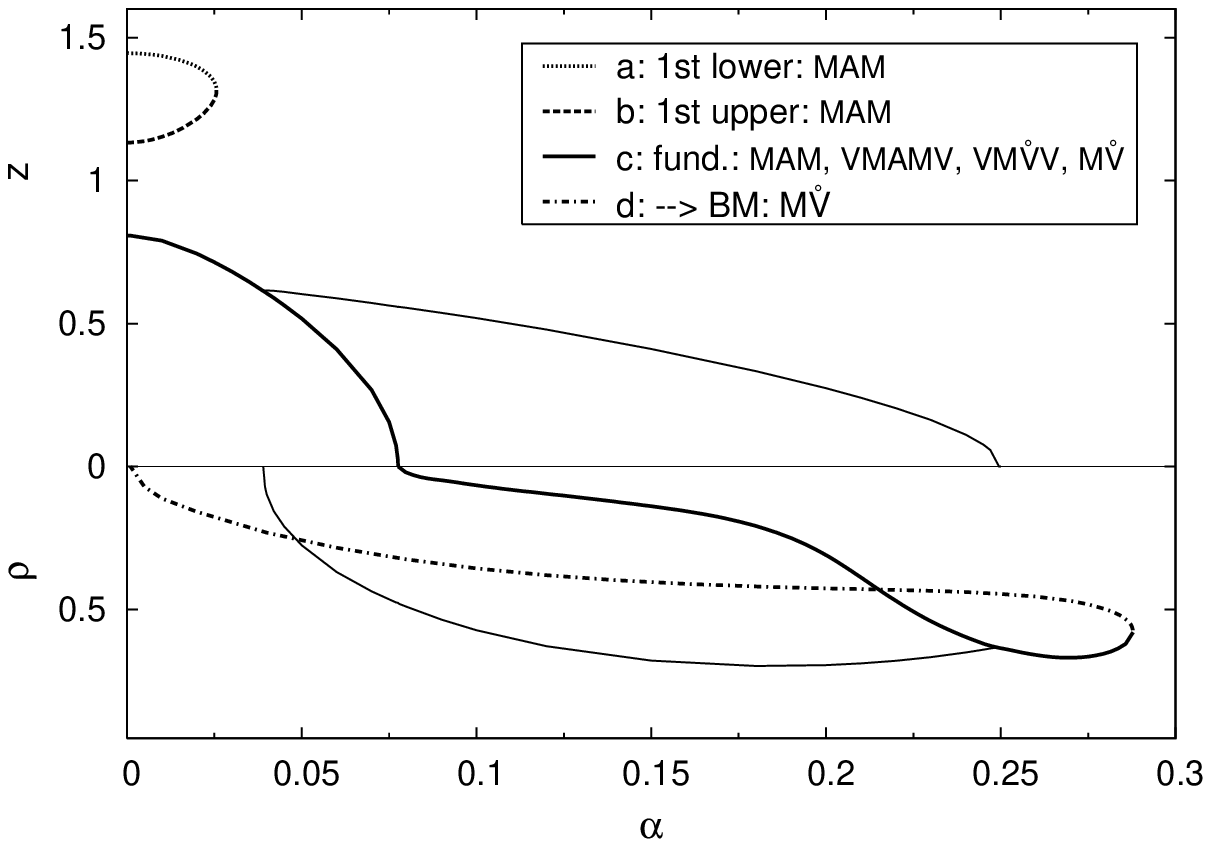}
\end{center}
\vspace{-0.5cm}
\caption{\small
(a)
The location of the isolated nodes $z$ and vortex rings $(\rho, z)$ 
of the Higgs field of $m=3$, $n=3$ solutions
versus $\alpha$
for several values of the scalar coupling $\lambda$
in the critical range $\lambda_{c_1} \le\lambda\le \lambda_{c_2}$ and beyond.
(b)
The isolated nodes and vortex rings of the Higgs field for the complete set of 
solutions for $\lambda=0.815$.
(See Table~2  for the notation of the node and vortex ring configurations of
the Higgs field.)}
\end{figure}

While the left part of the figure shows the evolution of the zeros
of the Higgs field with increasing scalar coupling $\lambda$,
the right part gives for a particular value, $\lambda=0.815$,
the assignment of the branches, to which the zeros belong.
The $\alpha$-branch emerging from the fundamental $\lambda$-branch 
is labeled {\sf c}, and the $\alpha$-branches emerging
from the new lower $\lambda$-branch and the new upper $\lambda$-branch
are labeled {\sf a} and {\sf b}, respectively.

It is intuitively clear
that in the complicated pattern of interaction between the monopoles
and antimonopoles,
the gravitational attraction plays a role similar to the scalar interaction,
thus one can expect that coupling to gravity may provide a compensation
for the weakening of the scalar interaction.
This is indeed, what we here observe 
regarding the existence of the solutions
and their node structure.
For increasing $\alpha$, 
the branches {\sf a} and {\sf b} merge and disappear,
while the branch {\sf c} persists, but changes character,
reversing the steps seen for the fundamental flat-space $\lambda$-branch
in Fig.~\ref{f-7}.

Indeed, at $\alpha=0$, all three branches begin as MAC solutions.
As $\alpha$ increases, the two MAC branches {\sf a} and {\sf b} merge and end
at a small $\alpha_{\rm max}$.
The branch {\sf c}, however, extends further and changes its node structure
accordingly, as $\alpha$ increases.
First a vortex ring emerges from each outer node.
Next the nodes move towards each other, merging at the origin.
Then a central vortex ring appears in the $xy$-plane, and finally
the outer rings merge with the central ring.

The branch {\sf c} still persists to somewhat larger values of $\alpha$,
and then merges at an $\alpha_{\rm max}$ with a second branch
of vortex ring solutions labeled {\sf d}. 
As expected, this branch extends backwards to $\alpha=0$,
where it connects to a generalized BM solution.

The evolution of the solutions with increasing $\lambda$
can now be understood from the left part of the figure.
As $\lambda$ increases, the branches {\sf a} and {\sf b} extend to larger
values of $\alpha$, before they merge.
At the same time the nodes of the branches {\sf b} and {\sf c} approach each
other. At the critical value $\lambda_{c_2}$
the two flat-space branches merge, and their nodes coincide.
As $\lambda$ increases further, this critical point
(where two branches merge) is retained,
but it moves to finite values of $\alpha$.
Note, that the pattern of nodes of the gravitating solutions 
in the interval $[0.03 < \alpha < 0.07]$ for $\lambda=0.8385$
exhibited in Fig.~\ref{f-8},
nicely reflects the (reversed) pattern of nodes of the flat-space solutions
in the interval $[0.8 < \lambda < 0.85]$
exhibited in Fig.~\ref{f-7}.

{\sl Emergence of the new flat-space branches at $\lambda_{c_3}$}

\begin{figure}
\lbfig{f-9}
\begin{center}
        \hspace{0.0cm} (a) \hspace{-0.7cm}
                \includegraphics[height=.25\textheight, angle =0]{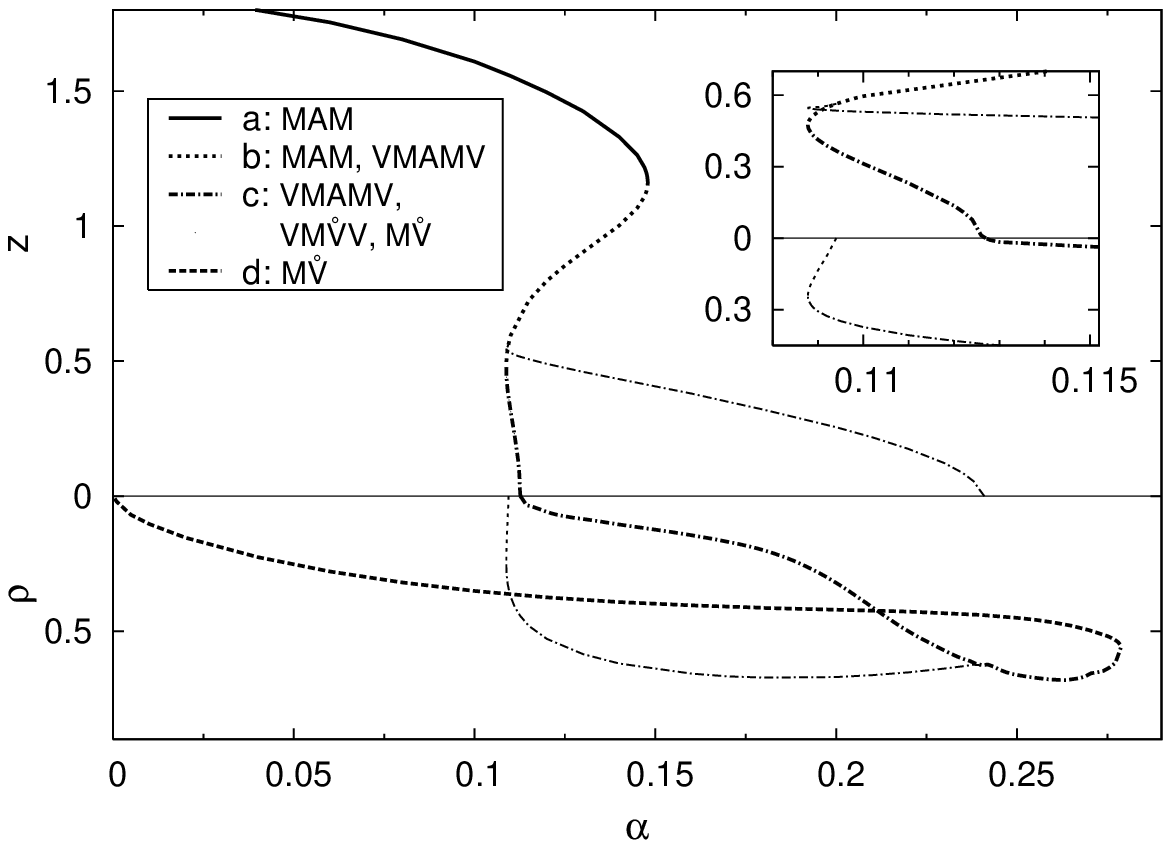}
        \hspace{0.0cm} (b) \hspace{-0.7cm}
                \includegraphics[height=.25\textheight, angle =0]{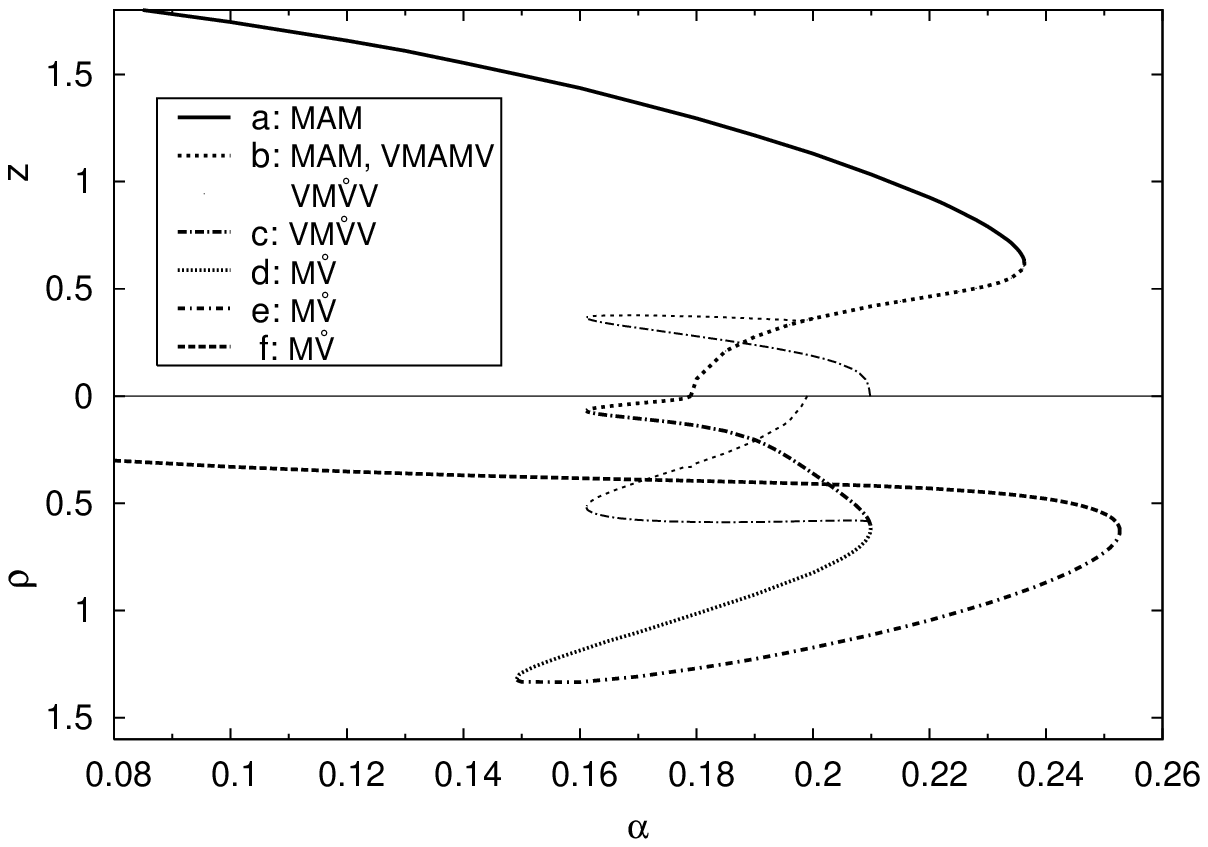}
        \hspace{0.0cm} $\phantom{(b)}$ \hspace{-0.7cm}
\\
\hspace{0.0cm} (c) \hspace{-0.7cm}
\begin{minipage}[b]{15.2cm}
\begin{minipage}[t]{15.0cm}
  \includegraphics[height=.084\textheight, angle =0]{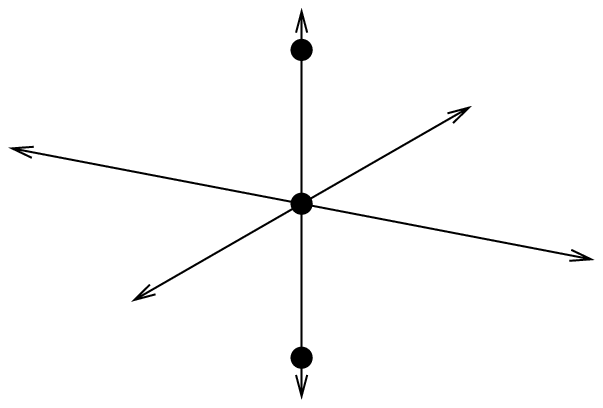}
  \includegraphics[height=.084\textheight, angle =0]{MAM2.eps}
  \includegraphics[height=.084\textheight, angle =0]{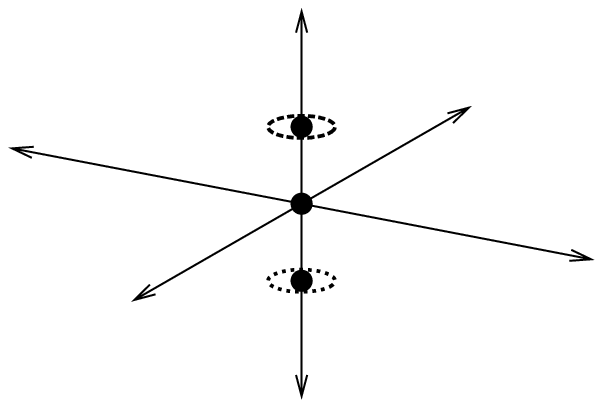}
  \includegraphics[height=.084\textheight, angle =0]{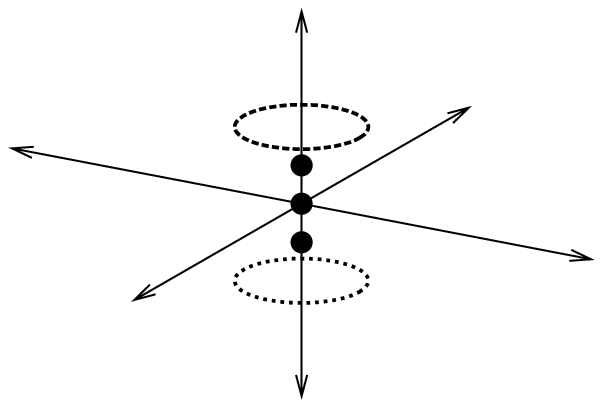}
  \includegraphics[height=.084\textheight, angle =0]{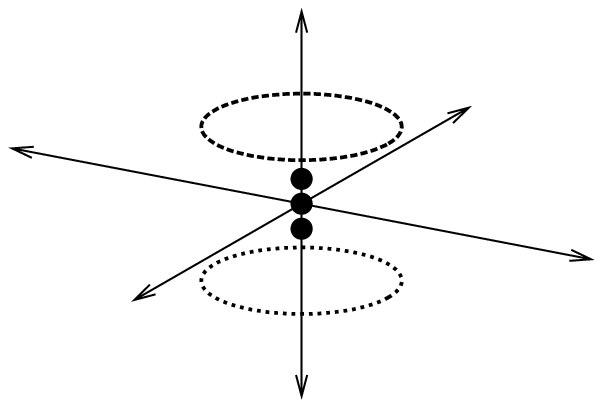}
  \includegraphics[height=.084\textheight, angle =0]{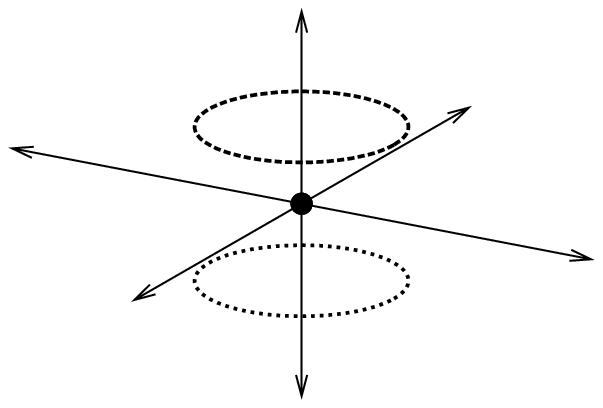}
        \end{minipage}\hfill
        \vspace{7mm}
        \begin{minipage}[b]{15.0cm}
  \includegraphics[height=.084\textheight, angle =0]{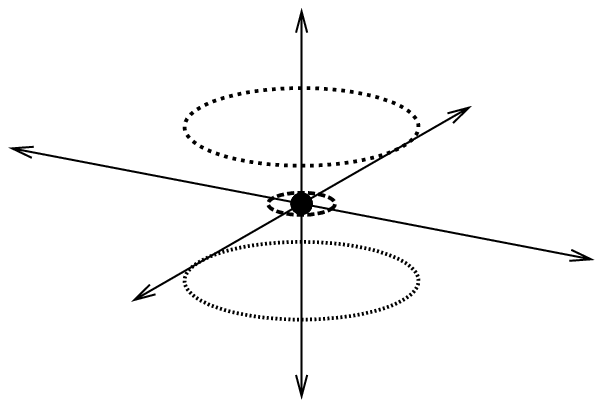}
  \includegraphics[height=.084\textheight, angle =0]{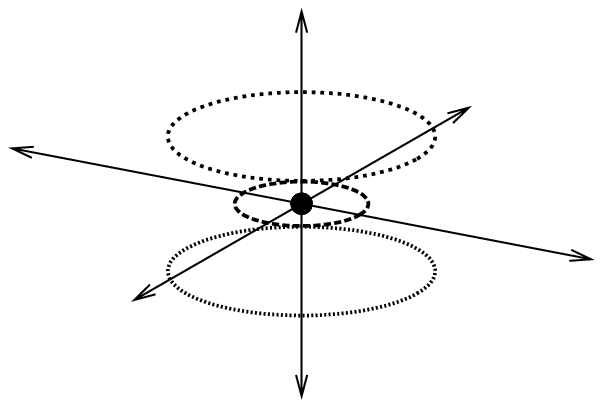}
  \includegraphics[height=.084\textheight, angle =0]{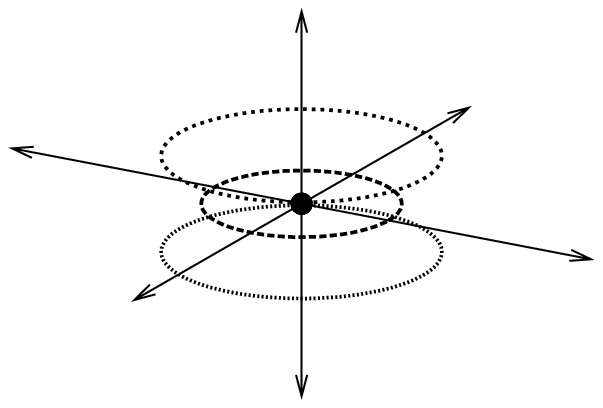}
  \includegraphics[height=.084\textheight, angle =0]{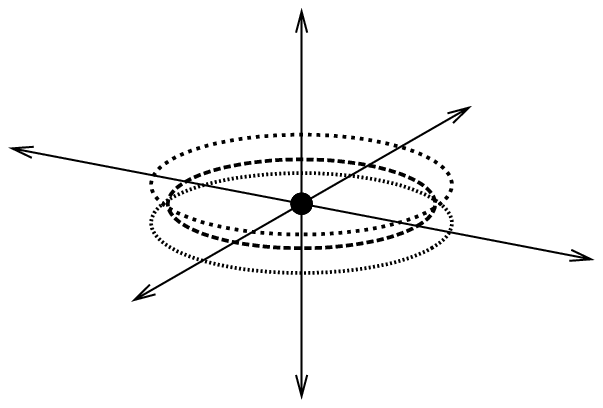}
  \includegraphics[height=.084\textheight, angle =0]{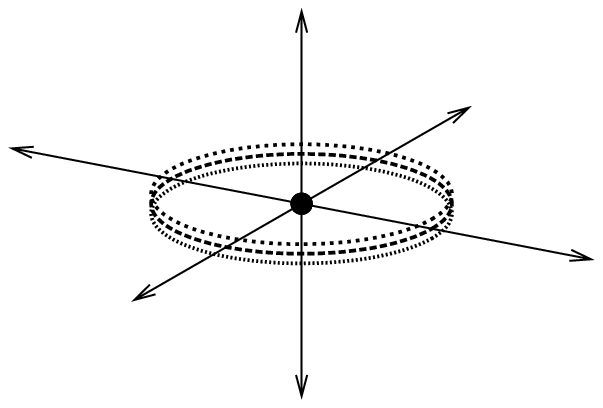}
  \includegraphics[height=.084\textheight, angle =0]{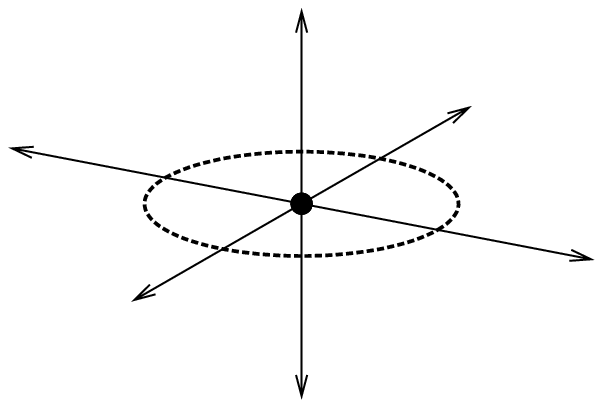}
        \vspace{1mm}
        \end{minipage}
\end{minipage}
\\
\vspace{0.5cm}
\hspace{0.0cm} (d) \hspace{-0.7cm}
\includegraphics[height=.25\textheight , angle =0 ]{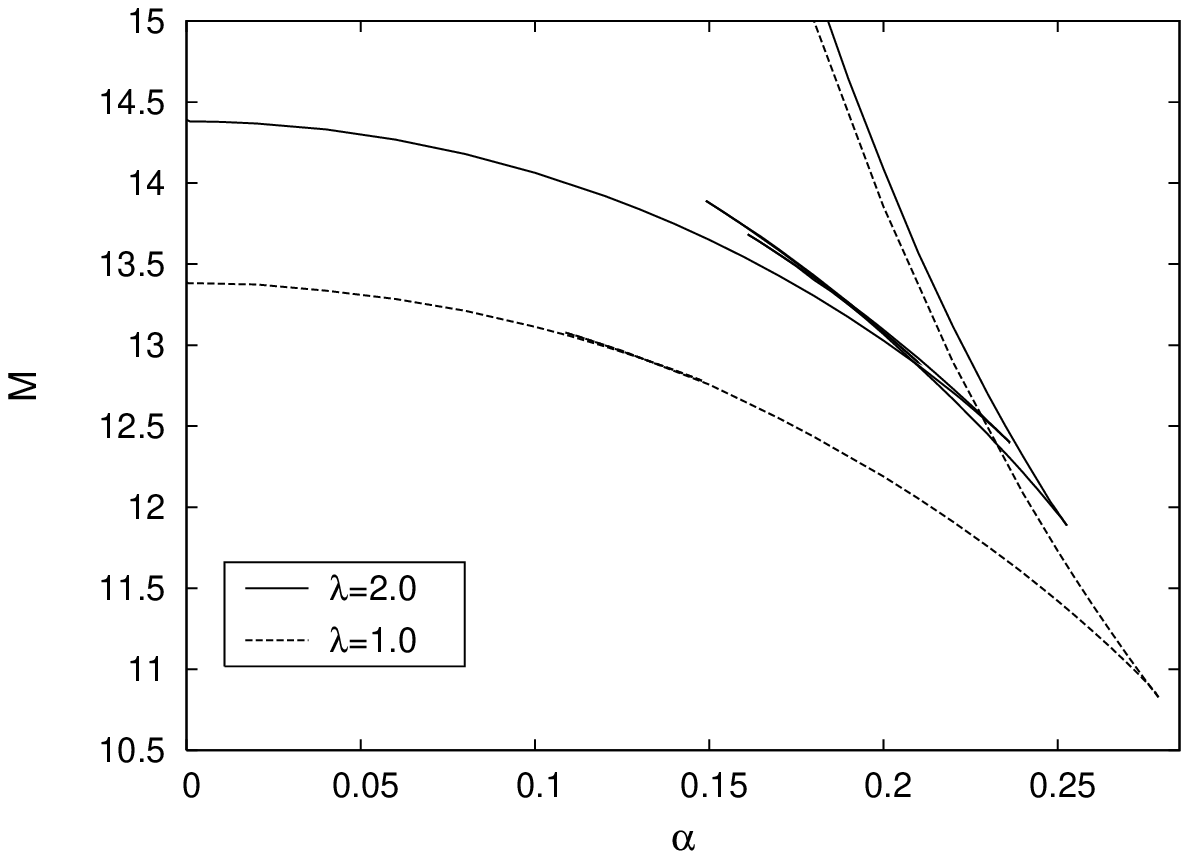}
\hspace{0.5cm} (e)\hspace{-0.7cm}
\includegraphics[height=.25\textheight , angle =0 ]{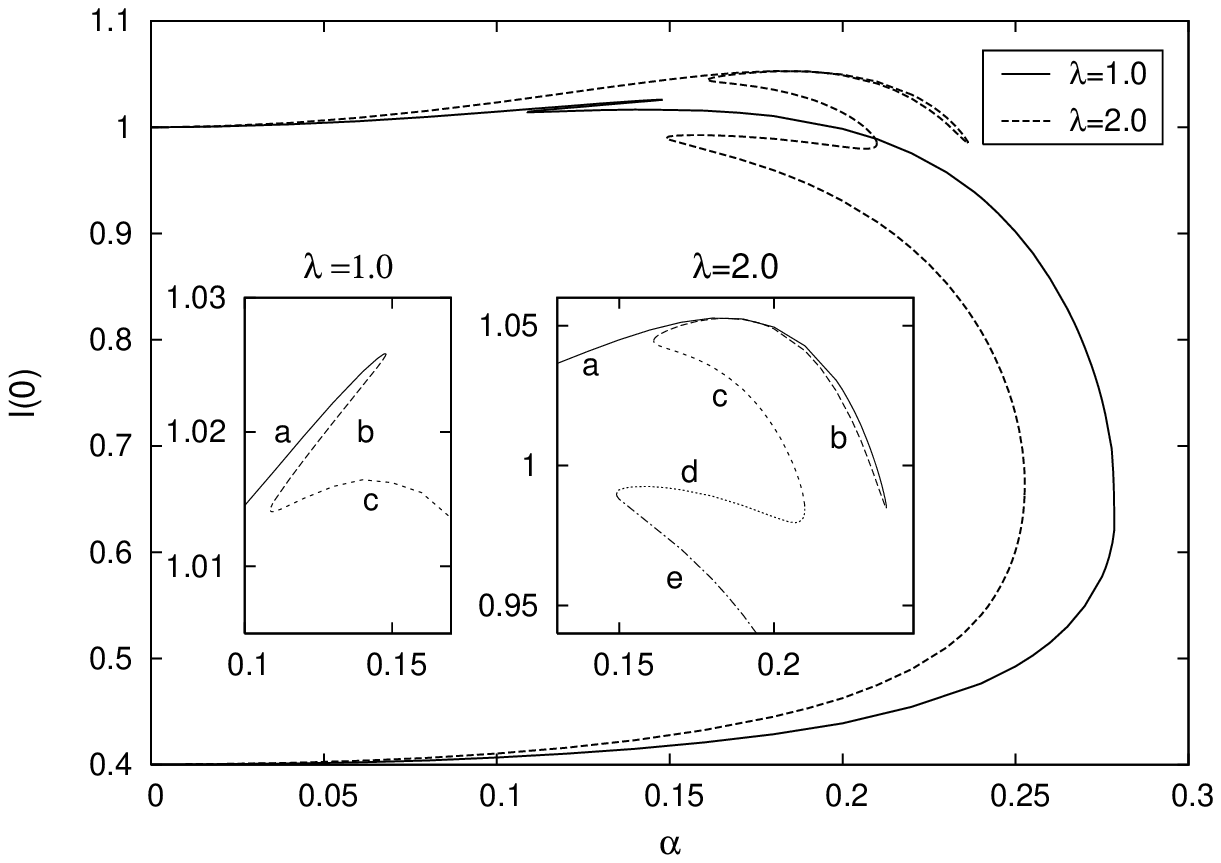}
\end{center}
\caption{\small
The location of the isolated nodes $z$ and vortex rings $(\rho, z)$
 of the Higgs field for the complete set of
solutions for $\lambda=1$ (a) and $\lambda=2$ (b) versus $\alpha$.
Change of the structure of the nodes of the Higgs field following 
the evolution of the $\alpha$-branches (c).
The mass $M$ (d) and the value of metric function at the origin $l(0)$ (e) 
for the same set of solutions.
}
\end{figure}

As seen above, the bifurcations of the flat-space branches at
$\lambda_{c_1}$ and $\lambda_{c_2}$ are reflected by bifurcations
in curved space. 
In particular, the critical values of the
gravitational coupling strength $\alpha$,
where bifurcations arise 
for a given value of the scalar coupling strength $\lambda$,
move towards smaller values of $\alpha$
with decreasing $\lambda$. 
As the minimal critical value of $\alpha$ reaches zero,
the flat-space bifurcation point $\lambda_{c_2}$ arises.

It is thus obvious, that the new flat-space $\lambda$-branches
arising at $\lambda_{c_3}$, should also have precursors
in curved space. These precursors indeed led to the
detection of the new flat-space branches.
In the following we analyze the branches of gravitating solutions,
obtained with increasing scalar coupling strength $\lambda$,
to clarify the emergence of the new flat-space branches.

Let us first consider the sets of solutions at $\lambda=1$ and $\lambda=2$,
exhibited in Fig.~\ref{f-9}.
While the node structure nicely reveals the types of solutions
present at the respective value of $\lambda$,
and their evolution with $\alpha$, 
the values of the metric functions at the origin are instructive
to get an overview of the sets of solutions present.
For better identification,
we again label the various branches of solutions {\sf a}, {\sf b}, 
{\sf c}, etc.

For $\lambda=1$ there are four $\alpha$-branches, just like for the smaller
$\lambda$ values of Fig.~\ref{f-8}.
When moving continuously along these $\alpha$-branches,
the node structure changes from MAC, present for {\sf a} and most of {\sf b},
to MAC plus vortex rings, shortly before the bifurcation with {\sf c}.
The size and the location of the nodes and the rings then evolves
along {\sf c} until a single ring is left in the $xy$-plane (together with the
always present node at the origin).
This node structure is then retained on branch {\sf d}, 
where the solutions evolve
towards the generalized BM solution as $\alpha \rightarrow 0$.

As $\lambda$ increases further, interestingly, another bifurcation arises,
and two more $\alpha$-branches appear.
Consequently, there is now an interval of $\alpha$ with six $\alpha$-branches.
The branches labeled {\sf a} to {\sf f} 
are exhibited in the figure for $\lambda=2$.
Moving continuously along the branches, one again observes the
structure of the solutions change from MACs to MACs plus vortex rings,
and then to a monopole plus vortex ring(s).

Most important for understanding the emergence of the
new flat-space branches is the evolution of
the branches {\sf d} and {\sf e} with $\lambda$,
with particular emphasis on the critical value
of $\alpha$, where these two branches merge.
We therefore consider in Fig.~\ref{f-10} sets of
solutions with still higher values of $\lambda$.

\begin{figure}[h!]
\lbfig{f-10}
\begin{center}

\hspace{0.0cm} (a)\hspace{-0.8cm}
\includegraphics[height=.23\textheight, angle =0]{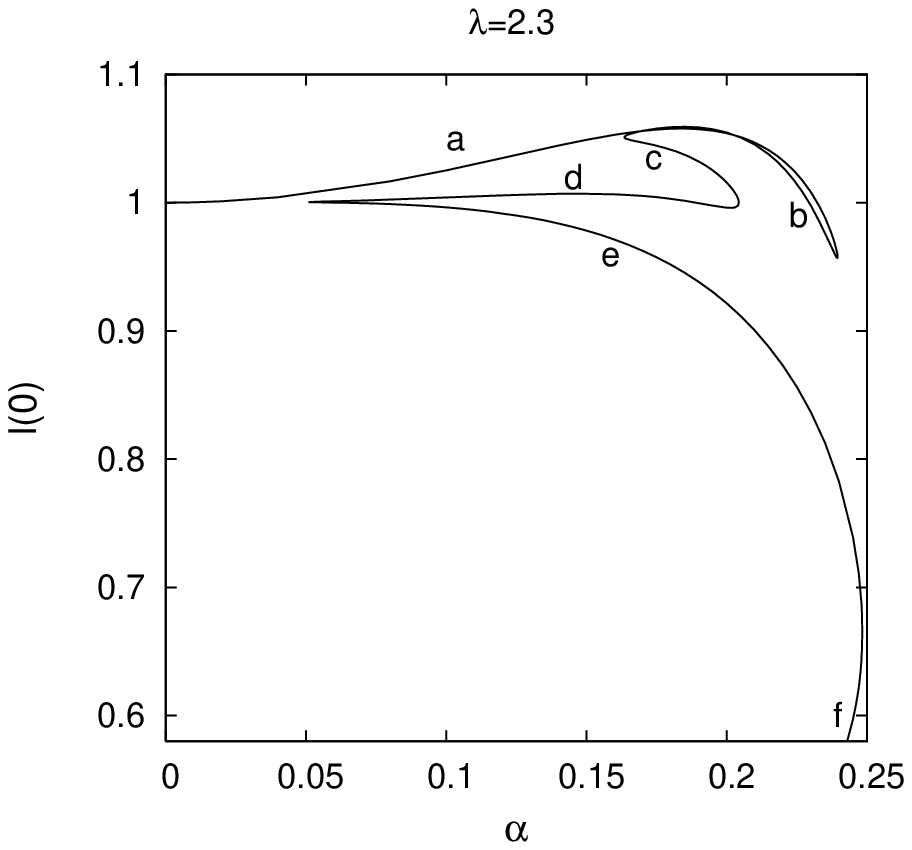}
\hspace{0.1cm} (b)\hspace{-0.8cm}
\includegraphics[height=.23\textheight , angle =0 ]{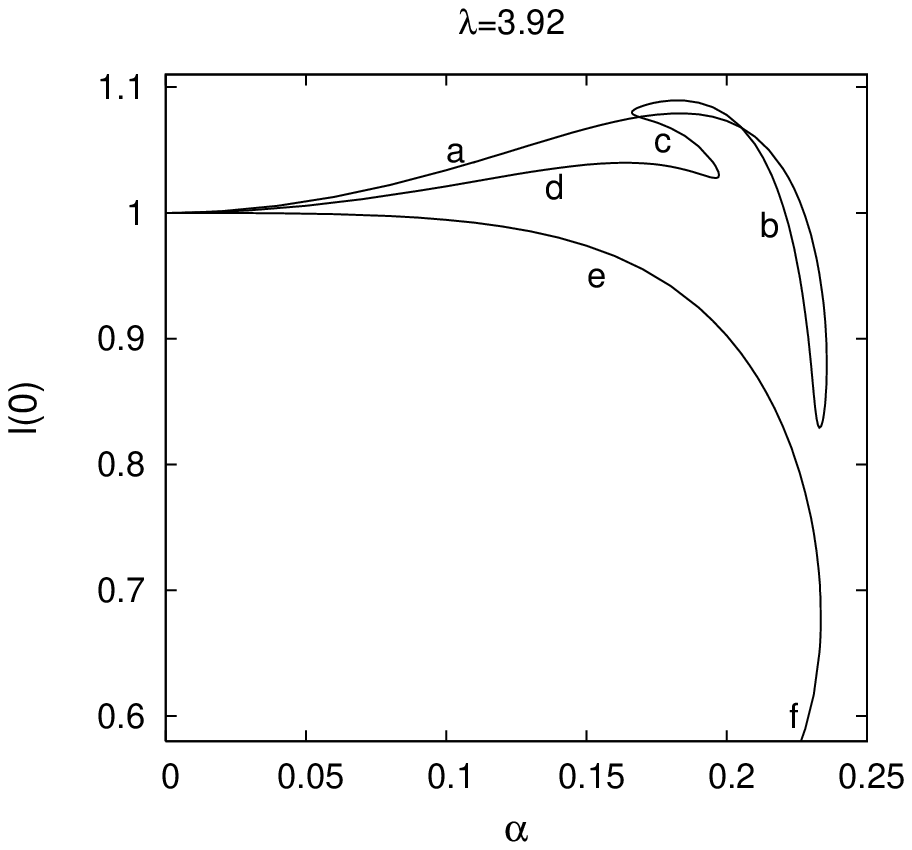}
\hspace{0.1cm} (c)\hspace{-0.8cm}
\includegraphics[height=.23\textheight , angle =0 ]{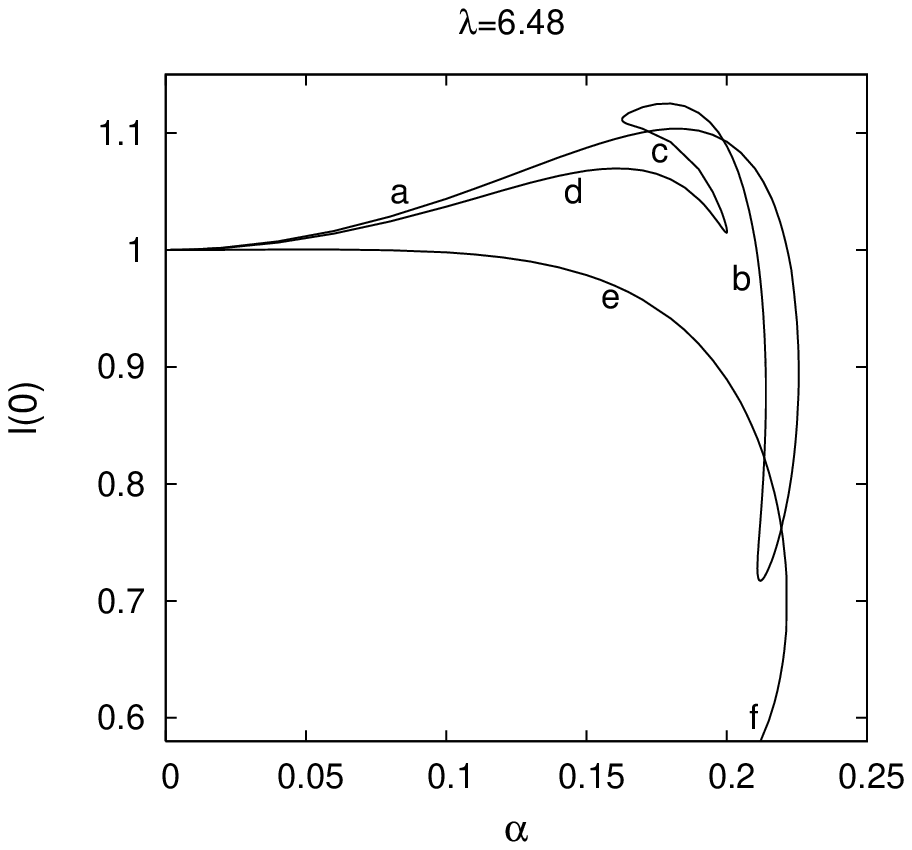}
\end{center}
\vspace{-0.5cm}
\caption{\small
The value of metric function at the origin, $l(0)$,
for $\lambda=2.3$ (a) and $\lambda=3.92$ (b)
and for  $\lambda=6.48$ (c).
}
\end{figure}

As $\lambda$ increases further, the branches {\sf d} and {\sf e}
extend to increasingly smaller values of $\alpha$.
At the critical value $\lambda_{c_3}$ of the scalar coupling,
the critical value of $\alpha$, where the branches {\sf d} and {\sf e}
merge, then precisely reaches zero,
giving rise to the bifurcation point
of the new flat-space solutions.

For values of the scalar coupling beyond $\lambda_{c_3}$,
the set of solutions then splits into two disconnected parts.
The first part emerges with branch {\sf a} 
from the respective fundamental flat-space solution,
evolves via branch {\sf b} and branch {\sf c}, and reaches along branch {\sf d} 
the new upper mass flat-space solution
in the limit $\alpha \rightarrow 0$.
The second part emerges with branch {\sf e}
from the new upper mass flat-space solution,
and evolves with branch {\sf f} towards the limiting generalized BM
solutions in the limit $\alpha \rightarrow 0$.
This is demonstrated in Fig.~\ref{f-10} for the set of solutions with
$\lambda=3.92$. 

{\sl Solutions at large $\lambda$}

In the case of the $m=2$, $n=3$ solutions we observed the appearance 
of new $\alpha$-branches of solutions for large values of $\lambda$,
exhibiting an intriguing behaviour:
In the limit $\lambda \rightarrow \infty$
the minimal value $\alpha_{\rm min}$,
beyond which these new branches exist,
decreases towards zero,
while the critical solutions themselves
approach the generalized BM solution
as $\alpha_{\rm min} \rightarrow \infty$.

We here observe an analogous pattern for the $m=3$, $n=3$ solutions.
To see this pattern arising, we reconsider the evolution
of the branches {\sf a} and {\sf b} with $\lambda$, exhibited
in Fig.~\ref{f-9} and Fig.~\ref{f-10}.
As $\lambda$ increases,
the branches {\sf a} and {\sf b}
first elongate towards larger values of $\alpha$.
Then at a critical value of $\lambda$, 
a new bifurcation arises, and two additional branches appear.
The minimal value $\alpha_{\rm min}$ possible for these new branches
now decreases with increasing $\lambda$,
analogous to the case of the respective
branches of the $m=2$, $n=3$ solutions.

\boldmath
\subsection{Topologically trivial sector:  $m=4$, $n=3$}
\unboldmath

\subsubsection{Flat space solutions}

We now turn to the $m=4$, $n=3$ solutions,
which reside in the topologically trivial sector.
As before we first address the flat-space solutions,
reviewing and supplementing previous results \cite{KNS}.

We exhibit in Fig.~\ref{f-11} the mass and the node structure
of the $m=4$, $n=3$ flat-space $\lambda$-branches.
The fundamental $\lambda$-branch is present for arbitrary
scalar coupling strength.
Its solutions possess two vortex rings,
located symmetrically with respect to the $xy$-plane. 

\begin{figure}[h!]
\lbfig{f-11}
\begin{center}
\hspace{0.0cm} (a)\hspace{-0.6cm}
\includegraphics[height=.25\textheight, angle =0]{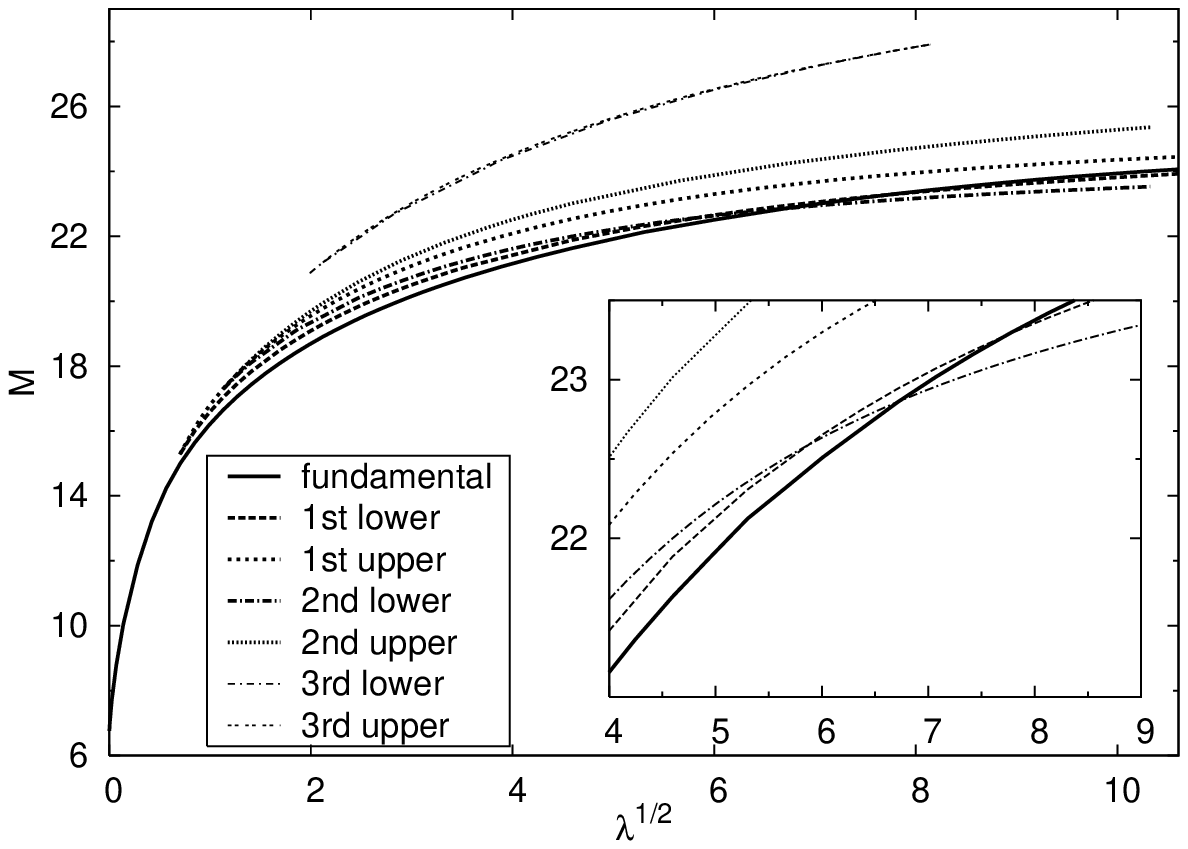}
\hspace{0.5cm} (b)\hspace{-0.6cm}
\includegraphics[height=.25\textheight, angle =0]{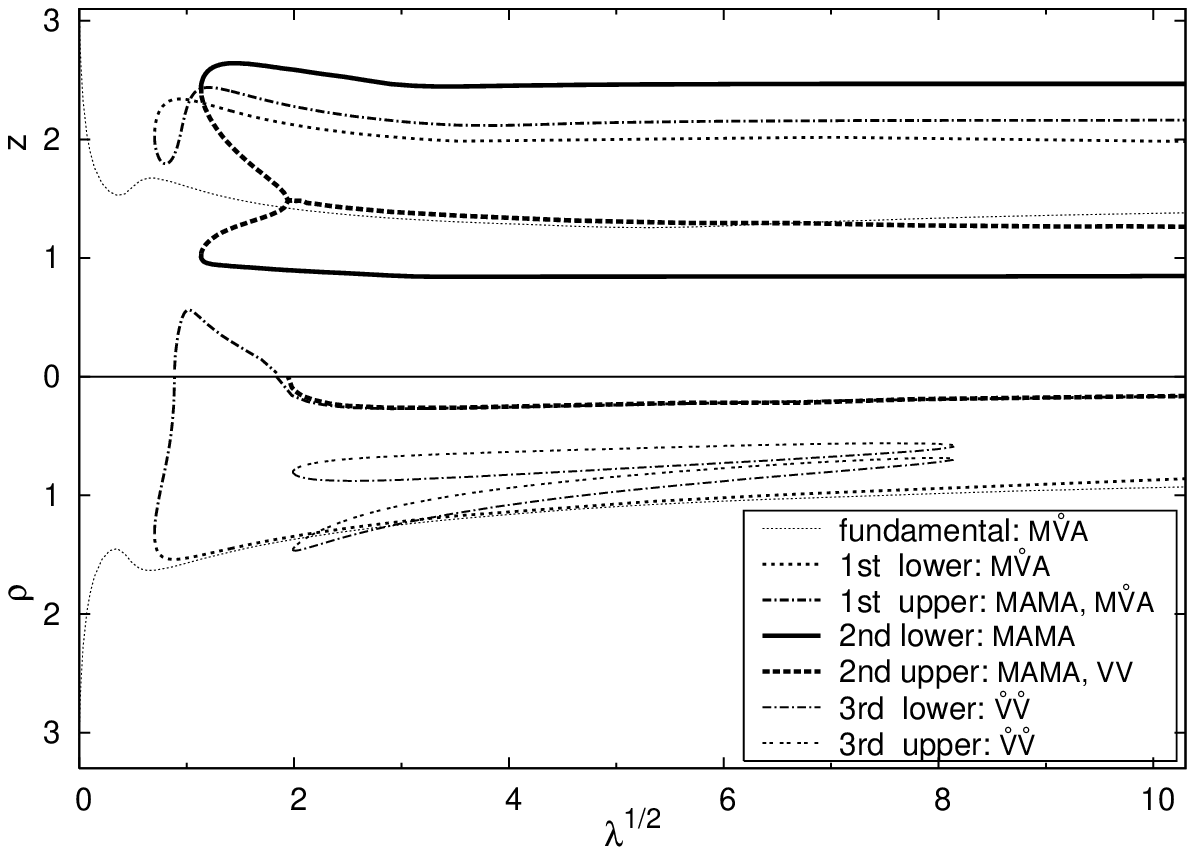}
\end{center}
\vspace{-0.5cm}
\caption{\small
(a)
The mass $M$ of the flat-space solutions on the fundamental $\lambda$-branch
as well as on the additional $\lambda$-branches for $m=4$, $n=3$ solutions.
(b)
The location of the isolated nodes $z$ and vortex rings $(\rho, z)$
 of the Higgs field for the same set of solutions.
(See Table~3  for the notation of the node and vortex ring configurations of
the Higgs field.)}
\end{figure}

\begin{table}[h!]
\label{t-3}
\vspace{3mm}
{\small
\begin{center}
\begin{tabular}{|p{1.6cm}|p{3cm}|p{8cm}|}\hline
\begin{center}\vspace{1mm}{\sf MAMA}\end{center}
		& 
		\vspace{-8mm}\begin{center}\includegraphics[height=.104\textheight, angle =0]{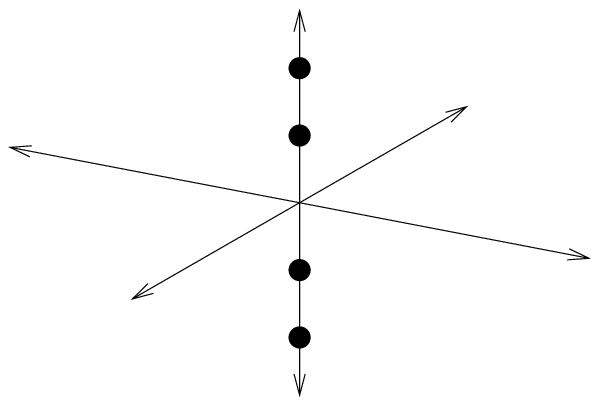}\end{center}\vspace{-11mm}
		& 
\begin{center} \vspace{-4mm}
		A chain with two monopoles and two antimonopoles 
                in alternating order
                on the symmetry axis.
\vspace{-7mm} \end{center}
\\ \hline 
\begin{center}\vspace{0mm}{\sf M\r{V}A}\end{center}
		& 
		\vspace{-8mm}\begin{center}\includegraphics[height=.104\textheight, angle =0]{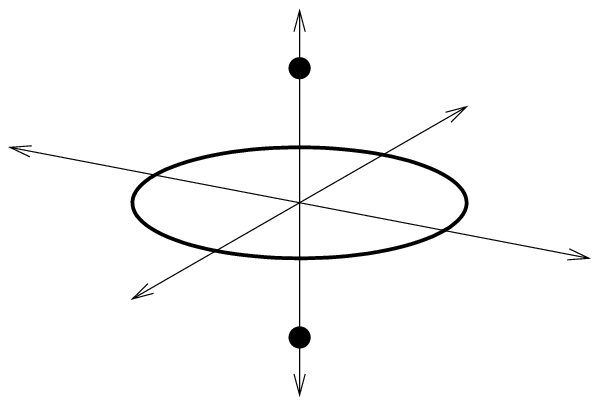}\end{center}\vspace{-11mm}
		&
\begin{center} \vspace{-4mm}
		A monopole above and an antimonopole below the $xy$-plane on the symmetry axis
		  and a vortex ring in the $xy$-plane.
\vspace{-2mm} \end{center}
\\ \hline
\begin{center}\vspace{1mm}{\sf VV}\end{center}
		& 
		\vspace{-8mm}\begin{center}\includegraphics[height=.104\textheight, angle =0]{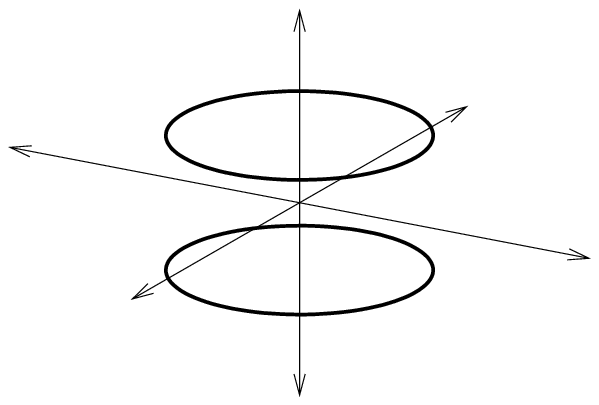}\end{center}\vspace{-11mm}
		& 
\begin{center} \vspace{-2mm}
		Two vortex rings, one above and one below the $xy$-plane.
\vspace{-1mm} \end{center}
\\ \hline
\begin{center}\vspace{0mm}{\sf \r{V}\r{V}}\end{center}
		& 
		\vspace{-8mm}\begin{center}\includegraphics[height=.104\textheight, angle =0]{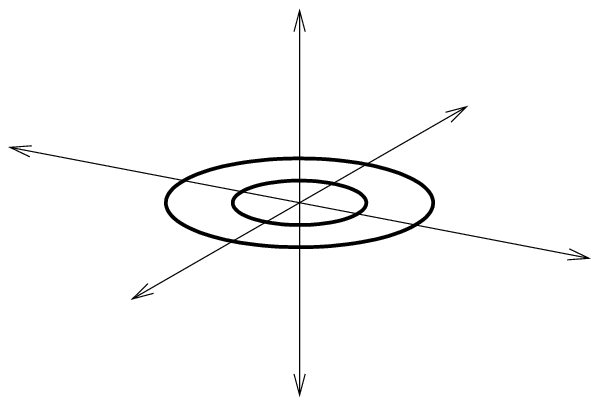}\end{center}\vspace{-11mm}
		& 
\begin{center} \vspace{-0mm}
		Two concentric vortex rings in the $xy$-plane.
\vspace{-1mm} \end{center}
\\ \hline
\end{tabular}
\end{center}
}
\vspace{-3mm}
\caption{\small Configurations of the nodes and vortex rings
 of the Higgs field for $m=4$,
 $n=3$ solutions.}
\end{table}

At $\lambda_{c_1}=0.491$ a first bifurcation arises,
where a pair of new $\lambda$-branches appears,
possessing higher mass and a mixed node structure,
with two outer nodes on the symmetry axis
and a vortex ring in the $xy$-plane \cite{KNS}.
With increasing $\lambda$ the 
solutions on the lower mass branch retain their node structure, 
while the upper mass branch solutions 
transform first into MAC solutions with 
four isolated nodes on the symmetry axis,
and then back to the mixed node structure.

Concerning the mass of these solutions,
a transition between the fundamental branch and the new lower mass branch
is observed at $\lambda_m^1 \approx 59.8$,
making the new lower branch solutions energetically favourable.
As for the $m=2$ and $m=3$ solutions, 
it appears energetically advantageous for the $m=4$ solutions
to exchange vortex rings for isolated nodes,
when the scalar coupling is large.

We therefore argued before \cite{KNS}, that a further 
bifurcation was likely to exist, where
another pair of branches would appear, representing MAC solutions
with four isolated nodes on the symmetry axis.
The solutions on this (conjectured) lower (mass) branch should then
become the energetically most favourable configurations
for large values of $\lambda$.

Indeed, as seen in Fig.~\ref{f-11}, such a second bifurcation
arises at $\lambda_{c_2}=1.288$, and a pair of MAC solutions appears.
With increasing $\lambda$ the
solutions on the lower mass branch retain the MAC node structure,
while the node structure of the upper mass branch solutions
changes and forms two vortex rings.
Concerning the mass of the solutions we also observe
the conjectured transition: 
beyond $\lambda_m^2 \approx 45.2$ 
the new MAC branch has the lowest mass 
and thus becomes the energetically favoured branch.

Surprisingly,
at $\lambda_{c_3}=3.962$ a third bifurcation arises
and a third pair of branches appears
with a new type of node structure.
These solutions possess two concentric vortex rings in the $xy$-plane.
Consequently their mass is considerably higher than the mass 
of the other configurations.

But these two branches of solutions have another interesting feature.
Unlike the other branches of solutions,
this third pair of branches exists only in a relatively small
range of the scalar coupling.
At the bifurcation point $\lambda_{c_4}=66.354$, the pair of branches
merges and disappears again.

Since this third pair of branches of solutions appeared unexpectedly,
the existence of further branches of flat-space solutions 
in certain ranges of the scalar coupling seems possible.

\subsubsection{Gravitating solutions}

When gravity is coupled, branches of gravitating solutions arise
from all of these $m=4$, $n=3$ flat-space configurations.
But as seen above, such $\alpha$-branches can bifurcate many times,
leading to a plethora of gravitating solutions
for larger values of $\lambda$.
We therefore refrain from obtaining the complete picture
for these $m=4$, $n=3$ solutions,
and present only results for a few selected values of $\lambda$.
In particular, we exhibit in Fig.~\ref{f-12}
and Fig.~\ref{f-13} the mass, the node structure
and the metric function value $l(0)$ versus $\alpha$,
for sets of gravitating $m=4$, $n=3$ solutions
with $\lambda=1$, $2$ and $\lambda=4.5$.

\begin{figure}[p!]
\lbfig{f-12}
\begin{center}
\hspace{0.0cm} (a)\hspace{-0.6cm}
\includegraphics[height=.25\textheight, angle =0]{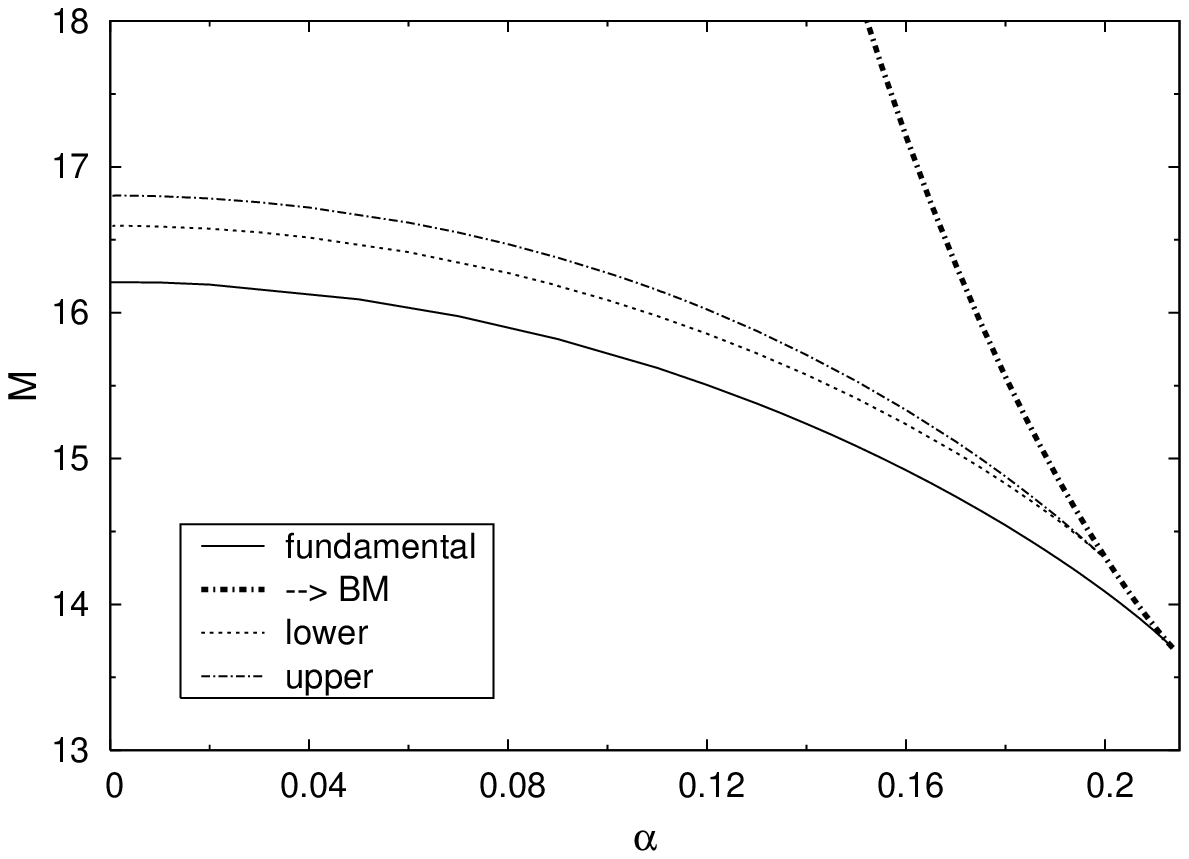}
\hspace{0.5cm} (b)\hspace{-0.6cm}
\includegraphics[height=.25\textheight, angle =0]{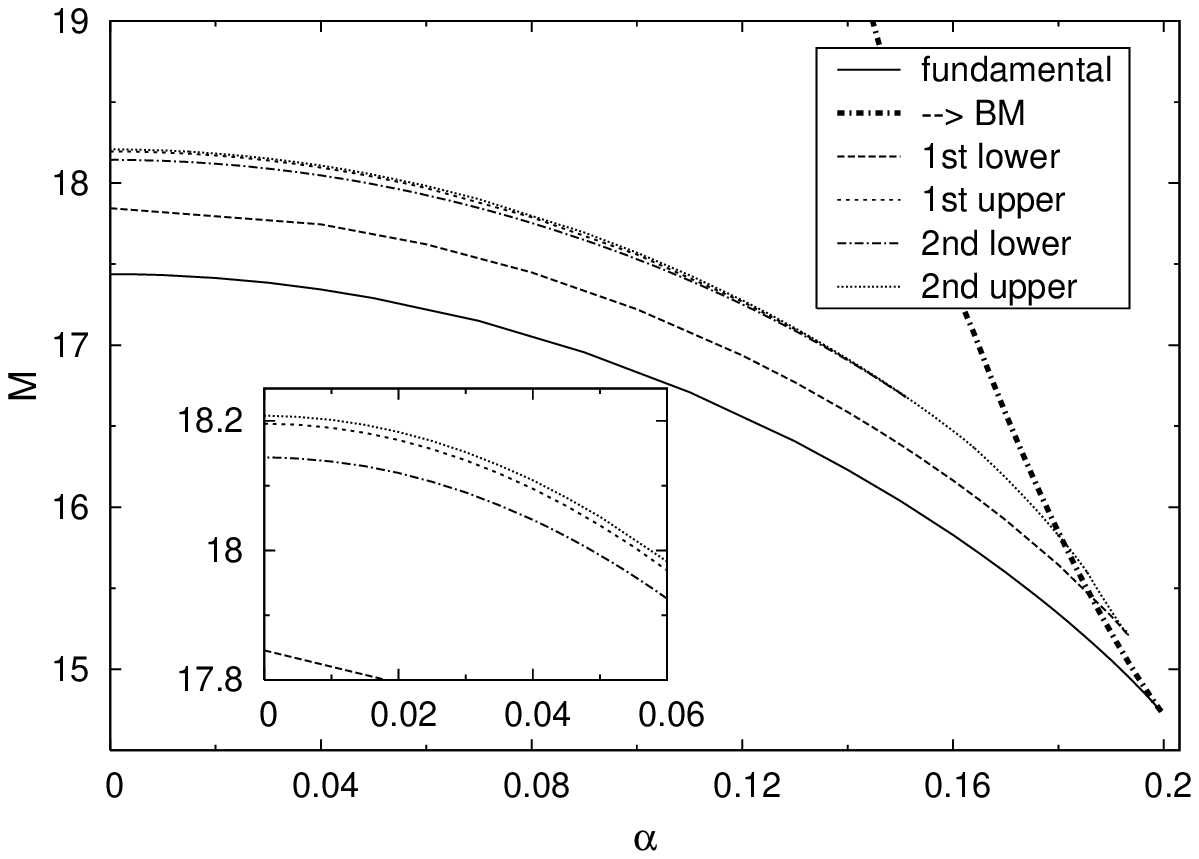}
\\
\hspace{0.0cm} (c)\hspace{-0.6cm}
\includegraphics[height=.25\textheight, angle =0]{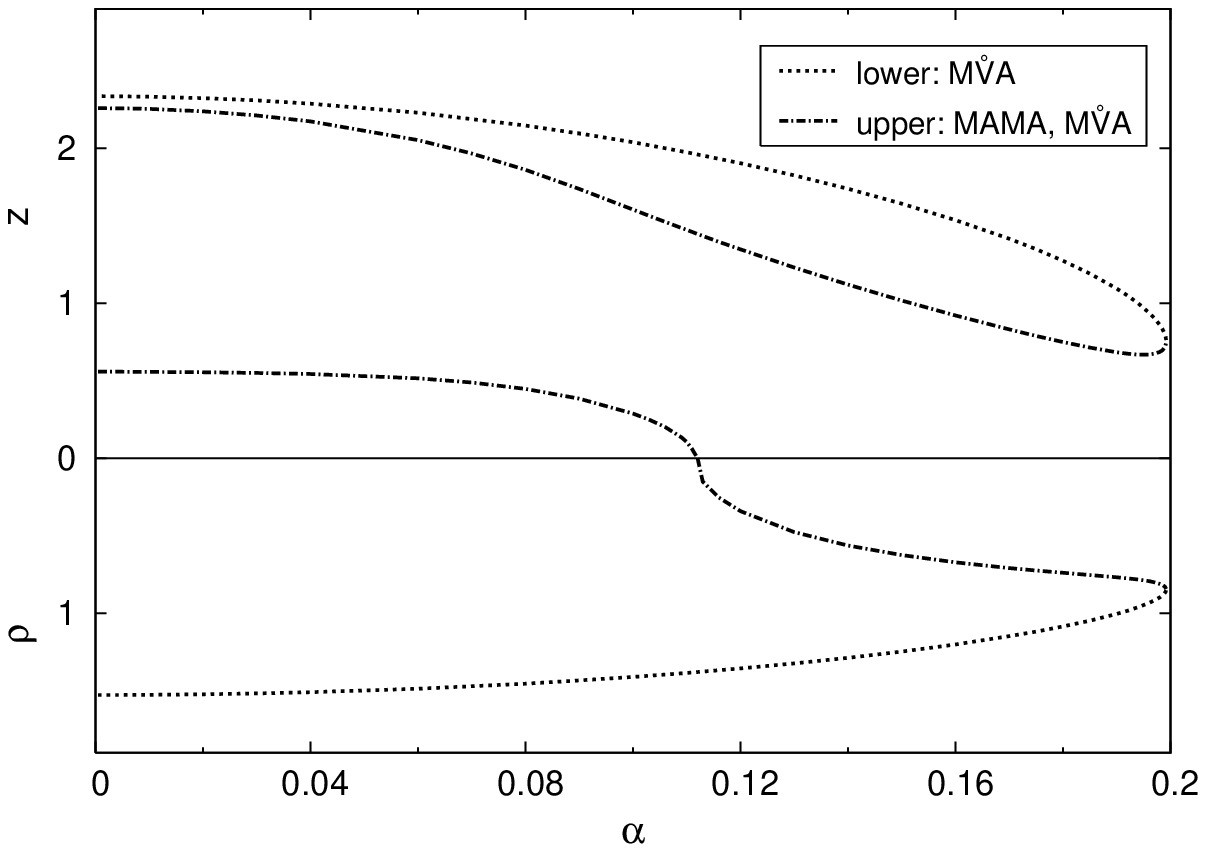}
\hspace{0.5cm} (d)\hspace{-0.6cm}
\includegraphics[height=.25\textheight, angle =0]{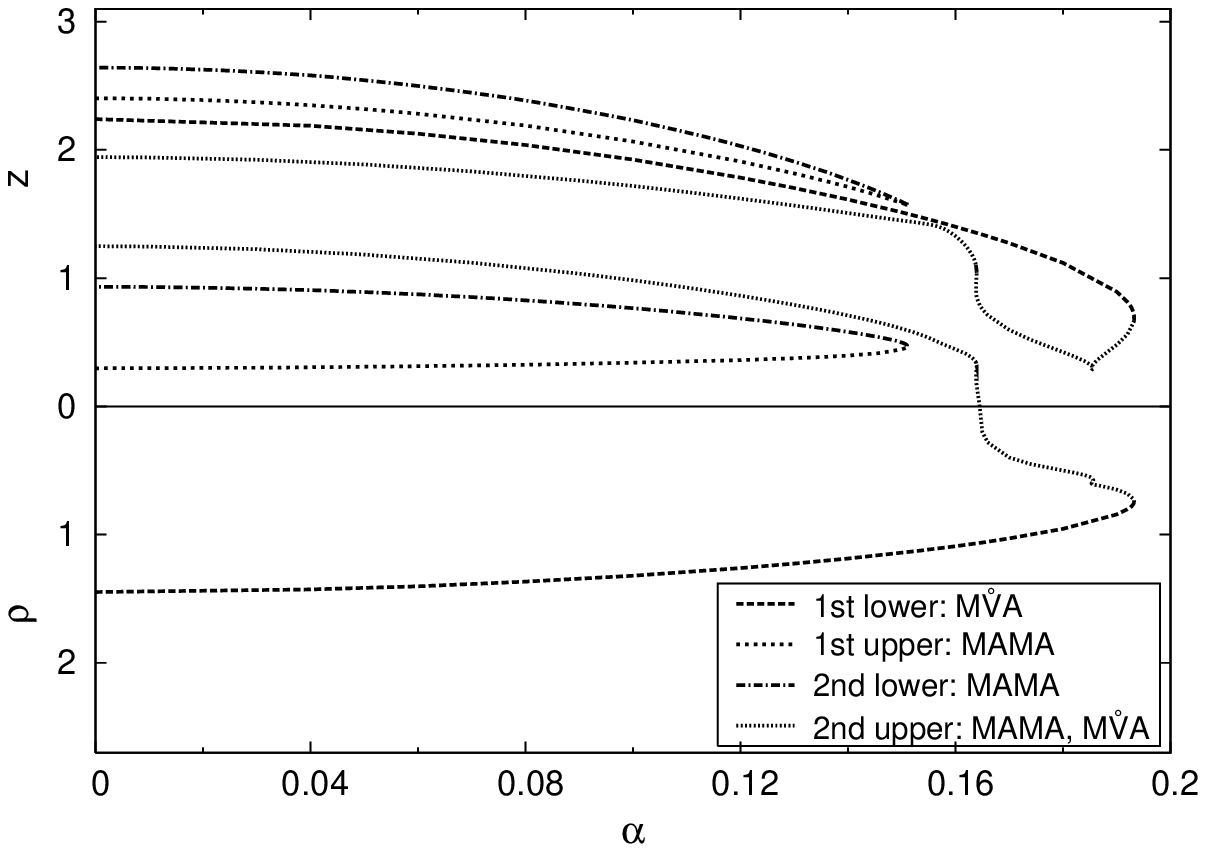}
\\
\vspace{0.5cm}
\hspace{0.0cm} (e)\hspace{-0.6cm}
\includegraphics[height=.25\textheight, angle =0]{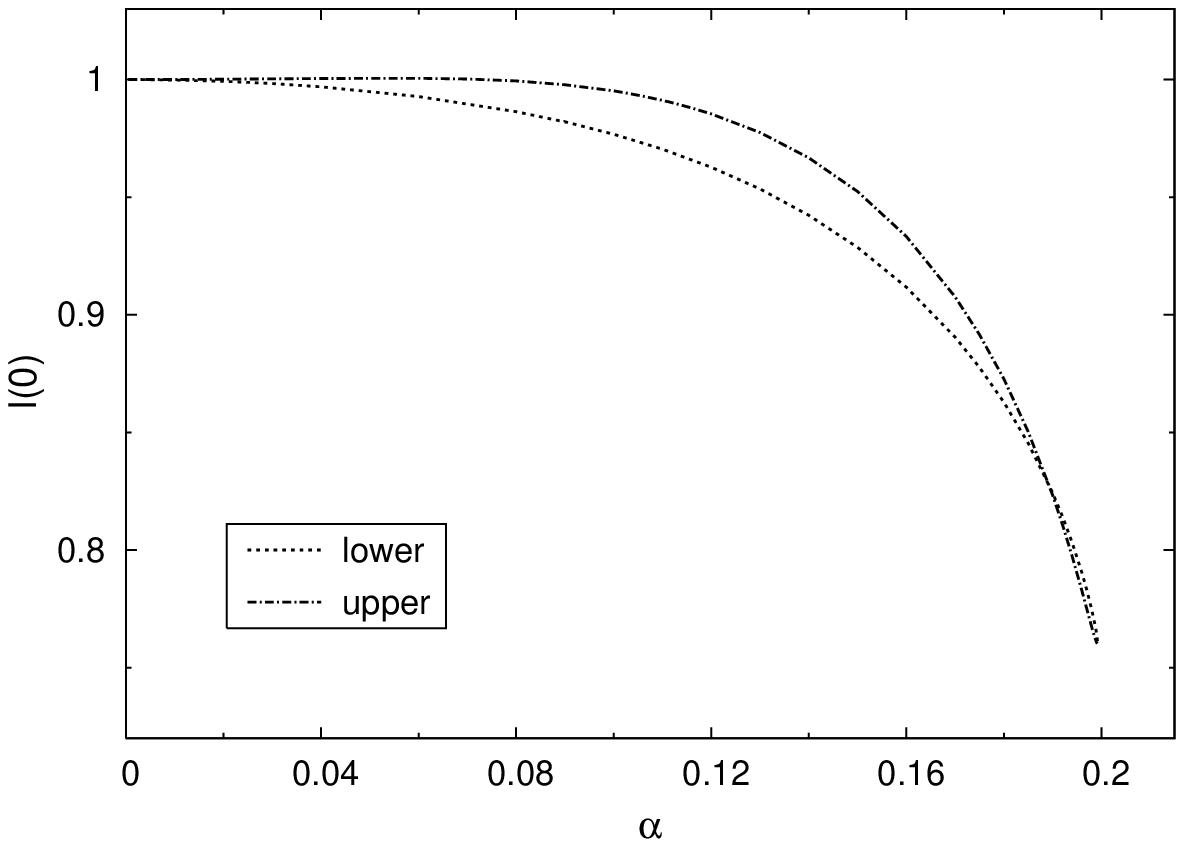}
\hspace{0.5cm} (f)\hspace{-0.6cm}
\includegraphics[height=.25\textheight, angle =0]{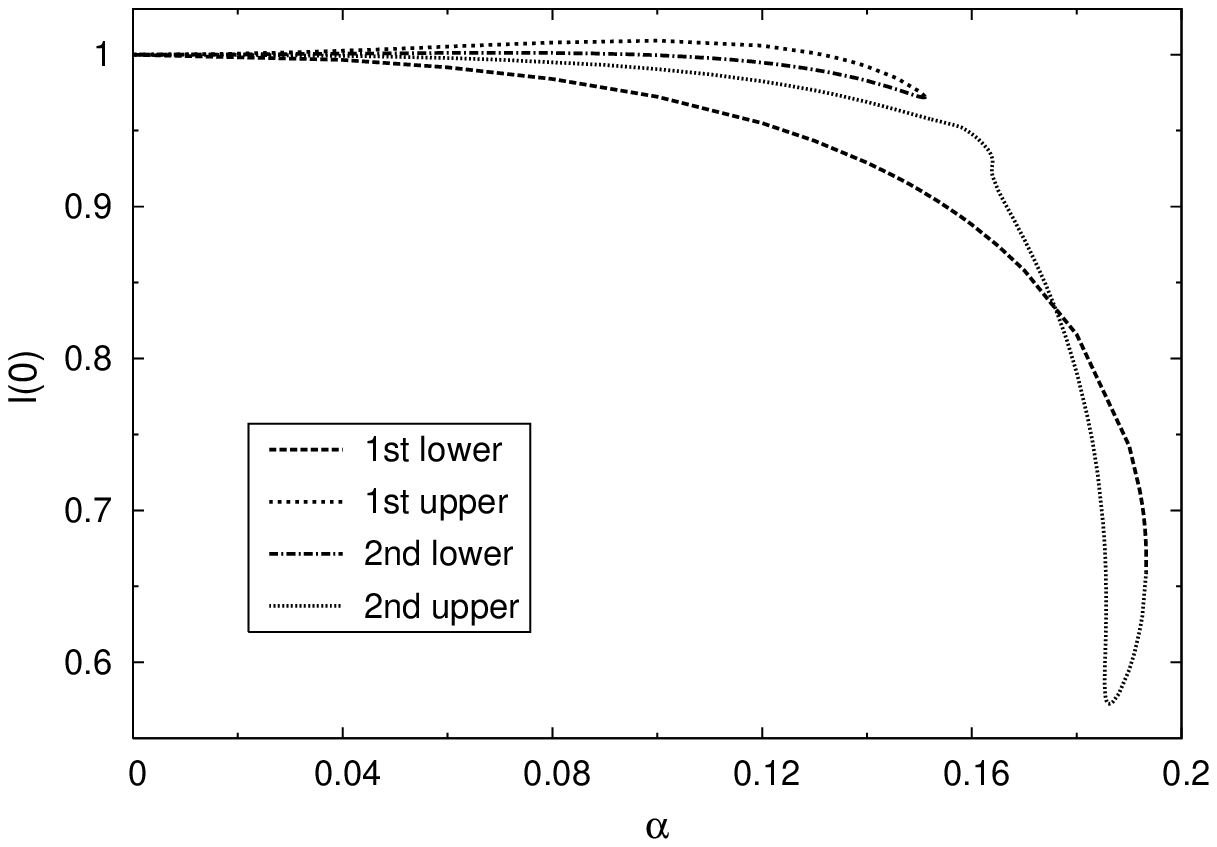}
\end{center}
\vspace{-0.5cm}
\caption{\small
The mass $M$ of $m=4$, $n=3$ solutions
versus $\alpha$ for $\lambda=1$ (a), $\lambda=2$ (b).
The location of the isolated nodes $z$ and vortex rings $(\rho, z)$
of the Higgs field for the same
set of solutions for $\lambda=1$ (c), $\lambda=2$ (d),
and likewise the value of the metric function at the origin $l(0)$
for $\lambda=1$ (e), $\lambda=2$ (f).
(See Table~3  for the notation of the node and vortex ring configurations of
the Higgs field.)}
\end{figure}

The first value $\lambda=1$ is chosen in the interval
$\lambda_{c_1} < \lambda < \lambda_{c_2}$,
after the first bifurcation. So there are three flat-space solutions,
each giving rise to an $\alpha$-branch.
The $\alpha$-branch emerging from the fundamental flat-space solution
connects via a second $\alpha$-branch to the generalized BM solution,
while the $\alpha$-branches emerging from the first new pair
of flat-space solutions merge with each other.
By considering the evolution of the nodes on the new upper mass branch
with $\lambda$ and with $\alpha$, respectively,
we note again, that an increase of the scalar coupling can lead to
an analogous effect as the decrease of the gravitational coupling.
In both cases, the MAC node structure can change into a mixed
node structure at a critical value of the coupling.

\begin{figure}[p!]
\lbfig{f-13}
\begin{center}
\hspace{0.0cm} (a)\hspace{-0.6cm}
\includegraphics[height=.25\textheight, angle =0]{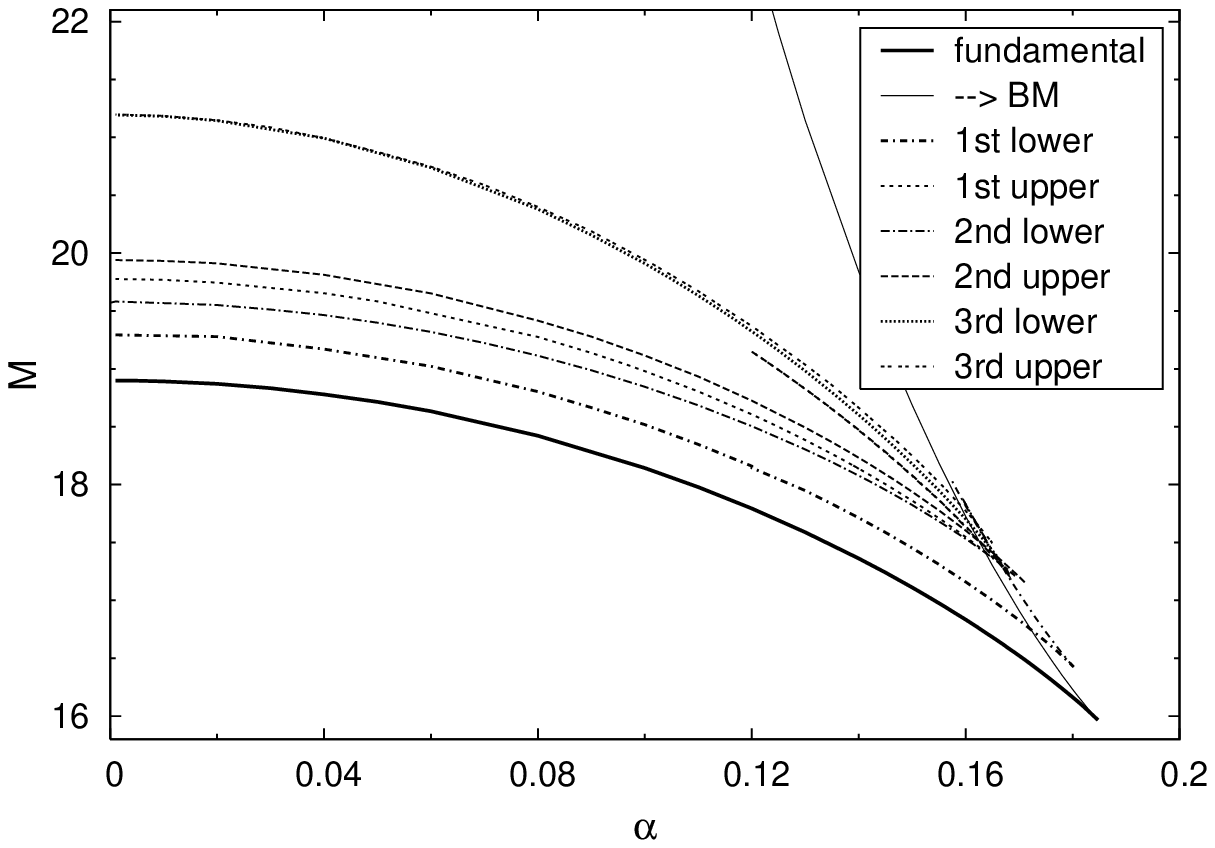}
\hspace{0.5cm} (b)\hspace{-0.6cm}
\includegraphics[height=.25\textheight, angle =0]{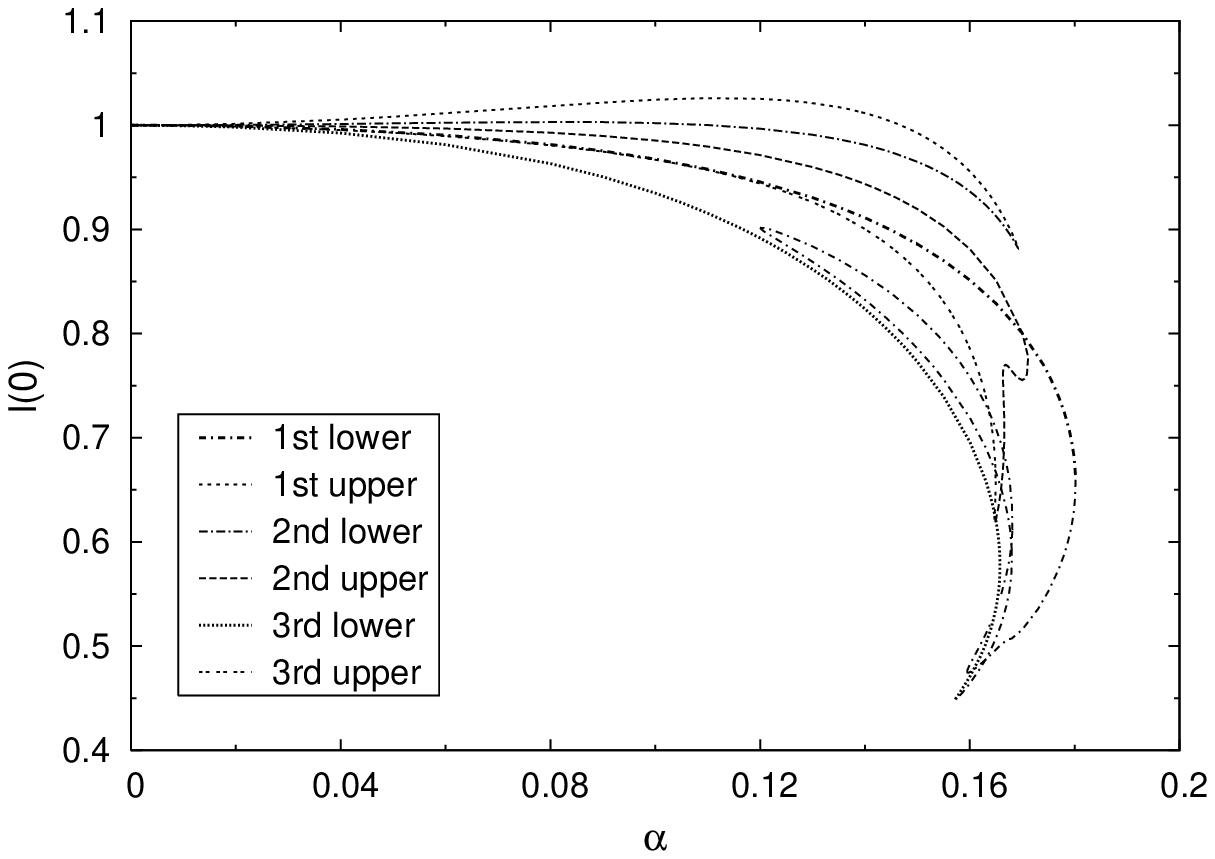}
\\
\vspace{0.5cm}
\hspace{0.0cm} (c)\hspace{-0.8cm}
\includegraphics[height=.22\textheight, angle =0]{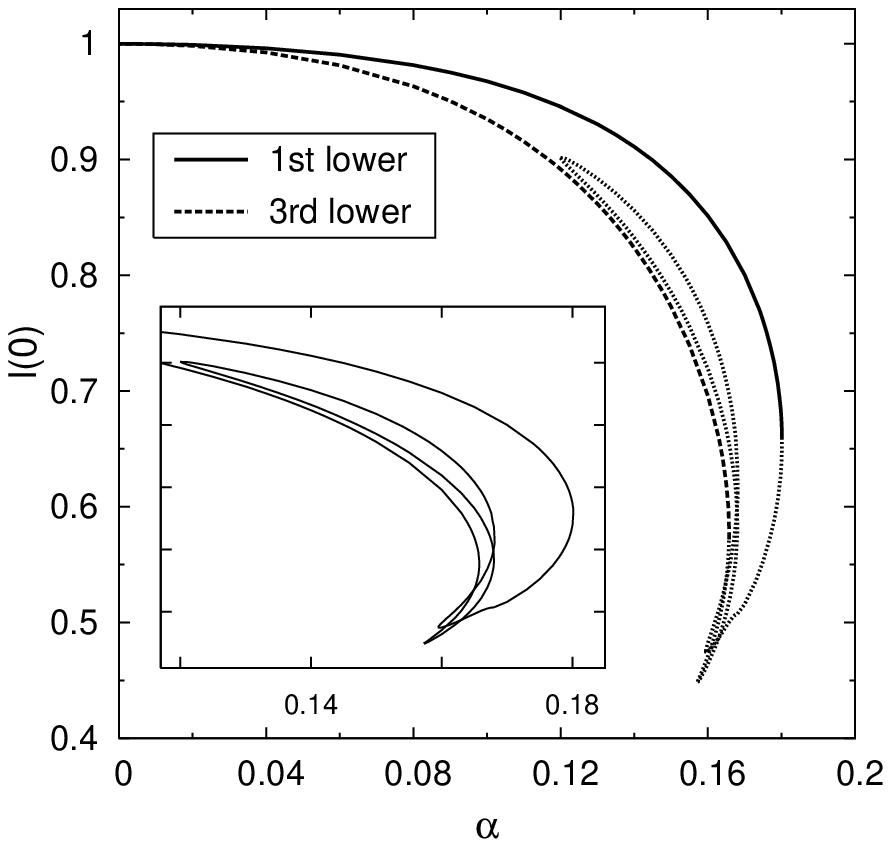}
\hspace{0.1cm} (d)\hspace{-0.8cm}
\includegraphics[height=.22\textheight, angle =0]{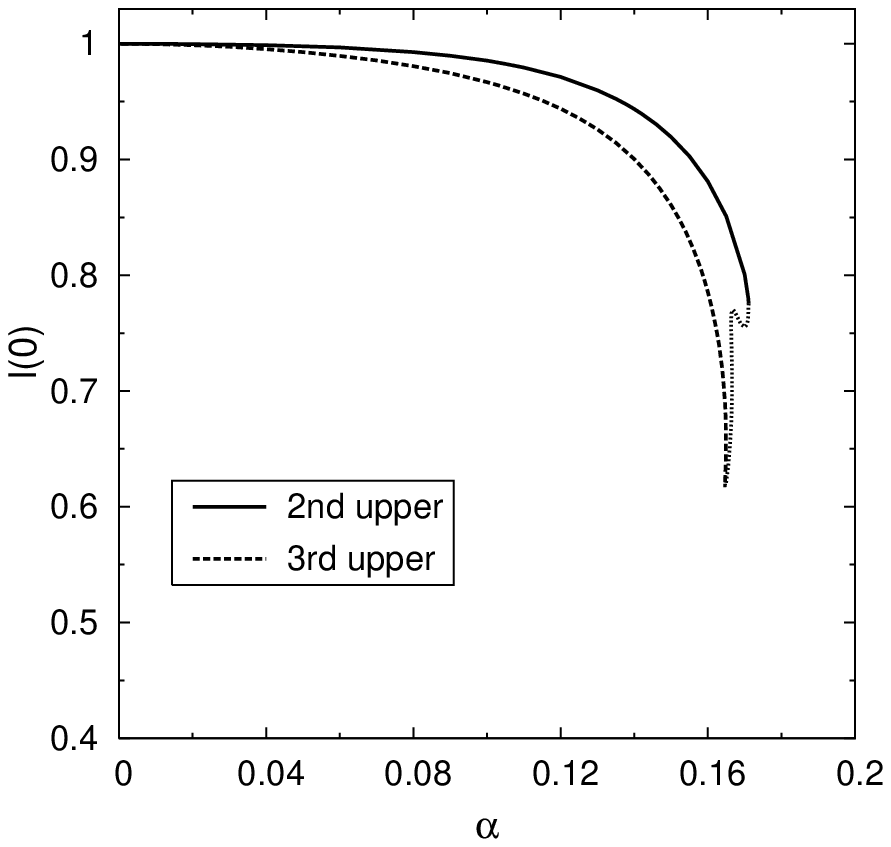}
\hspace{0.1cm} (e)\hspace{-0.8cm}
\includegraphics[height=.22\textheight, angle =0]{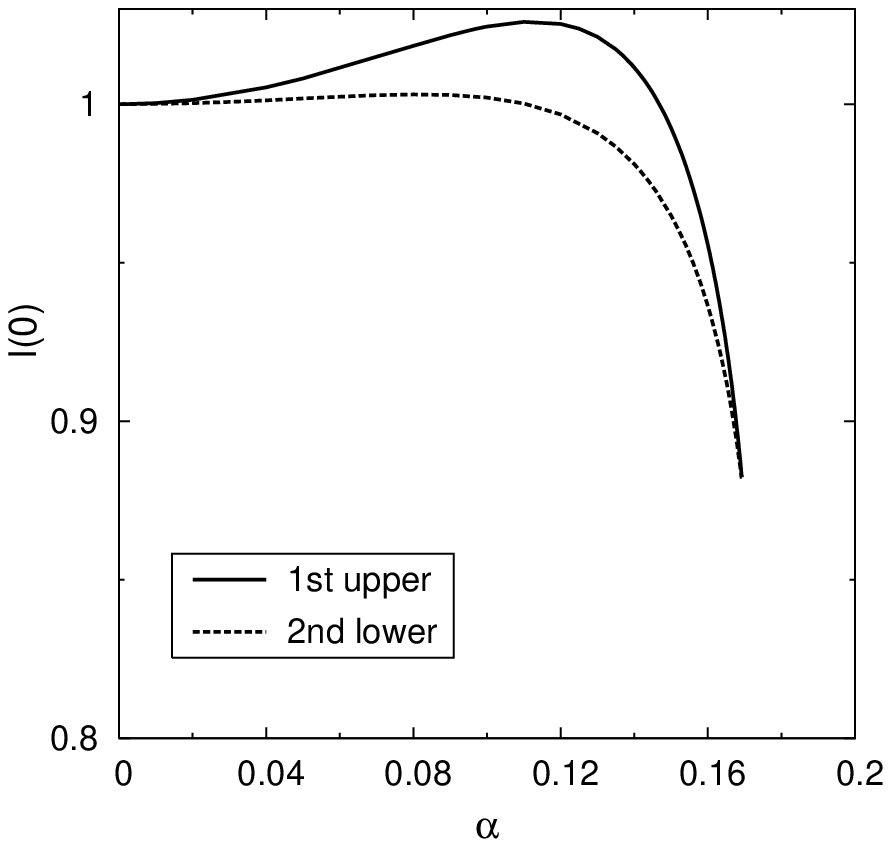}
\\
\vspace{0.5cm}
\hspace{0.0cm} (f)\hspace{-0.8cm}
\includegraphics[height=.22\textheight, angle =0]{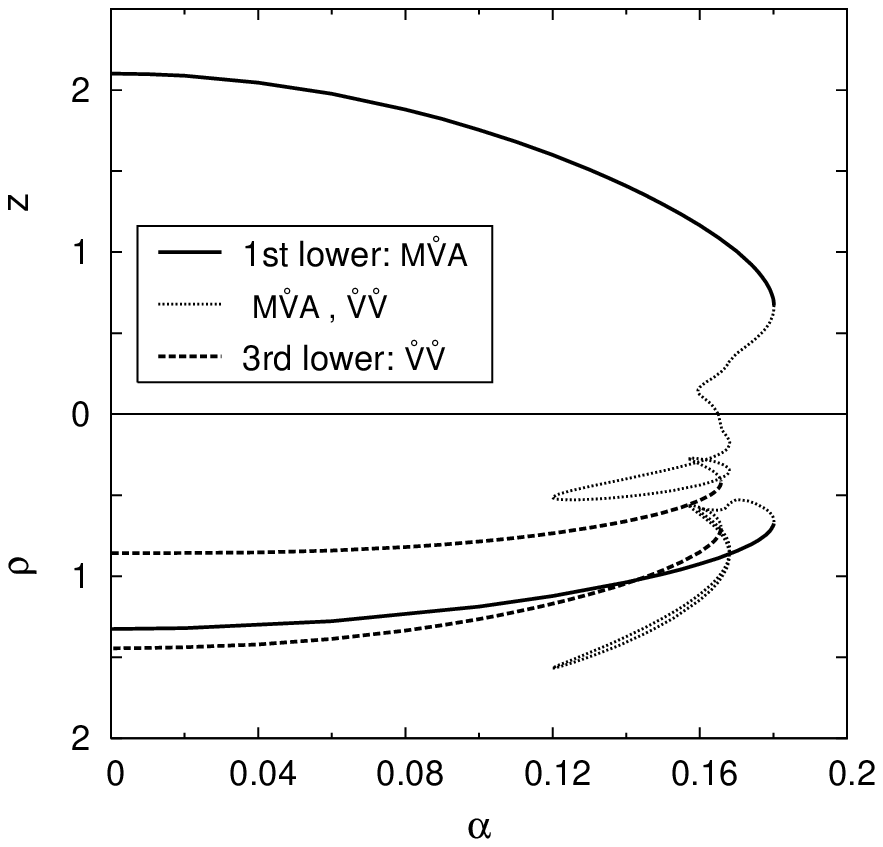}
\hspace{0.1cm} (g)\hspace{-0.8cm}
\includegraphics[height=.22\textheight, angle =0]{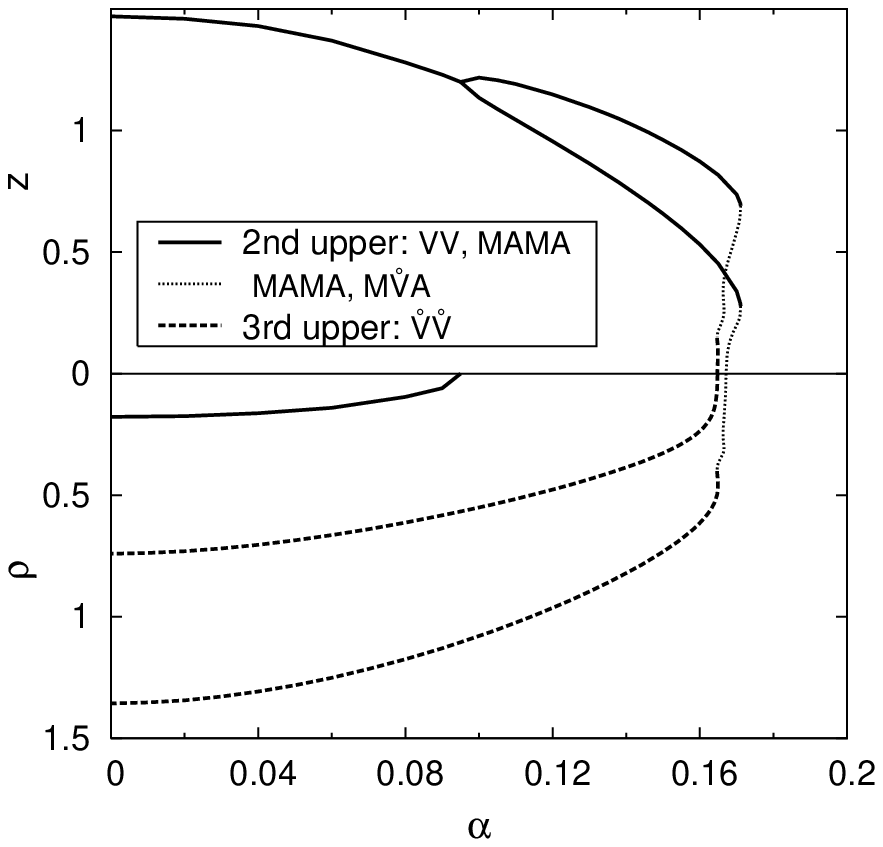}
\hspace{0.1cm} (h)\hspace{-0.8cm}
\includegraphics[height=.22\textheight, angle =0]{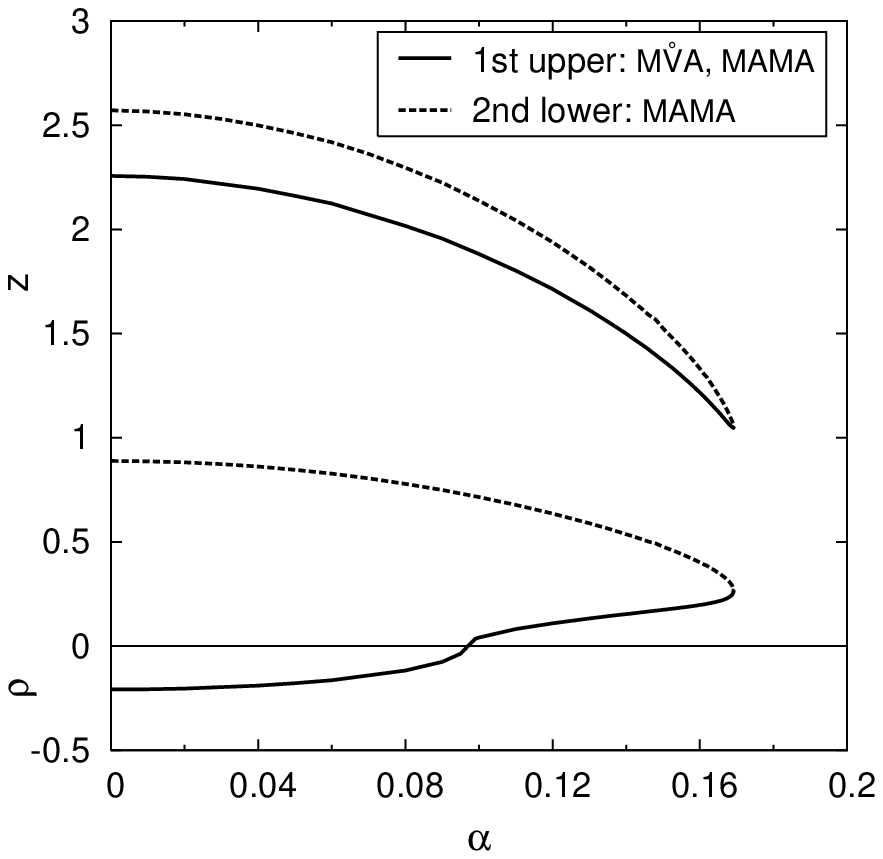}
\end{center}
\vspace{-0.5cm}
\caption{\small
The mass $M$ of $m=4$, $n=3$ solutions
versus $\alpha$ for $\lambda=4.5$ (a).
The value of metric function at the origin, $l(0)$, for the complete 
set of solutions
related to the new flat space $\lambda$-branches (b)
and for the first (c), second (d) and third (e) pair of  
$\alpha$-branches emerging from the
flat space $\lambda$-branches.
The location of the isolated nodes $z$ and vortex rings $(\rho, z)$
 of the Higgs field for the solutions
emerging the first (d), second (e) and third (f) pair of $\alpha$-branches.
(See Table~3  for the notation of the node and vortex ring configurations of
the Higgs field.)}
\end{figure}

The second value $\lambda=2$ resides in the interval 
$\lambda_{c_2} < \lambda < \lambda_{c_3}$,
after the second bifurcation. Thus there are five flat-space solutions,
each giving rise to an $\alpha$-branch.
As in the previous case,
the $\alpha$-branch emerging from the fundamental flat-space solution
connects via a second $\alpha$-branch to the generalized BM solution,
and this feature of the fundamental branch solutions is retained,
independent of $\lambda$.

However, the $\alpha$-branches emerging from the first new pair
no longer merge with each other. 
Instead the first new lower branch merges with the second new upper branch,
and the first new upper branch merges with the second new lower branch.
The reason for this can be inferred from the evolution
of the $\alpha$-branches with $\lambda$,
since the emergence of the second new pair of flat-space solutions
can be traced back to the appearance of a bifurcation
on one of the $\alpha$-branches of the first new pair of solutions,
for a value of $\lambda > 1$.
As $\lambda$ then increases, the two additional $\alpha$-branches
grow in size, while their bifurcation point $\alpha_{\rm min}$
moves towards smaller values of $\alpha$, 
reaching $\alpha=0$ at $\lambda_{c_2}$.
Thus beyond $\lambda_{c_2}$, the $\alpha$-branches of the first
pair of solutions are disconnected from each other
and merge instead with those of the second pair.

As seen in the figure, at $\lambda=2$
the solutions on the $\alpha$-branch emerging from the first lower mass solution
possess a mixed node structure, those connected to the second upper
mass solution have a node structure changing from MAC to mixed,
while the solutions on the other branches have MAC structure.

The third value of $\lambda$
is chosen in the interval $\lambda_{c_3} < \lambda < \lambda_{c_4}$, 
where three pairs of flat-space branches are present.
Their corresponding $\alpha$-branches are exhibited in Fig.~\ref{f-13}.
At this value of $\lambda$ a plethora of $\alpha$-branches is present,
caused by numerous further bifurcations.
The node structures of these solutions range from MACs to
double vortex rings.
Thus one may speculate about the existence of solutions
with still more complicated node structures,
which might arise from new $\alpha$-branches,
as the scalar coupling is further increased.

\section{Conclusions}

We have investigated static axially symmetric solutions of 
$SU(2)$ Einstein-Yang-Mills-Higgs theory,
representing monopole-antimonopole pairs, chains, vortex rings,
and new types of configurations with mixed node structure.
Such new configurations appear for larger values of the scalar coupling,
when the subtle interplay between repulsive and attractive forces
allows for more than one non-trivial equilibrium configuration
of these systems.

In flat space, at critical values of the scalar coupling,
bifurcations arise, where pairs of new branches of solutions appear,
which possess a different node structure than the 
solutions on the fundamental branches. 
In particular, new solutions appear, where
vortex rings present in the fundamental solutions,
are replaced by isolated nodes on the symmetry axis.
For high values of $\lambda$
the new solutions with monopole-antimonopole chain
structure have the lowest mass.

While we have studied here in detail
only the monopole-antimonopole systems with $m=2,~3,~4$ and $n=3$,
we conjecture, that this phenomenon is not restricted
to these particular systems but that it is of a more general
nature, implicating an enormous richness of configuration space
for high values of $\lambda$ and larger $m$ and $n$.
For $m=4$ and $n=3$, for instance, we observed 
already seven distinct flat-space solutions
in a certain range of the scalar coupling.

The coupling to gravity leads to additional attraction 
between the components of these monopole-antimonopole systems,
shifting the balance of forces and thus affecting the possible
equilibrium configurations.
Whereas for small scalar coupling only two gravitating
branches of solutions are associated with a flat-space solution,
there is a surge of further branches
when the scalar coupling becomes large.

Particular of these branches can be shown to give
rise to new pairs of flat-space solutions. Thus 
study of the Einstein-Yang-Mills-Higgs
configuration space gives valuable insight
into the configuration space of Yang-Mills-Higgs theory.
Indeed, by exploiting this interrelationship, we were able
to find the new flat-space solutions presented.

But the Einstein-Yang-Mills-Higgs configurations 
are also related to Einstein-Yang-Mills solutions.
First of all,
in the limit $\alpha \rightarrow 0$, one of the
$\alpha$-branches of a particular set of solutions,
characterized by $m$, $n$ and $\lambda$,
always connects to a generalized Bartnik-McKinnon solution
\cite{BM,KK,IKKS}.
But another connection arises in the limit of
infinite scalar coupling.
In this limit the bifurcation point $\alpha_{\rm min}$
of two of the $\alpha$-branches
is seen to approach $\alpha=0$, while the solutions at $\alpha_{\rm min}$
approach the respective generalized Bartnik-McKinnon solution.

\vspace{0.5cm}
{\bf Acknowledgement}

We would like to acknowledge valuable discussions with Burkhard Kleihaus,
Eugen Radu, Tigran Tchrakian, and Michael Volkov.
Ya.S. would like to thank the Department of Mathematical Physics, NUI
Maynooth, and the SPhT CEA/Saclay, for their kind hospitality.

\labelsep20pt

\end{document}